\documentclass[12pt,a4paper]{article}            
 \usepackage[skins,theorems]{tcolorbox}
\tcbset{highlight math style={enhanced,
  colframe=red,colback=white,arc=0pt,boxrule=1pt}}
  \usepackage[bookmarksopen, bookmarksnumbered, bookmarksopenlevel=2]{hyperref}
  \usepackage{tikz}
  \usepackage{tikz-3dplot}
 \usetikzlibrary{calc}
 \usetikzlibrary{decorations} %
 \usepackage[UKenglish]{babel}
 \usepackage[toc,page]{appendix}
 \usepackage{amsmath}
 \usepackage{amssymb}
 \usepackage{graphicx}
 \usepackage{hhline}
 \usepackage[bf]{caption}
\usepackage{cite}
\usepackage[vcentermath]{youngtab}
\usepackage{geometry}
\usepackage{slashed}
\usepackage{color}
\usepackage{stackrel}
\usepackage{tikz-cd} 
\usepackage{cancel} 

\usepackage{empheq}
\usepackage{arydshln}

\newcommand{\bqa}{\begin{eqnarray}}
\newcommand{\eqa}{\end{eqnarray}}

\newcommand{\nn}{\nonumber}

\def\la{\lambda}

\def\gA{g_{\rm IIA}}
\def\gB{g_{\rm IIB}}

\def\MCS{{\cal M}_{CS}}
\def\QMK{{\cal QM}_{K}}
\def\hT{{\hat T}}

\def\RRC{{\mathbf C}}

\hypersetup{
    pdftitle={},
    pdfauthor={},
    pdfsubject={}
}
\numberwithin{equation}{section}
\numberwithin{table}{section}

\makeatletter

\newtheorem{theorem}{Theorem}

\newtheorem{Lemma}{Lemma}
\newtheorem{conjecture}{Proposal}

\newtheorem{proposition}{Proposition}

\newcommand{\be}{\begin{equation}}
\newcommand{\ee}{\end{equation}}
\newcommand{\beq}{\begin{equation}}
\newcommand{\eeq}{\end{equation}}
\newcommand{\ba}{\begin{aligned}}
\newcommand{\ea}{\end{aligned}}

\newcommand{\bea}{\begin{eqnarray}}
\newcommand{\eea}{\end{eqnarray}}

\newcommand{\cO}{\mathcal{O}}

\newcommand{\cE}{\mathcal{E}}

\newcommand{\cC}{\mathcal{C}}

\newcommand{\cS}{\mathcal{S}}

\newcommand{\cB}{\mathcal{B}}

\newcommand{\cI}{\mathcal{I}}

\newcommand{\cV}{\mathcal{V}}

\def\Im{\mathop{\mathrm{Im}}\nolimits}

\def\unit{{1\kern-.65ex {\rm l}}}
\def\1{{1\kern-.65ex {\rm l}}}

\def\IZ{\mathbb{Z}}
\def\IP{\mathbb{P}}

\def\CE{{\cal E}}

\def\CI{{\cal I}}

\def\CL{{\cal L}}

\def\CN{{\cal N}}
\def\CO{{\cal O}}

\def\IR{{\mathbb{R}}}

\newcount\hour \newcount\minute
\hour=\time \divide \hour by 60
\minute=\time
\def\now{%
\ifnum \hour<13
  \ifnum \hour=0 \advance \hour by 12 \number\hour:\else \number\hour:\fi%
     \ifnum \minute<10 0\fi%
     \number\minute%
\ A.M.%
\else \advance \hour by -12 \number\hour:%
  \ifnum \minute<10 0\fi%
  \number\minute%
  \ P.M.%
\fi%
}

\def\fnote#1#2{\begingroup\def\thefootnote{#1}\footnote{#2}
     \addtocounter{footnote}{-1}\endgroup}

\makeatother

\begin{document}

\begin{flushright}
{\tt\normalsize CERN-TH-2019-159}\\
\end{flushright}

\vskip 40 pt
\begin{center}
{\large \bf Emergent Strings from Infinite Distance Limits 
} 

\vskip 11 mm

Seung-Joo Lee${}^{1}$, Wolfgang Lerche${}^{1}$,
and Timo Weigand${}^{1,2}$

\vskip 11 mm

\small ${}^{1}${\it CERN, Theory Department, \\ 1 Esplande des Particules, Geneva 23, CH-1211, Switzerland} \\[3 mm]
\small ${}^{2}${\it PRISMA Cluster of Excellence and Mainz Institute for Theoretical Physics, \\
Johannes Gutenberg-Universit\"at, 55099 Mainz, Germany}

\fnote{}{seung.joo.lee, wolfgang.lerche,  
timo.weigand @cern.ch}

\end{center}

\vskip 7mm

\begin{abstract}

As a refinement of the Swampland Distance Conjecture,
we propose that a quantum gravitational theory in an infinite distance limit of its moduli space either decompactifies,
or reduces to an asymptotically tensionless, weakly coupled string theory.
We support our claim by classifying, as special cases, the behaviour of M-Theory and Type IIA string theory compactifications on Calabi-Yau 
three-folds at infinite distances in K\"ahler moduli space.

The analysis comprises three parts: We first classify the possible infinite distance limits in the classical K\"ahler moduli space of a Calabi-Yau three-fold. Each such limit at finite volume is characterized by a universal fibration structure, for which the generic fiber shrinking in the limit is either an elliptic curve, a K3 surface, or an Abelian surface. 

In the second part we focus on M-Theory and investigate the nature of the towers of 
asymptotically massless states that arise from branes wrapped on the shrinking fibers.
Depending on which of the three classes of fibrations are considered, we obtain decompactification to F-Theory, or a theory with a unique asymptotically tensionless, weakly coupled heterotic or Type II string, respectively. 
The latter probes a dual D-manifold which is in general non-geometric. In addition to the intrinsic string excitations, towers of states from M2-branes along non-contractible curves become light and correspond to further wrapping and winding modes of the tensionless heterotic or Type II string.

In the third part of the analysis, we consider Type IIA string theory on Calabi-Yau three-folds and show that quantum effects obstruct taking finite volume infinite distance limits in the K\"ahler moduli space.
The only possible infinite distance limit which is not a decompactification limit involves   K3-fibrations with string scale fiber volume and gives rise to an emergent tensionless heterotic string.

\end{abstract}

\vfill

\thispagestyle{empty}
\setcounter{page}{0}
\newpage

\tableofcontents

\thispagestyle{empty}
\setcounter{page}{1}


\section{Introduction and Summary} \label{sec_Intro}

Investigating a physical theory at the boundaries of its moduli space oftentimes reveals interesting insights about its dynamics.
For example, in a supersymmetric theory with gauge degrees of freedom, strong coupling regimes are typically located at finite distance loci of the moduli space. 
In such regimes,
either a dynamically generated scale far below the Planck scale emerges and the physics below this scale decouples from gravity, or the theory approaches a superconformal fixed point decoupled from gravity.
Asymptotic weak coupling regimes, on the other hand, are expected to lie at infinite distances in moduli space.
Studying the dynamics of such limits within a gravitational theory should similarly lead to valuable information about the nature of the degrees of freedom of quantum gravity.

According to the so-called Swampland Distance Conjecture \cite{Ooguri:2006in}, infinite distance limits in a theory  of quantum gravity are necessarily accompanied by an infinite tower of asymptotically massless degrees of freedom.
This conjecture has recently been confirmed quantitatively and further discussed in various different classes of string compactifications \cite{Klaewer:2016kiy,Heidenreich:2016aqi,Palti:2017elp,Heidenreich:2017sim,Heidenreich:2018kpg,Andriolo:2018lvp,Grimm:2018ohb,Blumenhagen:2018nts,Lee:2018urn,Lee:2018spm,Grimm:2018cpv,Gonzalo:2018guu,Corvilain:2018lgw,Lee:2019tst,Joshi:2019nzi,Marchesano:2019ifh,Font:2019cxq,Lee:2019xtm,Erkinger:2019umg,Grimm:2019wtx}.
Apart from being interesting by itself, the appearance of a light  tower of states at infinite distance is also linked to other recent conjectures on the nature of quantum gravity such as  \cite{Ooguri:2018wrx,Klaewer:2018yxi,Lust:2019zwm,Kehagias:2019akr}. Detailed reviews of the Swampland Programme are provided in \cite{Brennan:2017rbf,Palti:2019pca}.

An obvious example for a tower of asymptotically massless states at infinite distance are the Kaluza-Klein (KK) states, which become light as some of the directions in the compactification space of a higher-dimensional theory are taken to be large such that the total internal volume diverges (in units of the higher-dimensional Planck scale). The importance of these KK states has been stressed in the swampland context in particular in \cite{Ooguri:2006in,Blumenhagen:2018nts,Font:2019cxq}. 
Such a decompactification limit to a higher dimensional theory leads to a diverging Planck scale as viewed from the original theory, and indeed the dominant Kaluza-Klein spectrum is of purely field theoretical nature.
To go beyond situations dominated by such a tower of light, field theoretical KK excitations we must, by contrast, focus on {\it equi-dimensional infinite distance limits} 
leading to a gravitational theory in the same number of effective space-time dimensions 
as the starting point of the trajectory in moduli space. Only then can
the dominant lowest mass states be interpreted in terms of an intrinsically gravitational theory.

\subsection{Emergent strings}

What could such a weakly coupled theory in presence of gravity be? One natural expectation would be that this is a string theory and the tower of states appearing are related, one way or another, to a weakly coupled and light fundamental string. 
Clearly this pattern has long been familiar in ten dimensions, however more recently it
has been observed in \cite{Lee:2018urn} also for the most general\footnote{Subject to that requirement that the volume of the internal space is kept fixed thus to avoid decompactification.} weak coupling limits at infinite distance of
geometric F-Theory/heterotic compactifications to 
  six dimensions, with $N=(1,0)$ supersymmetry. Here the light modes at infinite distance turn out to be the excitation modes of a solitonic heterotic string,
   which turns into a fundamental heterotic string upon passing to a weakly coupled duality frame. A similar pattern has been found, barring possible non-perturbative quantum obstructions, in F-Theory compactifications to four dimensions \cite{Lee:2019tst} with $N=1$ supersymmetry.
   Likewise, for Type IIB string theory compactified on a K3 surface to six dimensions, an infinite distance limit in K\"ahler moduli space leads to a light Type II string furnishing a tower of particle excitations~\cite{Lee:2019xtm}. 
   
   In these situations, the limits are equi-dimensional in the sense that the tower of KK states appear parametrically at the same mass scale as the string states, and not, as for a decompactification limit, with masses suppressed by additional powers of 
  the large parameter that governs the large distance limit. The interplay of the KK scale and the mass scales associated with solitonic strings, as well as with domain walls from wrapped branes, has recently been studied in \cite{Font:2019cxq} in the context of Type II compactifications on Calabi-Yau three-folds (and their orientifolds).
  
A most basic characteristic of the spectrum of a weakly coupled string is that it is much denser than a field theoretic KK spectrum, with a level-mass relation given by $M_n^2 \sim n M^2_{\rm string}$ for sufficiently large $n$, rather than by $M_n^2 \sim n^2 M^2_{\rm KK}$. This changes
qualitatively the behaviour of the theory compared to a conventional decompactification limit.
 A question is now if any other type of particle tower could appear at a similar parametric mass scale (up to finite factors), but with a denser spectrum than the high-level KK states.
  From experience with attempts of defining theories of quantum gravity in terms of higher-dimensional fundamental objects different from strings, one might expect that this is not the case. This would suggest that the KK spectrum either dominates the infinite distance limit completely, in which case the limit is not equi-dimensional, or that there emerges a light fundamental string.\footnote{Here and in the remainder of this article the
 term `emergence' is used in a weaker sense than in the `emergence proposals' of \cite{Palti:2017elp,Heidenreich:2017sim,Heidenreich:2018kpg,Grimm:2018ohb,Palti:2019pca}, which relate the polynomial behaviour of couplings at infinite distance in moduli space, to a tower of asymptotically light particles.}
 
In this paper we will provide further support for this picture by classifying all possible equi-dimensional infinite distance limits in the vector multiplet moduli space of M-Theory and Type IIA string theory, compactified on a Calabi-Yau three-fold.
Maybe more surprising than the very fact {\it that} in such limits Type II or heterotic strings emerge, is the way {\it how} they are realized as solitonic objects that become light in suitable regions in moduli space.
As we will see, this hinges upon highly non-trivial geometrical properties of the K\"ahler moduli space. The uniqueness of the emergent weakly coupled strings points to an intriguing level of consistency in the geometry as probed by M- and string theory. The appearance of further infinite towers of particles that can be interpreted as wrapping modes of these emerging strings, can also be independently confirmed in
terms of non-vanishing BPS invariants that figure as Fourier coefficients of certain modular forms.

Encouraged by this circumstantial evidence we make the following

\begin{conjecture}[String emergence at infinite distance]
Any equi-dimensional infinite dist\-ance limit in the moduli space of a $d$-dimensional theory of quantum gravity,
 reduces to a weakly coupled string theory.
In particular, there appears an infinite tower of asymptotically massless states which form the particle excitations of a unique weakly coupled, asymptotically tensionless Type II or heterotic string in $d$ dimensions.
\end{conjecture}

An important point in sharpening the above conjecture from a pure effective field theory perspective concerns the precise definition of the term {\it equi-dimensional limit}.
In the Einstein frame we can always set the value of the Planck scale to one and measure all masses and volumes with respect to this scale.
Partial decompactification occurs if there appears a tower of asymptotically light KK states (with respect to this scale) that is parametrically lighter 
than any other tower of states. In the following we take as our working definition that a KK tower is a tower which for high values of $n$ scales as
\be \label{KKbehav}
\frac{M^2_n}{M^2_{\rm Pl}}   \sim n^2 \frac{M^2_{\rm KK}}{M^2_{\rm Pl}}  \qquad \quad \text{for} \quad  n \to \infty \,.
\ee
Clearly this is the behaviour of field theoretic KK modes associated with the mass operator on a flat internal manifold, and at sufficiently small length scales the details of the internal space should not matter, as far as the spectrum is concerned. 

If the above proposal is true, all infinite distance limits in the classical moduli space of a gravitational theory for which no light fundamental string degrees of freedom emerge must be either (at least partial) decompactification limits, or be obstructed by quantum effects. An example where quantum effects majorly affect the dynamics has been studied already in \cite{Marchesano:2019ifh}. 

Certainly all this is in line with general expectations based on dualities, and in particular conforms with the intuition gained in 10 dimensions \cite{Witten:1995ex} (famously an 11th dimension can arise as a strong coupling decompactification limit). However, a priori it might have been that exotic gravitational theories could arise, 
in analogy with the  zoo of exotic strongly coupled non-critical string and superconformal theories without gravity
 that arise at finite distance loci in the moduli space. 
 For example, one might have discovered the heterotic string via dualities if it had not been known before. Furthermore, the conjecture could easily fail if for instance several Type II or heterotic strings became light at the same parametric rate: In this case we would encounter a highly non-perturbative theory without a clear interpretation in terms of a definite, weakly coupled string theory. That this does not happen in the framework that we will study explicitly is a consequence of some rather intricate uniqueness results concerning the K\"ahler geometry at infinite distance that we will establish in this paper. 

While proving or disproving our conjecture more generally is beyond the scope of this article, we will instead analyze in detail how the conjecture is realized in the vector moduli space of 
M-Theory and Type IIA string compactifications on Calabi-Yau three-folds.
If we consider an equi-dimensional limit in the compactification of a higher dimensional theory, a necessary condition is that the volume of the internal manifold, measured with respect to the higher-dimensional Planck scale,
stays finite in the limit. Otherwise, a tower of KK modes emerges from the large extra dimensions that is parametrically leading.
The criterion of arriving at finite volume is however only necessary, not sufficient:  As is well known,
 towers of BPS particles arising from wrapped branes (as analysed in infinite distance limits in \cite{Grimm:2018ohb,Grimm:2018cpv,Corvilain:2018lgw}) can scale as in (\ref{KKbehav}), and hence can mimic, and so effectively implement, decompactification as well.
We will indeed encounter this phenomenon frequently.

\subsection{Summary of results}

Our analysis comprises three major parts:
The first step consists of a purely mathematical classification of  {\it finite volume limits at infinite distance} in the classical K\"ahler moduli space of Calabi-Yau three-folds; we will denote such spaces summarily by $Y$.
The requirement of finite total volume may also easily be dropped, which then leads
to a complete picture of infinite distance limits (see Section \ref{sec_infvolume}).
As described in Section~\ref{subsec_Kahlercone}, all finite volume limits are characterized by two different types of scaling behavior for the K\"ahler parameters of $Y$, which we call {\it limits of $J$-Class A and $J$-Class B}.
Our analysis is a refinement of the investigation of finite volume limits for K\"ahler three-folds put forward in \cite{Lee:2019tst}, using similar reasoning as previously for K\"ahler surfaces in \cite{Lee:2018urn}.
Though following a very different method, our conclusions are in agreement with the classification of infinite distance limits (generally not of finite volume) in the K\"ahler moduli space of Calabi-Yau three-folds presented in ref.~\cite{Corvilain:2018lgw} (see also \cite{Grimm:2019wtx}), based on the analysis of infinite distance limits in complex structure moduli space given in \cite{Grimm:2018ohb,Grimm:2018cpv}. 

\begin{figure}[t!]
\centering
\includegraphics[width=12cm] {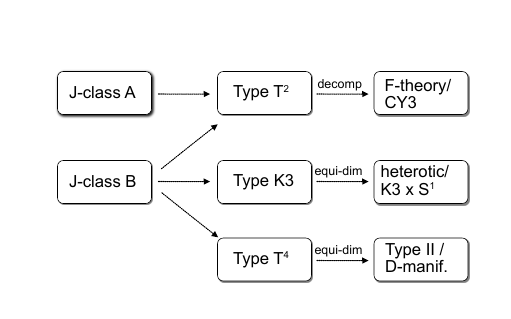}
\caption{Classification of large distance finite volume limits in classical K\"ahler moduli space,
 in terms of fibration types of Calabi-Yau three-folds, $Y$.
Shown are also the corresponding dual stringy geometries that emerge from M-Theory on $Y$ in the respective limits. Thus
only limits of Type K3 and $T^4$ are equi-dimensional. As will be discussed later, for Type IIA strings on $Y$  the situation is
quite different.}
\label{f:FigMlimits}
\end{figure}

In Section \ref{classification-result} we present the main results of our geometric analysis: As summarized in Theorem 1, all  infinite distance finite volume limits are characterized by some specific fibration structure of a given Calabi-Yau three-fold, $Y$.
The fiber is either a genus-one curve $T^2$, a K3-surface, or an Abelian surface $T^4$. In the limit, the respective fiber is precisely the cycle that shrinks at the fastest rate, while the base of the fibration expands such as to keep the overall volume finite. If the three-fold $Y$ exhibits several topological fibrations at the same time, we prove that the fibration type with the fastest shrinking fiber is in fact unique. As noted already, this is crucial in order to establish an unambiguous physical interpretation
of the limits and hence forms one of our central results.
 The proof of this structure is rather elaborate and presented in the two technical Appendices \ref{pf_thm1} and \ref{unique-pf}. 
A~schematic overview of the types of limits which can occur is given in Figure \ref{f:FigMlimits}.

 After this general geometric analysis we investigate, as the second main part of the paper, in Section~\ref{sec_Mtheorylimits}
  the physics of M-Theory compactified on a three-fold $Y$ undergoing one of the possible types of infinite distance limits.
 Indeed, M-Theory probes the classical K\"ahler moduli space and hence the geometry is not subject to quantum corrections.
 
 We find that each of the three types of finite volume limits realizes one of the possible outcomes for infinite distance finite volume limits:
 Partial decompactification, the emergence of a {\it unique} light heterotic string, or the emergence of a {\it unique} light Type II string. Infinite {\it volume} limits, on the other hand, always imply decompactification in M-Theory.
 
 Limits of Type $T^2$ are decompactification limits despite the fact that the total three-fold volume is fixed in terms of the M-Theory scale.
 The reason is that a tower of M2-branes along the torus fiber effectively acts as a KK tower that implements decompactification from five to six dimensions. This type of infinite distance limit is identical to taking a conventional F-Theory  \cite{Vafa:1996xn} limit,
 as had already been pointed out in \cite{Lee:2019xtm} (see also \cite{Corvilain:2018lgw}). 
 
 Limits of Type K3 and $T^4$, on the other hand, are truly equi-dimensional in the sense of our definition. 
 For limits of Type K3, we identify two competing towers of light particle states sitting parametrically at the same scale 
 as the KK tower, which is weakly suppressed by the finite Calabi-Yau volume as measured in units of the M-Theory scale. 
 The first tower of particles are the excitations of a light and weakly coupled heterotic string associated with an M5-brane wrapped around the shrinking K3 fiber. 
 The second type of tower arises from M2-branes wrapping curves of non-negative self-intersection in the fiber. As we discuss, invoking general facts from heterotic/Type IIA duality, this tower of particles can be understood as wrapping modes of a dual fundamental heterotic string on a circle. Geometrically, the fact that M2-branes along said fibral curves give rise to an infinite tower of states, rather than a finite number of states,
is reflected by the infinitely many non-zero Gopakumar-Vafa invariants which correspond to Fourier coefficients of certain meromorphic forms \cite{Harvey:1995fq,Henningson:1996jf,Dijkgraaf:1996xw}.
 In this sense, both towers manifest the emergence of an asymptotically light heterotic string.
 
 An analogous  picture is found for limits of Type $T^4$. Such limits are governed by the emergence of a light string (including
  its wrapping modes) that arises from an M5-brane along the $T^4$ fiber.
 In a suitable dual weakly coupled frame, this string can be interpreted as a Type IIB string
 probing a (generically non-geometric)  D-manifold in the sense of \cite{Bershadsky:1995sp}, with spacetime-filling 5-branes backreacting on the geometry. 

 In section \ref{sec_infvolume} we drop the requirement of working at finite volume and will indeed confirm, based on our classification of finite volume limits, the expectation that all infinite volume limits in M-Theory are pure decompactification limits.
 
\begin{figure}[t!]
\centering
\includegraphics[width=12cm] {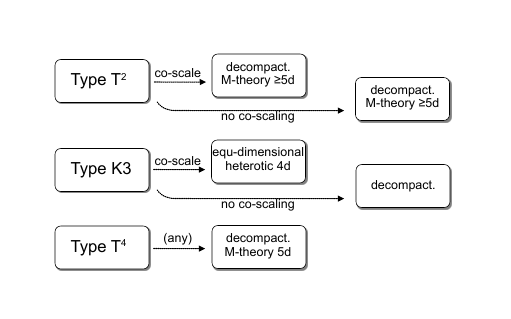}
\caption{Basic infinite distance limits in the quantum K\"ahler moduli space for Type IIA string theory on a Calabi-Yau three-fold. 
All further possible limits are decompactifications limits.}
\label{f:FigIIlimits}
\end{figure}

 In the third part of our analysis, presented in Section \ref{sec}, we investigate how quantum effects modify the appearance of equi-dimensional limits, as compared
 to the classical moduli space.
 To this end, we pass from M-Theory to Type IIA string theory compactified on the three-fold~$Y$. 
 In Type IIA string theory, we must distinguish between limits leaving the volume of $Y$ finite,  and limits keeping the four-dimensional Planck scale, 
 \be \label{PlscaleIntro}
 \frac{M_{\rm Pl}^2}{M_s^2}  = \frac{4 \pi}{g_{\rm IIA}^2} \, {\cal V}_Y \,,
 \ee
finite. 
 Since a finite volume limit at infinite distance necessarily involves the shrinking of some cycles in~$Y$, non-perturbative quantum geometry effects 
 are generically expected to play a crucial role.
 As we will recall in Section \ref{sec_gencons1}, the only way to sensibly describe the regime of small fiber volumes is to analyze the complex structure moduli space of the three-fold~$X$ that is mirror-dual to~$Y$.
 Key is to analytically continue the period integrals of $X$ from the large complex structure point to the regime of interest, 
 where --naively-- some cycle volumes become small.
We will argue that in the quantum corrected K\"ahler moduli space of Type IIA string theory, no {\it finite volume} limits exist at infinite distance. 
Nonetheless the Planck scale (\ref{PlscaleIntro}) can be kept finite by suitably co-scaling the ten-dimensional dilaton $g_{\rm IIA}$, and the interesting question is whether such co-scalings can lead to an equi-dimensional limit in the sense of our definition. A schematic overview of our results is given in Figure \ref{f:FigIIlimits}.

Specifically, for limits of Type $T^2$, the limit of vanishing fiber volume is in fact T-dual to the large K\"ahler regime, as noted already in \cite{Aspinwall:2002nw,Alim:2012ss,Klemm:2012sx,Huang:2015sta,Corvilain:2018lgw}.
 We will see that such limits always imply a partial decompactification to five-dimensional M-Theory on $Y$, irrespective of whether (\ref{PlscaleIntro}) is kept finite by co-scaling $g_{\rm IIA}$ or not; in the latter case, the theory is guaranteed to decompactify even further.
  For limits of Type K3, on the other hand, we observe a quantum obstruction against taking the limit of zero fiber volume, by carefully computing the quantum volume via mirror symmetry. This can also be understood in terms of a 1-loop correction in M-Theory, which reflects the
 vacuum energy of the solitonic heterotic string
that becomes light semi-classically. As a result, the total volume of $Y$ diverges. In fact such limits of Type K3 realize 
\cite{Kachru:1995fv}  the rigid, weak coupling limit of Seiberg-Witten gauge theory \cite{Seiberg:1994rs}, and we carefully analyze the towers of states that become light with respect to the Planck scale.
 If $g_{\rm IIA}$ is not co-scaled such as to achieve a finite Planck scale (\ref{PlscaleIntro}),
  the KK tower is found to be parametrically lighter than all other states, and we again run into a decompactication limit.
  Interestingly, this conclusion can be avoided by forcing a finite Planck scale by co-scaling $g_{\rm IIA}$: In this case we will find a tower of massless D0-branes, KK states {\it and} heterotic string excitations at the same parametric mass scale, leading to an equi-dimensional quantum limit.

  The quantum limit of Type $T^4$ behaves similar as the limit of Type $T^2$, in that T-duality relates small fiber volume to the large volume regime at strong coupling. Unlike for limits of Type $T^2$, such limits are automatically at finite Planck scale.
Thus the theory is found to decompactify to M-Theory, however without an additional decompactification beyond the M-Theory limit. 

We conclude in Section \ref{sec_Conclusions} and put our findings into a broader perspective of how the emergent string conjecture could be realized in other types of compactifications, commenting in particular on mirror dual
infinite distance limits in complex structure moduli space.

\section{Large Distance Limits in the Classical K\"ahler Moduli Space of Calabi-Yau three-folds}
\label{sec_classical}

As explained in the introduction, our goal is to study equi-dimensional infinite 
distance limits in the K\"ahler moduli space of a Calabi-Yau three-fold $Y$,
probed by M-Theory or Type IIA string theory.
These are infinite distance limits for which the gravitational interactions of the original five- or, respectively, four-dimensional effective supergravity theory remain dynamical.
It is generally expected - and will be confirmed - that as a necessary condition for this to be the case, the five- or four-dimensional Planck scale must stay finite (in fundamental units of M-Theory or Type II string theory).
For compactifications of M-Theory on $Y$ such limits leave the volume  ${\cal V}_Y$  of $Y$ finite.\footnote{In the context of Type IIA string compactifications infinite volume limits can lead to a finite Planck scale as long as the string coupling $g_{\rm IIA}$ is suitably co-scaled with the K\"ahler moduli. We will consider such double-scaling limits in Section \ref{sec}.}

Our first task is therefore to characterize finite volume limits in the {\it classical} K\"ahler moduli space of $Y$. 
The classical K\"ahler moduli space is the moduli space probed by M-Theory compactified on $Y$, which yields
 a five-dimensional supergravity theory with $8$ supercharges. 
Our results are summarized in Theorem 1 in Section \ref{classification-result}, and we encourage the reader who is interested more in the physics applications rather than in the technical details of the geometry to  jump directly to this section.
From a mathematical perspective, this analysis leads to a classification of all infinite distance limit in the classical K\"ahler moduli space (including the infinite volume limits) if we drop, in a second step, the restriction on the total volume to remain finite.

There are two ways in which the result of the classification can be phrased.
The first is in terms of intersection numbers of the K\"ahler cone generators whose associated K\"ahler parameters are taken to infinity. This analysis has been presented already in \cite{Lee:2019tst} for general K\"ahler three-folds (following the same techniques applied previously in \cite{Lee:2018urn} to complex K\"ahler surfaces.) In Section \ref{subsec_Kahlercone} we will summarize and specialise these results to Calabi-Yau three-folds.
The outcome of this classification scheme is closely related to the classification of infinite distance limits in the K\"ahler moduli space of Calabi-Yau three-folds as obtained in \cite{Corvilain:2018lgw}. We will comment on the relation of both approaches at the end of Section \ref{subsec_Kahlercone}.

To analyze the physical implications of the infinite distance limits  for M-Theory or Type IIA string theory, 
it is important to reorganize the different classes of infinite distance limits according to which type of cycles shrink at the fastest rate compared to the total volume. 
We will find 
three qualitatively different classes of finite volume limits at infinite distance, according to this physically relevant criterion. 
These limits are summarized in Theorem~\ref{classify} of Section~\ref{classification-result}. 
Strikingly, in all finite volume limits at infinite distance the fiber of a certain uniquely distinguished fibration vanishes while its base expands. This is key to unravelling the asymptotic physics of these limits in subsequent sections.

The derivation of Theorem~\ref{classify} based on the classification scheme of Section \ref{subsec_Kahlercone} is relegated to Appendices \ref{pf_thm1} and \ref{unique-pf}.

\subsection{K\"ahler form structure of infinite distance limits} \label{subsec_Kahlercone}

Consider a Calabi-Yau three-fold $Y$ and parametrise its K\"ahler form $J$ as 
\be \label{Jgens1}
J = \sum_{i \in \cI} {T}^i J_i  \,, \qquad {T}^i \geq 0 \, \, \, \, \forall i \, \in \, \cI \,, \\
\ee
where the divisor classes $J_i, i \in \cI$, generate the K\"ahler cone.
We are interested in infinite distance limits in the classical K\"ahler moduli space of $Y$. To take such a limit, at least one real K\"ahler parameter $T^i$ must be taken to infinity. If the classical volume 
\be
{{\cal V}_Y} = \frac{1}{3!} \int_Y {J}^3 
\ee
stays finite, we call this an {\it infinite distance finite volume limit}. 
Generically, however, taking some of the K\"ahler parameters to infinity will force the total volume ${\cal V}_Y$ to diverge. We can parametrise the scaling of ${\cal V}_Y$ as 
\be \label{VYinfinite}
{\cal V}_Y = \mu \, {\cal V}'_Y \qquad \text{with} \, \, \mu \to \infty \,, \qquad {\cal V}'_Y \, \, \text{finite} \,.
\ee 
Here $\mu$ denotes some combination of the K\"ahler parameters which scales to infinity,
and the finite volume contribution ${\cal V}'_Y$ is determined from the uniformly rescaled K\"ahler form
\be\label{JdeffromJ'}
J' = \mu^{-1/3} J =: \sum {T'}^i J_i  \,.
\ee
There are now two, fundamentally different possibilities:
The first option is that the K\"ahler parameters ${T'}^i$ of  $J'$ are either finite or some of them scale to zero in the limit $\mu \to \infty$. This can at best lead to the shrinking (with respect to $J'$) of {\it contractible} divisors or curves on $Y$.
Alternatively, at least one of the parameters ${T'}^i$ scales to infinity, but in such a way that the volume ${\cal V}'_Y$ remains finite.
Obviously, in this case,  ${\cal V}'_Y$  can only remain finite if other K\"ahler parameters ${T^j}'$ scale to zero with a suitable inverse power. This leads to the asymptotic shrinking of a non-contractible cycle.

The first possibility is what characterizes a finite distance limit in the moduli space after rescaling: A contractible cycle can shrink (at finite ${\cal V}'_Y$) without the volume of any other dual cycle diverging.
Branes wrapping contractible cycles do generally not give rise to a tower of weakly coupled states. 
 The second possibility refers to an infinite distance limit where some of the K\"ahler parameters ${T'}^i$ scale to infinity.  For 
 explicit recent
  computations of distances in moduli space with respect to the relevant K\"ahler metrics, see 
 \cite{Blumenhagen:2018nts,Grimm:2018ohb,Lee:2018urn,Lee:2018spm,Corvilain:2018lgw,Font:2019cxq}.

The important point to note is that it suffices to classify the behaviour of $J'$ which realizes the second, large distance limit type of behaviour.
These finite volume limits can then be superimposed with an overall rescaling of all K\"ahler parameters with scaling parameter $\mu^{1/3}$ on top of all the scalings of the ${T'}^i$ that we will describe. This  brings us back to the original K\"ahler form $J$ without the extra constraint that the volume remains finite. By including also those K\"ahler forms $J'$ for which all ${T'}^i$ are finite or scale to zero before rescaling with $\mu^{1/3}$ this reproduces all types of infinite distance limits.

With this general picture in mind, we now specialise to infinite distance limits at finite volume.
To avoid clutter of notation we omit the primes on $J'$ and on the K\"ahler parameters ${T'}^i$, with the understanding the total volume remains finite and fixed.
As mentioned already, every infinite distance finite volume limit requires now the scaling of  least one K\"ahler parameter to infinity. {Clearly there can be several K\"ahler parameters which scale to infinity, and to compare their parametric behaviour,
we will use the following notation: Given two functions $A(\la)$ and $B(\la)$,  we write
\beq \label{asimnotation}
A (\la) \sim B (\la) \quad \text{(resp. $A \prec B$ and $A\precsim B$)}\,, 
\eeq
to indicate in a certain limit for $\la$ that
\beq
\frac{A(\la)}{B(\la)} ~\to~ c \,, \qquad \text{with}\quad 0< c <\infty \quad\text{(resp. $c=0$ and $0\leq c < \infty$)}\,.
\eeq

Assume that the K\"ahler parameter scaling to infinity at the fastest rate scales as $\lambda$,
and define an index set $\cI_\lambda \subseteq \cI$ by stating that the K\"ahler parameters of all $J_i$ for $i \in \cI_\lambda$ scale at this rate $\lambda$.
In other words,
\be \label{TiIlambdadef}
T^i \sim \lambda \qquad   \forall i \,  \in \,  \cI_\lambda \,, \qquad \quad T^j \prec \lambda  \qquad \forall j \,  \in\,  \cI \setminus \cI_\lambda \,.
\ee
Here $\cI$ labels all the K\"ahler cone generators as in (\ref{Jgens1}).  Imposing finiteness of ${\cal V}_Y$ implies that 
\be
J_i^3 = 0 \qquad  \forall \, \,  i \in \cI_\lambda \,,
\ee 
and similarly for all other generators whose K\"ahler parameter scales as $T^j \succ 1$.
We furthermore split 
\be
\cI_\lambda  = \cI^{A}_{\lambda} \cup \cI^{B}_{\lambda}
\ee
according to the criterion that
\bea
J_i^2 &\neq& 0  \qquad    \text{if} \quad   i \in \cI^{A}_{\lambda} \\
J_i^2 &=& 0 \qquad  \text{if} \quad   i \in \cI^{B}_{\lambda} \,.
\eea
There are then two types of limits which can be denoted and distinguished as follows:\footnote{This definition is a slight refinement of the definition given in Ref.~\cite{Lee:2019tst}, to whose Appendix D we refer for more details.}
\bea
&\text{$J$-class A}:& \qquad \quad  \cI^{A}_{\lambda} \neq \emptyset \\
&\text{$J$-class B}:& \qquad \quad  \cI^{A}_{\lambda}  = \emptyset \,.
\eea

If  $\cI^{A}_{\lambda} \neq \emptyset $ we pick an arbitrary K\"ahler cone generator labelled by $\cI^{A}_{\lambda}$ and call it $J_0$.
If $\cI^{A}_{\lambda}  = \emptyset $, we pick instead an arbitrary K\"ahler cone generator labelled by $\cI^{B}_{\lambda}$ and call it $J_0$.
In either case, all the other K\"ahler parameters are of order $\la$ or less, i.e. $T^j \precsim \lambda$ for $j \neq 0$.}
Explicitly, we can split the index set $\cI$ which labels the K\"ahler cone generators as 
\beq\label{split}
\cI= \cI_0 \cup \cI_1 \cup \cI_2 \cup \cI_3 \,,
\eeq
and write the most general K\"ahler form as
\beq
J = \la J_0 +\sum_{\alpha \in \cI_1} a_\alpha J_\alpha + \sum_{\mu \in \cI_2} b_\mu J_\mu +  \sum_{r\in \cI_3} c_rJ_r \,.
\eeq
In the splitting~\eqref{split} of the index set, $\cI_0$ only labels the distinguished generator $J_0$ and the remaining three subsets are defined with respect to $J_0$ in such a way that 
\bea\label{alpha}
J_0^2 \cdot J_{\alpha} &\neq& 0 \,\quad \forall\alpha\in \cI_1 \,, \\
J_0^2 \cdot J_{\mu} &=& 0 \,\quad \forall\mu\in \cI_2 \,, \quad \text{and}\quad J_0 \cdot J_\mu \cdot J_{\nu} \neq 0\quad \text{for some}~\nu\in\cI_2\,, \\
J_0^2 \cdot J_{r} &=& 0 \,\quad \forall r\in \cI_3 \,, \quad \text{and}\quad J_0 \cdot J_r \cdot J_i = 0 \quad\text{for all}~i\in \cI_2 \cup \cI_3 \,.
\eea 

The two classes of limits  have the following properties.
  
\paragraph{$J$-class A: $J_0^3 = 0$, but $J_0^2 \neq 0$.}\quad
The index set $\cI_2$ is empty and the K\"ahler class $J$ takes the form
\beq\label{classA-sec2}
J =  \la J_0 +\sum_{\alpha \in \cI_1} {a_\alpha} J_\alpha +  \sum_{r\in \cI_3} c_rJ_r \,,
\eeq
where
\be\label{classA-details}
a_\alpha={\hat a_\alpha} \la^{-2} \,, \quad c_r = {\hat c}_r \, \la^{\gamma_r}   \,\,( \gamma_r \leq 1)\,,
\ee
with 
${\hat a}_\nu$ and ${\hat c}_r$ finite for $\la \to \infty$.  
The parametric behavior~\eqref{classA-details} was derived in Appendix D of~\cite{Lee:2019tst} for a general K\"ahler three-fold, and the results apply to Calabi-Yau three-folds as well.
The generators with indices in $\cI_3$ satisfy 
\be
 J_r \cdot J_s = n_{r s} \,  J_0 \cdot J_0 \,, \qquad \quad n_{r s} \geq 0 \,,
\ee
and hence also $J_r \cdot J_s \cdot J_t =0$ for any triple $(r, s,t)$. Note that the latter is a necessary condition
for finiteness of ${\cal V}_Y$, for those cases where $c_r c_s c_t \succ 1$.

\paragraph{$J$-class B: $J_0^2 = 0$.}\quad 
The index sets $\cI_1$ and $\cI_3$ are empty and the K\"ahler class $J$ takes the form
\beq\label{classB-sec2}
J = \la J_0 + \sum_{\mu \in \cI_2} b_{\mu} J_{\mu} \,,
\eeq
where
\be\label{classB-details}
b_\mu = {\hat b}_\mu \, \la^{\beta_\mu}   \,\,( \beta_\mu \leq 1)\,,
\ee
with 
${\hat b}_\mu$ staying finite for $\la \to \infty$.  
The parameters $\hat b_\mu$ are further constrained by demanding that $\cV_Y$ be finite as  we discuss in detail in Appendix~\ref{unique-pf}. 

It is interesting to compare the above results to the classification of \cite{Corvilain:2018lgw} of infinite distance limits in Calabi-Yau three-folds in the large volume regime.
The methodology of both approaches is very different: While our approach, based on \cite{Lee:2018urn,Lee:2019tst}, starts directly from the K\"ahler form of $Y$, Ref. \cite{Corvilain:2018lgw} uses the classification of degenerations in the complex structure moduli space of a mirror three-fold based on the theory of limiting mixed Hodge structures \cite{Grimm:2018ohb,Grimm:2018cpv}.
Ref \cite{Corvilain:2018lgw} finds three types of infinite distance limits at large K\"ahler volume which are distinguished by the structure of intersection numbers of the K\"ahler cone generators which scale to infinity. They are called limits of Type II, III and IV,
respectively. A priori, no distinction is made as to whether or not the total volume diverges. The limit of Type IV of \cite{Corvilain:2018lgw} always leads to a diverging volume. The cases of Type II and Type III are closely related to our $J$-class B and $J$-class A limits, respectively, possibly upon rescaling by the overall volume.

\subsection{Finite volume limits at infinite distance as vanishing fiber limits}\label{classification-result}

In this section we characterize the finite volume limits of the previous section in a way that most directly reveals the physics,
 as probed by M-Theory or Type IIA string theory.
Understanding the physics requires us to analyze the cycles of $Y$ whose volume vanishes at the fastest rate in the infinite distance limit. These cycles give rise to light towers of BPS states via wrapped branes, whose properties reflect the possible weakly coupled
theories that  emerge at the boundary of moduli space.

Our main result, summarized in Theorem 1 below, is that the limits of $J$-class A and B are only possible if  $Y$ admits 
some definite fibration structure and that the relevant cycles of $Y$ that vanish at the fastest rate lie in the fiber.

More precisely, in order to allow for a limit of $J$-class A or B, the Calabi-Yau three-fold $Y$ must be a fibration with generic fiber 
being either a genus-one curve $T^2$, a K3-surface, or an Abelian surface, $T^4$.
As explained in Appendix \ref{pf_thm1}, this follows by noting that the K\"ahler cone generator $J_0$ that appears in (\ref{classA-sec2}) or (\ref{classB-sec2}) satisfies one of the three criteria of~\cite{oguiso}, which then implies a corresponding fibration structure for $Y$.
 In a first approximation, limits of $J$-class A imply that $Y$ must admit a $T^2$-fibration while limits of $J$-class B imply that $Y$ admits a K3- (or $T^4$-) fibration if $J_0 \cdot c_2(Y) \neq 0$ (or $=0$).
However, the fiber topology by itself does not unambiguously distinguish what we will define as Type $T^2$, Type $K3$ and Type $T^4$ limits in Theorem 1. This is because a three-fold can for instance be K3-fibered with an additional
 compatible $T^2$-fibration, in which case the physics of the limit depends on 
additional details of the scaling behavior of various cycles.
The definitions in Theorem~\ref{classify} below are carefully chosen such that they give rise to three different and mutually exclusive physical limits, which cover all possible situations compatible with finite volume limits at infinite distance.

Using the notation (\ref{asimnotation}), we are now ready to state our main classification theorem as follows. 

\begin{theorem}\label{classify}
Consider an infinite distance limit in the classical K\"ahler moduli space of a Calabi-Yau three-fold $Y$, subject to the constraint that the total volume
\be
{\cal V}_Y = \frac{1}{3!} \int_Y J^3
\ee
remains finite. The K\"ahler form must be of $J$-class A or $J$-class B, as summarized in Section \ref{subsec_Kahlercone}. Furthermore, the geometry of $~Y$ necessarily falls into one of the following three classes, which are mutually exclusive and hence well-defined:
\begin{enumerate}
\item {\bf Type $T^2$} \\[1mm]
$Y$ is a genus-one fibration\footnote{Throughout this article, our notion of a ``genus-one'' fibration assumes the existence of at least a $k$-section $\sigma$ which embeds the base as a cycle into the full space. For notational simplicity we do not distinguish between the base and its image under this (multi-)section.} over some two-dimensional base, $B_2$,
and the volumes of the fiber and the base scale in the limit as
\beq \label{TypeT2scaling1}
{\cal V}_{T^2} \sim \la^{-2} \,, \qquad {\cal V}_{B_2} \sim \la^2 \,,  
\eeq
where the parameter $\la \to \infty$ characterizes the infinite distance limit.\footnote{If the limit arises from a $J$-class A limit, $\lambda$ is the parameter appearing in (\ref{classA-sec2}). If the limit arises from a $J$-class B limit, we identify $\lambda$ with the parameter $\mu$ in Eq (\ref{VT^2beginningApp}), whose relation to the parameter appearing in (\ref{classB-sec2}) is given in (\ref{lambdavsmu}).}

Suppose $Y$ admits in addition a compatible K3- or $T^4$-fibration, i.e. suppose the base
 $B_2$ is itself a fibration over base $\mathbb P^1_b$ 
with a generic  fiber ${\cal F}$ being either  
a rational curve, $\mathbb P^1_f$, or a genus-one curve, $T^2_f$.
Then in order for the limit to lie in Type $T^2$,
the respective volumes scale as
\bea \label{Ellipticextracond}
{\cal V}_{T^2} \sim \la^{-2} \,, \qquad \la^{-2} \prec {\cal V}_{\mathbb {\cal F}} 
 \,, \qquad  {\cal V}_{\mathbb P^1_b}  \sim  \frac{\la^{2}}{\cV_{\cal F}}   \,, \qquad  \la^{-4} \prec {\cal V}_{K3/T^4} \,.
\eea

\item {\bf Type $K3$} \\[1mm]
$Y$ is a K3-fibration over a base $\mathbb P^1_b$,
such that their volumes scale in the limit as 
\bea \label{VK3JB}
{\cal V}_{K3} \sim \la^{-1} \,, \qquad   {\cal V}_{\mathbb P^1_b}  \sim \la\,,   
\eea
where $\la \to \infty$,  
and the volume of every curve class $C$ lying in the K3-fiber of self-intersection $ C \cdot_{K3} C \geq 0$ scales as
\bea \label{VCK3Type}
{\cal V}_{C} \sim \la^{-1/2} \,.
\eea

\item {\bf Type $T^4$} \\[1mm]
$Y$ is a $T^4$-fibration over a base $\mathbb P^1_b$
such that their volumes scale in the limit as  
\bea \label{VT4sec2}
{\cal V}_{T^4} \sim \la^{-1} \,, \qquad   {\cal V}_{\mathbb P^1_b}  \sim \la\,,   
\eea
where $\la \to \infty$,  
and the volume of every curve class $C$ lying in the $T^4$-fiber of self-intersection $ C \cdot_{T^4} C \geq 0$ scales as
\bea \label{VCT4Type}
{\cal V}_{C} \sim \la^{-1/2} \,.
\eea

\end{enumerate}
For each of the three classes above, the corresponding fibration with the parametrically smallest fiber volume is unique, even when there exist multiple fibrations of the same fiber topology.

\end{theorem}

As alluded to before,  a priori an ambiguity can arise if a $K3$-fibration admits an additional compatible $T^2$-fibration.
The physics depends on whether the base $\mathbb P^1_f$ of the K3-fiber or the genus-one fiber $T^2$ vanishes at a faster rate. 
We call the fibration of Type $T^2$ if the $T^2$ fiber vanishes at the faster rate. This is taken care of by the requirement ~\eqref{Ellipticextracond}.
Indeed, the requirements~\eqref{Ellipticextracond} and~\eqref{VCK3Type} are mutually exclusive if the $K3$ fibration admits a compatible genus-one fibration.
Analogous statements hold for $T^4$-fibrations with a compatible $T^2$-fibration, see~\eqref{Ellipticextracond} and~\eqref{VCT4Type}. 
This resolves all ambiguities and ensures that the three Types as defined in Theorem 1 are mutually exclusive.

We leave a careful derivation of the above classification to Appendix~\ref{pf_thm1}. 
In Appendix \ref{AppA_subTypeT2} we show that a K\"ahler form limit of $J$-class A gives rise to an infinite distance limit of Type $T^2$, for which in particular the scaling behaviour (\ref{TypeT2scaling1}) holds. 
In Appendix \ref{SubsecTypeB}  we prove that limits of $J$-class B imply the existence of a $K3$ (or $T^4$) fibration, and that one of the following two situations can occur: Either all fibral curves of non-negative self-intersection scale as in (\ref{VCK3Type}) (or (\ref{VCT4Type})), in which case the limit is of Type K3 (or $T^4$), respectively.
Or, in every other situation, the $K3$ (or $T^4$) fibration must admit a compatible $T^2$-fibration and the scaling (\ref{Ellipticextracond}) occurs.
This abstract structure is illustrated in a concrete example in Appendix \ref{App_EllK3fibraEx}.

In Appendix~\ref{unique-pf} we show that if $Y$ admits several types of incompatible fibration structures, there is a unique fibration which is singled out by the fact that its volume vanishes at the fastest rate in the infinite distance limit. 
The precise statements are given in Propositions
\ref{Prop-uniqueT2}, \ref{prop-uniqueK3} and \ref{propT2fromB}.
This is pivotal for identifying the relevant physics of this limit.

Just to give an idea how non-trivial these uniqueness results are, suppose a three-fold $Y$ admits two different K3 (or $T^4$)-fibrations  such that the volume of the two generic fibers scales at the same rate $\lambda^{-1}$ in a finite volume limit.
If the limit is of $J$-class A, this is not a problem because such limits are always of Type $T^2$, i.e. the K3-fibers in question scale at a rate slower that twice of the rate of the vanishing genus-one fiber whose existence is guaranteed in
 $J$-class A limits (see Appendix \ref{AppA_subTypeT2} and \ref{AppT2ClassAunique}). On the other hand, if the limit is of $J$-class B, we will prove in Appendix \ref{K3unique} that in this case $Y$ necessarily admits a genus-one fibration whose fiber vanishes at a rate $\lambda^{-2}$ or faster, i.e. the limit is again really a limit of Type $T^2$.
Furthermore all limits of Type $T^2$ from $J$-class B limits are unique, as proven in Appendix \ref{T2Unique-JClassB}.

\subsection{Infinite distance limits as weak coupling limits} \label{subsec_weakcoupling}

Even before delving into a more detailed analysis of the physical implications of infinite distance limits described in the previous section,
we recall that such limits correspond to the weak coupling regime for some of the gauge fields in a compactification of string or M-Theory on the three-fold $Y$.
Naively, since by definition we keep the volume finite, gravity remains dynamical, but this statement will need to be carefully revisited in the next section.
The relation between weak coupling and infinite distance limits is a general theme which has recently
played an important role in the context of various weak gravity conjecures \cite{Grimm:2018ohb,Lee:2018urn,Lee:2018spm,Grimm:2018cpv,Corvilain:2018lgw,Marchesano:2019ifh,Font:2019cxq,Lee:2019tst,Lee:2019xtm,Grimm:2019wtx}.

The following discussion applies morally equally well 
to the effective action of Type IIA string theory and M-Theory on a Calabi-Yau three-fold, $Y$. For definiteness we phrase the discussion in M-Theory
and start from the bosonic part of the 11-dimensional effective action
\beq
S=\frac{2\pi}{\ell_{11}^9} \left(\int_{\IR^{1,10}} \sqrt{-g} R - \frac12 \int_{\IR^{1,10}} d {\RRC}_3 \wedge * d  {\RRC}_3 + \cdots\right) \,,
\eeq
where the ellipses indicate the flux contribution as well as the higher curvature term, both of which will not concern us. Upon compactifying on the three-fold $Y$, equipped with some basis $\omega_\alpha$ of $H^{1,1}(Y)$, the 3-form $ {\RRC}_3$ decomposes into abelian gauge potentials as $ {\RRC}_3 = \ell_{11} A^\alpha \wedge \omega_\alpha + \ldots$, and 
we read off the relevant kinetic terms in the five-dimensional effective action as 
\beq
S_5 = \int_{\IR^{1,4}} \frac{1}{2} {M_{\rm Pl}^3} \,  R *1 -  {\frac{2\pi}{2}} \int_{\IR^{1,4}}g_{\alpha \beta} F^\alpha \wedge *F^\beta + \cdots \,. 
\eeq
Here the $5d$ Planck mass is given as the volume of the internal manifold in units of $\ell_{11}= M^{-1}_{11}$,
\beq \label{MplM11} 
\frac{M_{\rm Pl}^3}{M^3_{11}} = {4\pi} {\cal V}_Y \,,
\eeq
and the dimensionless quantity $\ell_{11} g_{\alpha\beta}$ is determined as 
\bea
\ell_{11} \,  g_{\alpha\beta} &=& \frac{1}{\ell^6_{11}}\int_Y \omega_\alpha \wedge *\omega_\beta \\ \nn
&=&  \frac32 \frac{\int_Y  J^2 \wedge \omega_\alpha \int_Y  J^2 \wedge \omega_\beta}{\int_Y J^3} - \int_Y J\wedge \omega_\alpha \wedge \omega_\beta  \,.
\eea
Note that the latter can be rewritten as
\beq
\ell_{11} \, g_{\alpha\beta} = \frac{\cV_\alpha \cV_\beta}{\cV} - \cV_{\alpha\beta} \,,
\eeq
where 
\beq
\cV = \frac{1}{3!} \int_X J^3 \,, \quad\quad 
\cV_\alpha = \frac{1}{2!} \int_{S_\alpha} J^2 \,, \quad\quad
\cV_{\alpha \beta} = \int_{C_{\alpha \beta}} J \,, 
\eeq
denote the volume (in units of $\ell_{11}$) of the Calabi-Yau three-fold $Y$, and those of the surfaces $S_\alpha$ as well as the curves $C_{\alpha\beta}$. The classes of the latter are defined by
\beq
[S_\alpha]=\omega_\alpha\,,\quad\quad [C_{\alpha\beta}]=\omega_\alpha \wedge \omega_\beta \,.
\eeq
In order for (at least) one linear combination of the gauge fields to be weakly coupled, while five-dimensional gravity remains dynamical, there must exist a diverging eigenvalue of the matrix $\ell_{11} g_{\alpha\beta}$ of gauge kinetic terms. In addition, the Planck scale, $M_{\rm Pl}$, must remain finite with respect to $\ell_{11}$.
This can only be achieved if either the divisor volume $\cV_\alpha$ or the curve volume $\cV_{\alpha\beta}$ (both measured with respect to $\ell_{11}$) tends to infinity for some $S_\alpha$ or $C_{\alpha\beta}$. 
In other words, one must reach a limiting regime in the moduli space where the volume of a subvariety, be it a surface or a curve, scales to infinity while the total volume of the three-fold $Y$ stays fixed. 
These are precisely the limits studied in the previous section.

\section{M-Theory on Calabi-Yau 3-folds in finite volume infinite distance limits} \label{sec_Mtheorylimits}

We are now ready to analyze the physical implications of the limits in the classical K\"ahler geometry of a Calabi-Yau three-fold, as studied in Section \ref{classification-result}. This classical geometry 
 is the geometry probed by M-Theory compactified on $Y$ to $d=5$ dimensions, with
 $8$ unbroken real supercharges (provided that $Y$ does not have reduced holonomy, which we assume).

 With the exception of our analysis in Section \ref{sec_infvolume}, we will consider limits for which the total volume of $Y$ remains finite.
Hence the Planck scale as given (\ref{MplM11}) is fixed, and we obtain a theory with dynamical gravity, which at the same time is at weak coupling, in the sense described in Section \ref{subsec_weakcoupling}.
 Of course a first guess would be that a weakly coupled quantum gravitational theory should be a string theory - as otherwise one would have found a new perturbative theory of quantum gravity. 
 Indeed this expectation will be confirmed: Modulo an important caveat, the effective theory indeed reduces, in the geometric large distance
 limit, to a theory of weakly coupled, asymptotically tensionless strings of either heterotic type or of Type~II.
 This is analogous to the emergence of heterotic and Type II strings as discussed
 in \cite{Lee:2018urn,Lee:2019tst} and \cite{Lee:2019xtm}, respectively.
 
 The caveat is that even when the total volume stays finite,  a distinguished tower of light modes from wrapped branes may appear that mimic a (partial) decompactification of the theory.
In such a case,  the theory flows  to a gravitational theory in one or several dimensions higher, which need not be weakly coupled and hence not be a theory of light strings.

We will analyze the situation explicitly for the three types of finite volume limits at infinite distance,
as distinguished in Section \ref{classification-result}.
Our findings can be summarized as follows:

\begin{enumerate}
\item
In a limit of Type $T^2$, the five-dimensional theory of M-Theory on $Y$ asymptotes to a six-dimensional compactification of F-Theory on $Y$, i.e. the limit of Type $T^2$ is a conventional F-Theory limit.
In this sense, limits of Type $T^2$ are not equi-dimensional limits. The light tower of modes responsible for the decompactification is the tower of M2-branes along the $T^2$-fiber, as is well-known from standard 
F/M-Theory duality. This will be discussed in Section \ref{sec_MtheoryEllipticType}.
\item
Limits of Type K3 are truly equi-dimensional weak coupling limits which lead to {\it emergent tensionless heterotic} strings.
As has been familiar since long \cite{Harvey:1995rn}, 
such heterotic strings arise from M5-branes that wrap the small K3-fiber and give rise to a tower of light particle excitations. Another tower of light particles arises from M2-branes on curves of non-negative self-intersection in the K3-fiber.
Their BPS-indices are counted by coefficients of meromorphic modular forms. These particles arise, when viewed in the proper duality frame, from the heterotic string wrapped on a circle. The limit is equi-dimensional in the sense that the Kaluza-Klein scale sits at the same scale as the scale of the emergent heterotic string and the M2-brane states. The analysis of this limit is performed in Section \ref{sec_TypeBK3M}.

\item
Limits of Type $T^4$ are likewise equi-dimensional weak coupling limits with an {\it emergent tensionless Type II string}.
This string probes a dual D-manifold (in general non-geometric), which subsumes the backreaction of certain 5-branes. In the M-Theory picture, it arises from an M5-brane that wraps the small $T^4$ fiber. Its tower of light BPS excitations is augmented by the spectrum of M2-branes wrapping fibral curves, which can analogously be understood as modes of a wrapped Type II string on a circle. This limit is the subject of Section \ref{sec_Dmanifolds}.

\item 
In Section \ref{sec_infvolume} we consider the most general infinite distance limits in the K\"ahler moduli space of M-Theory, without the additional requirement that the volume remains finite. 
This analysis shows explicitly how all such limits can be recovered from the three types of finite volume limits of our classification, via the procedure outlined at the beginning of section \ref{subsec_Kahlercone}.
We will confirm that all limits at infinite volume are decompactification limits, i.e. no light strings can compete with the KK scale.

\end{enumerate}

\subsection{The Type $T^2$ Limit as an F-Theory limit} \label{sec_MtheoryEllipticType}

In this section we explain how M-Theory compactified on a Calabi-Yau 3-fold, in a large distance limit of Type $T^2$, 
approaches partial decompactification from five to six dimensions. The resulting effective theory is described by F-Theory compactified on the base $B_2$ of the genus-one fibration. 
If the three-fold $Y$ admits in addition a compatible K3 or $T^4$-fibration with $\lambda^{-4} \prec {\cal V}_{K3/T^4} \prec \lambda^{-1}$,
this theory undergoes an additional weak coupling limit, where a tensionless Type II or heterotic string appears in six dimensions.

The analogue of this phenomenon has been pointed out already in \cite{Lee:2019xtm} for M-Theory compactified on a K3-surface, in a comparable finite volume limit in K\"ahler moduli space.
The physics of the F/M-Theory  limit has also been discussed from the perspective of infinite distance limits on Calabi-Yau 3-folds in \cite{Corvilain:2018lgw}.
Since the duality between M- and F-Theory is standard and well understood in the literature, we can be brief. Our main point is to explain under which conditions a limit at finite volume can nonetheless undergo a partial decompactification due to the appearance of an extra,
 distinguished tower of massless states.

The relevant scales for M-Theory compactified on $Y$, for a limit of Type $T^2$, are as follows:
The five-dimensional Planck mass (\ref{MplM11})  is by construction finite with respect 
to the 11-dimensional fundamental length scale, $\ell_{11} = M^{-1}_{11}$.
Hence as far as the spectrum of Kaluza-Klein modes associated with the overall volume ${\cal V}_Y$ is concerned, one would naively expect the limit to be equi-dimensional.
That the theory nonetheless undergoes a partial decompactification to six dimensions, is due to a tower of asymptotically massless  BPS states
from M2-branes wrapping the genus-one fiber arbitrarily many, say $n$, times. Their mass is
\be \label{M2onT2scale}
\frac{M_n}{M_{11}} = 2 \pi {\cal V}_{n \, T^2} \sim \frac{n}{\lambda^2} \,,  \qquad \lambda \to \infty \,,
\ee
where $\lambda$  the infinite distance scaling parameter defined in (\ref{TypeT2scaling1}).
In order for such a tower of states to mimic the Kaluza-Klein tower associated with decompactification on a large $S^1$ to six dimensions, three additional conditions must be met:

\begin{enumerate}
  \setcounter{enumi}{0}
  \item The tower of BPS states must be charged under an asymptotically weakly coupled abelian gauge symmetry. 
\end{enumerate}
This condition ensures that we can interpret the associated abelian gauge boson as a Kaluza-Klein vector boson; that it is satisfied 
follows from the discussion in Section \ref{subsec_weakcoupling}.

\begin{enumerate}
  \setcounter{enumi}{1}
  \item 
  The multiplicities of the states at every level $n$ must be equal.
  \end{enumerate}
This is indeed the case because the curve that is wrapped $n$ times is the $T^2$ fiber.
In fact, the {\it 5d index} of BPS particles that arise from M2-branes wrapping a torus fiber $n$ times, is given by certain 
 Gopakumar-Vafa invariants \cite{Gopakumar:1998ii,Gopakumar:1998jq}, and for a genus-one fibration $Y$ these take the form \cite{Klemm:1996hh,Klemm:2012sx} 
  \be \label{NT2}
 N_{n \,  T^2} = \chi(Y) \,, \qquad n \in \mathbb Z \setminus \{0\} \,,
 \ee
where  $\chi(Y)$ is the Euler characteristic of $Y$. The resulting M2-brane spectrum hence behaves exactly like the Kaluza-Klein spectrum associated with the circle reduction of a six-dimensional theory to five dimensions.

\begin{enumerate}
  \setcounter{enumi}{2}
  \item 
The BPS tower from M2-branes on $T^2$ is {\it parametrically leading} as $\lambda \to \infty$, as compared to any other tower of light BPS states {\it coupling only to a weakly coupled gauge sector}, in the sense that
the mass scale of any other tower scales as $\lambda^{-2+\Delta}$ for $\Delta > 0$. 
\end{enumerate}
To verify this condition, we have to discuss the two possible sources for such BPS states: M2-branes along curves different from $T^2$, and light strings from M5-branes along divisors and their potential excitations.\footnote{The supergravity KK states from curves or divisors can easily been shown to be subleading.}

\paragraph{Potential towers from M2-branes}

We begin with the M2-brane states and 
 assume first that the genus-one fibration is flat. We can then choose a basis of $H_2(Y,\mathbb Z)$ generated by curves lying
 either entirely in the fiber, called $C_{\rm f}$,  or by curves, $C_{\rm b}$, lying entirely in the base, $B_2$. Schematically:
\be
H_2(Y,\mathbb Z) = \langle T^2; \{C_{\rm b}\} ; \{C_{\rm f}\}  \rangle \,.
\ee
As is familiar from the F-Theory literature\footnote{See e.g. the review \cite{Weigand:2018rez} for details on the types of fibral curves and the original literature.}, the fibral curves other than $T^2$ are rational curves associated with the weight lattice of some Lie algebra.
M2-branes along such curves do not create a tower of infinitely many BPS states, but rather only a finite number of states which are in one-to-one correspondence with a finite set of representations of the Lie algebra. Mathematically, the reason is that given  a linear combination $C_{w}$ of fibral curves associated with a weight $w$ (not including the total fiber class), 
the Gopakumar-Vafa invariants for $n C_{w}$ vanish, $N_{n \, C_{w}} = 0$, if $|n| > 1$.

Similarly, M2-branes along curves $C_{\rm b}$ entirely in the base $B_2$ do not create towers of BPS particles that could compete with the states from the fiber, in a limit of Type~$T^2$. 
Of course, the base can always contain contractible curves, i.e. rational curves of self-intersection $C_{\rm b}\cdot_{B_2}C_{\rm b}\leq -1$, which may shrink at any rate, and in particular at rate $\lambda^{-2-\Delta}$ for $\Delta \geq 0$. However, the BPS invariants associated with multiples of such curves vanish for all but a finite number of wrappings, $n$. Thus
the only potential extra tower of BPS states can come from non-contractible curves with  ${C_{\rm b} \cdot_{B_2} C_{\rm b}\geq0}$.
By Lemma \ref{lemmaBlimit} of Appendix \ref{App_Surface-Limits}, if the volume of such a curve vanishes in a limit that keeps the volume of $B_2$ fixed, $B_2$ must be a rational or genus-one fibration and the curve $C_{\rm b}$ is the generic fiber, be it 
 $\mathbb P^1_f$ or $T^2_f$, of this fibration. The point is now that by the requirement (\ref{Ellipticextracond}), these curves vanish at a rate $\lambda^{-2 + \Delta}$ with $\Delta > 0$, if the limit is of Type~$T^2$. This is the reason why we  have defined the limits of Type $T^2$ via (\ref{Ellipticextracond}) for situations where $B_2$ admits a fibration structure: It guarantees that the physics is that of an F-Theory limit.

Finally, let us drop the restriction that the fibration $Y$ is flat. In this case, the dimension of the fiber can jump over points on $B_2$ such that 
\be \label{Spdivisor}
S_p = \pi^{-1}(p)
\ee
is a contractible K\"ahler surface. In general, $S_p$ does contain curve classes which give rise to infinite towers of BPS states. Contracting $S_p$ to a point is possible only at finite distance in the moduli space \cite{Witten:1996qb}. As this happens, a strongly coupled superconformal fixed point is approached \cite{Morrison:1996pp,Klemm:1996hh,Morrison:1996xf} (see in particular \cite{Apruzzi:2019opn} for a systematic study of such non-flat fibrations in this context). 
A priori it may not be clear if the tower of BPS states obtained in this way survive the strong coupling regime as stable, nearly
massless states, but in any case the appearance of a {\it  strongly coupled} tower of states per se does not indicate a decompactification limit as noted above.

\paragraph{Potential tensionless strings from M5-branes} We now analyze the nearly massless states due to asymptotically tensionless strings that arise from M5-branes that wrap a shrinking divisor, $D$.
The tension of such a solitonic string is given by
\be
\frac{T}{M^2_{11}} = 2\pi {\cal V}_{D} \,.
\ee
Concerning the spectra of pointlike excitations of the resulting strings,
not every shrinking divisor gives rise to a {\it tower} of {\it weakly coupled} BPS states. 
As long as the divisor is contractible on $Y$ at some finite distance in the moduli space,
the resulting tensionless string is a non-critical string, and this
indicates the appearance of a strongly coupled superconformal point. 
An example is the type of divisors we introduced in eq.~(\ref{Spdivisor}). 

Towers of weakly coupled BPS excitations are expected to arise only from a tensionless {\it critical} string. 
Apart from the fibral divisors $S_p$,
 the divisor class group of a genus-one fibration $Y$ is generated by the (multi-)sections, the exceptional divisors responsible for the resolution of singularities in the fiber, and the vertical divisors. Recall that the latter are of the form
\be \label{DCbsec3}
D_{C_{\rm b}} =\pi^* C_{\rm b} \,,   \qquad C_{\rm b}  \in H^2(B_2) \,.
\ee
 For our purpose of identifying nearly massless BPS towers from asymptotically tensionless weakly coupled strings, 
 it suffices to focus on the last class of divisors.
If $C_{\rm b} \cdot_{B_2} C_{\rm b} <0$, the M5-brane gives rise to a non-critical string which is expected not to lead to a tower of BPS particles relevant for our discussion. 
In fact, after flopping the curve $C_{\rm b}$ into the fiber, the string is described by an M5-brane associated with a fibral divisor of the type discussed around eq.~(\ref{Spdivisor}) \cite{Morrison:1996pp,Klemm:1996hh}.
It therefore suffices to check the vanishing rate of ${\cal V}_{D_{C_{\rm b}}}$ for  $C_{\rm b} \cdot_{B_2} C_{\rm b}  \geq 0$. If such a curve vanishes on $B_2$, no flop to a non-flat fibration is possible.
Importantly, Lemma \ref{lemmaVDblimit} in Appendix \ref{App_Surface-Limits} guarantees that in a  limit of Type $T^2$, the volume  ${\cal V}_{D_{C_{\rm b}}}$     vanishes at the  rate 
\be \label{VDCTh1}
{\cal V}_{D_{C_{\rm b}}}  \sim \la^{-4 + \Delta}\,, \qquad \text{with} \quad \Delta > 0  \qquad  \text{if} \qquad C_{\rm b} \cdot_{B_2} C_{\rm b} \geq 0\,,
\ee
and therefore
\be
\frac{T}{M^2_{11}} = 2\pi {\cal V}_{D_{C_{\rm b}}} \sim \frac{1}{\lambda^{4 -\Delta}}  \qquad \Delta >0 \,.
\ee
The associated mass scale is parametrically subleading with respect to (\ref{M2onT2scale}). We hence first reach the 6d Kaluza-Klein scale before these strings become relevant in the effective theory.\footnote{In the context of 5d $N=1$ theories, ref.~\cite{Rud} has independently observed that a BPS string has vanishing tension at every boundary of the extended K\"ahler cone, whereas for an F-Theory limit, the string tension vanishes relative to the five-dimensional Planck scale, but remains finite relative to the six-dimensional Planck scale.}

\paragraph{Nested limits in 6d }On top of this partial decompactification, the effective six-dimensional theory may undergo a further infinite distance limit which leads to
 emergent tensionless, weakly coupled strings in six dimensions.
To determine when this happens, note that 
in the six-dimensional theory the strings from an M5-brane along the divisors (\ref{DCbsec3}) map to strings from D3-branes wrapping
 the curves $C_{\rm b}$.
The question is therefore under which conditions the tension of these strings vanishes,
as measured in the 6d frame, for curves with $C_{\rm b} \cdot_{B_2} C_{\rm b} \geq 0$.\footnote{As recalled above, if $C_{\rm b} \cdot_{B_2} C_{\rm b} < 0$, the associated string can become tensionless, however only at finite distance in moduli space.}

To answer this, we first match the Planck scales in $d=5$ and $d=6$, i.e. we identify
\be
\frac{M^3_{\rm Pl,5}}{M^3_{11}} = \frac{M^4_{\rm Pl,6}}{M^4_s} \,,
\ee
where $M_s = \ell_s^{-1}$ is the 10d string scale.
The left is given by the Calabi-Yau volume in the five-dimensional frame (\ref{MplM11} ), while the right corresponds to the volume of $B_2$ in the six-dimensional F-Theory frame,
\be
\frac{M^4_{\rm Pl,6}}{M^4_s}  = 4 \pi {\cal V}_{B_2,{\rm F}} =: 2 \pi \int_{B_2} J^2_{B_2,{\rm F}} \,.
\ee
The K\"ahler form in the six-dimensional frame is related to the pullback of K\"ahler form from the base on $Y$ via
\be
J_{B_2,{\rm F}} = {\cal V}^{1/2}_{T^2} J_{B_2} = \lambda^{-1} J_{B_2} \,.
\ee
In the six-dimensional frame, the volumes in units of $M_s$ are hence
\be \label{VolFandM}
{\cal V}_{B_2,\rm F} = \lambda^{-2} \,  {\cal V}_{B_2} \sim 1 \,,\qquad {\cal V}_{C_{\rm b},\rm F} = \lambda^{-1} \, {\cal V}_{C_{\rm b}}   \qquad \forall \, \, C_{\rm b} \in H_2(B_2,\mathbb Z) \,,
\ee
In the six-dimensional theory, the tension of the string obtained by wrapping $C_{\rm b}$ by a D3-brane in the limit $\lambda \to \infty$ is to be measured in units of $\ell_s$ and reads
\be
\frac{T}{M^2_s} = 2 \pi {\cal V}_{C_{\rm b}, \rm F}  \,.
\ee
Depending on the details of the limit, we find the following pattern:
Suppose  first that there exists a curve $C_{\rm b} = C_0$ on $B_2$ with $C_0 \cdot_{B_2} C_0 \geq 0$ such that in the M-Theory frame
\be
{\cal V}_{C_0} \prec \lambda \qquad \Longrightarrow \qquad {\cal V}_{C_0,\rm F} \prec 1 \,.
\ee
This means that the volume of $C_0 $ as measured in the 6d F-Theory frame vanishes asymptotically, while the volume ${\cal V}_{B_2,\rm F} $ stays finite.
As explained in Appendix \ref{App_Surface-Limits}, in this case the base $B_2$ itself
is a fibration with fiber $\mathbb P^1_f$ or $T^2_f$, and the curve $C_0$ is the generic fiber of one of these two fibrations.
In the first case, a D3-brane along $\mathbb P^1_f$ gives rise to a tensionless, weakly coupled heterotic string \cite{Lee:2018urn}. In the second situation a D3-brane on $T^2_f$ describes a weakly coupled tensionless Type II string whose physics will be discussed in more detail in Section \ref{sec_Dmanifolds}.
Therefore the situations in which a weakly coupled string becomes tensionless in an infinite distance limit of Type $T^2$ are precisely the cases where the three-fold $Y$ has an extra compatible K3 or $T^4$-fibration and the fiber vanishes at a rate ${\cal V}_{K3/T^4} \prec \lambda^{-1}$.

To summarise, 
 in a limit of Type $T^2$  which is also K3 or $T^4$-fibered and obeys ${\cal V}_{\mathbb P^1_f/T^2_f} \prec \lambda$, M-Theory on $Y$ is effectively described by an F-Theory compactification to six dimensions,
 which undergoes an extra weak coupling limit where a tensionless  heterotic or Type II string appears.
 The condition ${\cal V}_{\mathbb P^1_f/T^2_f} \prec \lambda$ is equivalent to stating that $   {\cal V}_{K3/T^4} \prec \lambda^{-1}$ for the $K3$ or $T^4$ fibers.
For ${\cal V}_{\mathbb P^1_f/T^2} \succsim \lambda$, or in absence of a compatible K3/$T^4$-fibration, M-Theory on $Y$ decompactifies to a six-dimensional F-Theory 
in a Type $T^2$ limit that does not undergo any further infinite distance limit, and no weakly coupled tensionless string appears.

\subsection{The Type K3 limit and an Emergent Heterotic String} \label{sec_TypeBK3M}

We will now describe the implications of taking a limit of Type K3 for M-Theory on a suitably fibered Calabi-Yau three-fold, $Y$.
Certainly the appearence of heterotic strings from K3 surfaces is a classic result \cite{Harvey:1995rn},
and K3-fibrations are familiar from refs.~\cite{Klemm:1995tj,Vafa:1995gm,Aspinwall:1995vk,Aspinwall:1996mn}. However,
our main point is the emergence of an asymptotically tensionless, critical heterotic
string, as a consequence of taking a truly equi-dimensional infinite distance limit. 
This is signified by two universally present types of towers of asymptotically massless BPS states. All-in-all we have:
\begin{enumerate}
\item
The asymptotically tensionless, weakly coupled heterotic string itself emerges in the infinite distance limit from a single M5-brane wrapping the K3-fiber of $Y$.
Its excitations furnish a tower of asymptotically massless BPS states of mass scale
\be \label{hetstringgen1}
\frac{M^2_{n}}{M^2_{\rm Pl}} \sim  n \, \frac{T_{\rm het}}{M^2_{\rm Pl}}  =   2\pi \, n \, {\cal V_{\rm K3}}  \frac{M^2_{11}}{M^2_{\rm Pl}}  = 2\pi \frac{n}{\lambda} \, {\cal V}^{1/3}_Y\,,
\ee
where $T_{\rm het}$ denotes the tension of the heterotic solitonic string and $\lambda$ is the scaling parameter that appears in
eq.~(\ref{VK3JB}).
\item
M2-branes wrapping $n$ times any distinguished curve class $C_{0} \subset K3$, with the property $C_{0} \cdot_{K3} C_{0} > 0$,
give rise each to an infinite tower of asymptotically massless BPS states at scale
\be  \label{hetstringgen2}
\frac{M^2_{n}}{M^2_{\rm Pl}} \sim (2 \pi)^2 n^2 {\cal V}^2_{C_{0}} \frac{M^2_{11}}{M^2_{\rm Pl}}    = (2\pi)^2 \frac{n^2}{\lambda} \,  {\cal V}^{2/3}_{Y}  \,.
\ee 
The BPS indices of these states are determined by the coefficients of some meromorphic modular form. In less generic situations, even higher-dimensional lattices of BPS towers can become light, for which  $C_0$ then generates a one-dimensional sublattice.
\item In addition, there exists a tower of ordinary Kaluza-Klein states associated with the scale
\be \label{hetstringgen3}
\frac{M^2_{\rm KK}}{M^2_{\rm Pl}} \sim \frac{1}{\lambda}  \frac{1}{{\cal V}^{4/3}_Y} \,.
\ee
These KK states are only suppressed by powers of the finite volume ${\cal V}_Y$ with respect to the BPS towers from Point 1 and 2. Therefore, they do not signal a decompactification limit.
\end{enumerate}
The first point indicates that in a limit of Type K3, one can change the duality frame to that of a weakly coupled and asymptotically tensionless string.
This explicitly realizes the famous duality \cite{Kachru:1995wm} 
\begin{center}
\begin{minipage}{4cm}
M-Theory on \\
K3-fibration   $Y$
\end{minipage}
\begin{minipage}{1cm}
$\longleftrightarrow$
\end{minipage}
\quad \quad \begin{minipage}{4cm}
Heterotic on \\
$\widehat K3 \times S^1_A$
\end{minipage}
\end{center}
Under this duality, an M5-brane wrapping the K3-fiber of $Y$ precisely once maps to a fundamental heterotic string that probes some dual geometry. As we will furthermore explain, the BPS particles from the M2-brane that wraps multiples 
of the distinguished curves $C_{0}$ in the K3-fiber of $Y$
map to certain excitations of the dual heterotic string wrapped on $S^1_A$.

To understand this, let us first recall the well-known fact that the solitonic string obtained by wrapping an M5-brane along a K3-surface is indeed described by a heterotic worldsheet theory \cite{Harvey:1995rn}.
The tension of this solitonic string in units of the 11d Planck mass is given by
\be
\frac{T_{\rm het}}{M^2_{11}} = 2\pi  {\cal V}_{K3} \,.
\ee
Together with (\ref{VK3JB}) and (\ref{MplM11}), this indeed
reproduces the parametric behaviour (\ref{hetstringgen1}) for the excitations of the heterotic sotlitonic string at level $n$.

Importantly, this tower of string excitations scales in the same way with $\lambda$ as the Kaluza-Klein tower, whose scaling is set by the inverse volume of the largest cycle of $Y$. From the discussion in Section \ref{SubsecTypeB}, we note that this cycle equals the 
base $\mathbb P^1_b$, and this leads to
\be
\frac{M^2_{\rm KK}}{M^2_{11}} \sim  \frac{1}{{\cal V}_{\mathbb P^1_b}} \sim    \frac{1}{\lambda} \,.
\ee

It remains to explain the appearance of the second type of BPS states, as referred to in (\ref{hetstringgen2}). 
For this we recall some relevant facts about K3-fibrations \cite{Aspinwall:1995vk,Aspinwall:1996mn}:
Since the K3 fiber is a divisor of $Y$ with embedding 
 \be
  \iota: K3 \hookrightarrow Y \,,
  \ee
 it contains at least one algebraic curve; such K3 surfaces are called algebraic.
For any such algebraic K3, the Picard lattice 
\be
{\rm Pic}(K3) = H^{1,1}(K3) \cap H^{2}(K3,\mathbb Z)
\ee
of algebraic curves is a lattice of signature $(1,\rho-1)$ with $1\leq \rho \leq 20$.  
Not all of the homologically independent algebraic curves of K3, viewed as curves or their dual divisors of $Y$, need to be homologically independent on $Y$.
In this case one defines instead a lattice
\be \label{Lambdalatticedef}
\Lambda = [\iota^* H^2(Y,\mathbb Z)]^\vee \,,
\ee
where by $[\ldots]^\vee$ we mean the dual lattice. This is a rational lattice of signature $(1, r-1)$ with $r \leq \rho$.\footnote{One can always find a basis $\{D_0= J_0, D_\mu, D_A\}$ of $H^2(Y,\mathbb Z)$ with the property that $D_0 \cdot D_\mu \neq 0$ and $D_0 \cdot D_A = 0$, where $\mu \in \{1,\ldots,r \}$. Then $\eta_{\mu \nu}:= D_0 \cdot D_\mu \cdot D_\nu$ has signature $(1,r-1)$.}
Upon suitably rescaling $\Lambda$, it is the lattice of {algebraic} curve classes of K3 which are homologically independent in $Y$. 

In the limit where the K3-fiber shrinks, these fibral curves necessarily shrink as well. Importantly for us, since the signature of $\Lambda$ is $(1,r-1)$, it is furthermore guaranteed that the K3 fiber contains a curve class $C_{0}$ with $C_{0} \cdot C_{0} > 0$, i.e. there exists a one-dimensional sublattice
\be \label{Lkdef1}
L_k := \{   C_0 \,  | \, C_0 \cdot C_0 =k > 0 \}\subseteq \Lambda \,.
\ee
In limits of Type K3,
such curves shrink at a rate as indicated in (\ref{VCK3Type}) and as derived in Appendix (\ref{SubsecTypeB}). Together with (\ref{MplM11}), this establishes the scaling behaviour (\ref{hetstringgen2}).

The crucial point is that given such a  fibral curve class $C_{0}$, one obtains a {\it tower of BPS states} by wrapping M2-branes $n$ times around $C_{0}$ for {\it infinitely many} values of $n$. In fact, it can be shown that
the 5d BPS index  - or Gopakumar-Vafa invariant  - associated with an M2-brane wrapping $n C_{0}$ is
\be \label{Nnc0bigger1}
N_{n \,  C_{0}} \neq 0 \qquad \forall n \geq 1  \,,
\ee
because  $C_{0} \cdot C_{0} > 0$.
As will be elaborated in more detail below, (\ref{Nnc0bigger1}) is a consequence of the related duality between Type IIA string theory on $Y$ and the heterotic string compactified on $\widehat K3 \times T^2$ as studied in \cite{Harvey:1995fq}.
By comparing the quantum corrected prepotential in the weak coupling limit of both sides, one concludes that the BPS invariants for a curve inside the K3 fiber with class $0 < \alpha \in \Lambda$
are counted by \cite{Harvey:1995fq}
\be \label{Nalpha}
N_{\alpha} = c\left(\frac{\alpha^2}{2}\right)    \,.
\ee
Here the $c(m)$ are the Fourier coefficients of some meromorphic modular form of weight $-2$:
\be
\Theta(q) = \sum_{m \geq -1} c(m) q^m \,.
\ee
While the specifics of the modular form $\Theta(q)$ depends on the model $Y$, it is universally true by modularity that for $m \geq -1$ its Fourier coefficients $c(m)$ are non-zero, i.e.
\be
N_{\alpha} \neq 0 \qquad \quad {\rm for} \, \, \alpha^2 \geq -2 \,.
\ee
 In particular, as long the curve $C_{0}$ satisfies $C^2_{0} \geq 0$, the BPS invariants for any multiple $n \, C_{0}$ are guaranteed to be non-zero. 
This is to be contrasted with the behaviour of negative self-intersection curves on the K3-fiber of $Y$, for which the BPS invariants with multiplicity bigger than one manifestly vanish.

Note that for a generic K3 fiber, only a one-dimensional sublattice of its Picard group gives rise to independent curve classes in $Y$.
In this case, $L_k = \Lambda$, and hence
 represents the {\it full} lattice of infinitely many asymptotically massless BPS states associated with wrapped M2-branes in the K3-fiber. More generally, the lattice of BPS states may contain additional towers, in particular due to curves of self-intersection zero in $\Lambda$ which are not contained in $L_k$ as defined in (\ref{Lkdef1}).

To explain the origin of the tower of BPS states associated with the (sub-)lattice $L_k$,
 it is necessary to scrutinize in more detail the duality with the heterotic string that underlies eqs.~(\ref{Nalpha}) and (\ref{Nnc0bigger1}). We split the discussion in two parts, starting first with non-generic K3-fibrations, whose K3-fiber is itself genus-one fibered, and then generalize to more generic fibrations. The reader not interested in the rationale behind the counting in (\ref{Nalpha}) may want to skip this discussion on a first reading.

\subsubsection{Genus-one fibered K3} 

Consider first K3 fibrations whose K3 fiber admits in addition a compatible genus-one
 fibration
\be \label{K3ellfibration1}
\ba
r :\quad T^2 \ \rightarrow & \  \ K3 \cr 
& \ \ \downarrow \cr 
& \ \  {\mathbb P}^1_f    
\ea
\ee
Apart from the presumed
existence of such a fibration, the following discussion is general and not restricted to any specific model.

Of special importance for us are the three following independent curve classes on $Y$ denoted respectively by
\be
C_S: = \mathbb P^1_b \,, \qquad C_U: = {\mathbb P}^1_f  + T^2 \,,  \qquad C_T:=T^2 \,,
\ee
where we recall from (\ref{TypeBfibrationstructure}) that $\mathbb P^1_b$ is the base of the K3-fibration of $Y$. The respective volumes in units of $\ell_{11}$ are given by
\be \label{STUdef}
{\cal V}_{C_S} =: S \,, \qquad {\cal V}_{C_U} =: U  \,, \qquad  {\cal V}_{C_T} =: T \,.
\ee
The divisors $D_i$ dual to the curves $C_i$ generate three universal abelian gauge symmetries, denoted  by
$U(1)_S$, $U(1)_T$ and $U(1)_U$, respectively. In particular, the divisor $D_S$ coincides with the K3-fiber. 
The corresponding gauge bosons sit in three 5d $N=1$ vector multiplets,
together with the scalar fields that parametrize the volumes of the respective curve classes.

For definiteness we assume the existence of a single
section for the fibration (\ref{K3ellfibration1}).\footnote{Our analysis can be easily
generalized to genus-one fibrations with a $k$-section, in which case
${{\mathbb P}^1_f  \cdot_{K3} T^2 = k}$.} 
Hence ${\mathbb P}^1_f  \cdot_{K3} T^2 = 1$, or  
\be
C_U \cdot_{K3} C_U = 0 
 \, \qquad C_T \cdot_{K3} C_T = 0 \,,\qquad \qquad C_T \cdot_{K3} C_U = 1 \,.
\ee
Consequently, $C_U$ and $C_T$ generate a hyperbolic sublattice,
\be \label{HinLambda}
{\bf H} \subseteq \Lambda \subseteq {\rm Pic(K3)} \,,
\ee  
of signature $(1,1)$ within the lattice $\Lambda$ defined in (\ref{Lambdalatticedef}). Depending on the specific realisation of the elliptic fibration (\ref{K3ellfibration1}), there may appear additional curve classes in $\Lambda$ which populate the negative definite piece of $\Lambda$.
From our general discussion we know that curve classes lying entirely in this negative-definite, model-dependent sublattice do not generate additional towers of BPS states by themselves, but only finitely many BPS particles.
Without loss of generality we can therefore assume that $\Lambda={\bf H}$, or else focus on the relevant hyperbolic sublattice of  $\Lambda$.

Our goal is to understand the tower of BPS states arising from wrapped M2-branes along the fibral curve class
\be \label{Ckldefsect3}
C_{k,l} = k \, C_T + l \, C_U \qquad {\rm with} \qquad C_{k,l} \cdot_{\rm K3} C_{k,l}  = 2 \, k \, l
\ee
for $k l \geq 0$, in terms of suitable modes of the heterotic string wrapped on $S^1_A$.
To make contact with (\ref{Lkdef1}), we note that the curve $C_{1,1}$ generates a sublattice
\be \label{L2sublattice}
L_2 \subset \mathbf H \subseteq \Lambda \,.
\ee
The states associated with $C_{k,l}$ hence include precisely the tower of light BPS particles advertised as a characteristic feature of limits of Type K3 at the beginning of this section.

This tower of BPS states can be further characterized as follows:
\begin{enumerate}
\item
F/M-Theory duality implies the identification
\begin{center}
\begin{minipage}{5cm}
 5d BPS tower of M2-brane states on  \\
 $C_{\rm M2} = C_{k,l} \equiv k \, C_T + l \, C_U$
\end{minipage}
\qquad \begin{minipage}{1cm}
$\longleftrightarrow$
\end{minipage}
\quad \begin{minipage}{5cm}
Heterotic string  wrapped on 
$S^1_A$  with wrapping number $l$ and KK momentum $k$ 
\end{minipage}
\end{center}
\item
The BPS numbers of the tower of M2-branes states on $C_{k,l}$ are counted as in (\ref{Nalpha}) with $\alpha = 2 k l$. In particular, this implies an infinite tower of asymptotically massless BPS states.
\end{enumerate}
The first point is a consequence of the following chain of dualities for M-Theory on a K3-fibration $Y$ that has
 a compatible $T^2$ fibration induced by (\ref{K3ellfibration1}):
\begin{center}
\begin{minipage}{3cm}
Heterotic on \\
 $\hat K3 \times S^1_F$
\end{minipage}
\begin{minipage}{1cm}
$\longleftrightarrow$
\end{minipage}
\quad \begin{minipage}{3cm}
F-Theory on \\
 $Y \times S^1_F$
\end{minipage}
\begin{minipage}{1cm}
$\longleftrightarrow$
\end{minipage}
\begin{minipage}{2.5cm}
M-Theory  \\
on $Y$
\end{minipage}
\begin{minipage}{1cm}
$\longleftrightarrow$
\end{minipage}
\begin{minipage}{3cm}
Heterotic on \\
$\hat K3 \times S^1_A$
\end{minipage}
\end{center}

Before explaining the implications of these dualities for the BPS states, let us point out that 
 the heterotic string compactified on $\widehat K3 \times S^1_A$ gives rise to three {\it universal} types of abelian gauge groups, which parallel the appearance of the abelian gauge group factors in M-Theory reviewed above:
 The heterotic string on $S^1_A$ at generic radius  $R_A$ gives rise to a gauge group $U(1)_L \times U(1)_R$  that corresponds to the KK reduction of the 6d metric and 2-form field $B_2$ on $S^1_A$. A third universal gauge field
 is obtained by reduction of the dual of the 10-dimensional tensor field $B_6$ on $\widehat K3 \times S^1_A$.  Additional gauge fields correspond to the unbroken piece of the 10d heterotic $E_8 \times E_8$ gauge group, whose sublattice of charged BPS states maps to the negative-definite sublattice of $\Lambda$. As noted above, this plays
no role in our context. We henceforth assume that the only massless gauge fields on the heterotic side are given by the three universal abelian gauge factors. 
 They form 5d $N=1$ vector multiplets involving suitable combinations of the three independent real scalar fields, which are associated with the volume modulus ${\cal V}_{\widehat K3}$, the radius $R_A$,\footnote{Measured in units of the heterotic string scale, $\ell_{\rm het} = M^{-1}_{\rm het}$.} and the heterotic dilaton $g_{\rm het} = e^{\phi_{10}}$, respectively.

Now, the above identification of BPS states is a special case of the more general  relation \cite{Klemm:1996hh} between wrapped solitonic strings in F-Theory on $Y \times S^1_F$, and BPS particles in M-Theory on~$Y$:
On the F-Theory side, consider a solitonic string obtained by wrapping a D3-brane along a curve $C_{\rm b}$ on the base $B_2$ of $Y$. Wrapping this solitonic string $l$ times along $S^1_F$,  with KK momentum $k$, yields a BPS particle that is obtained 
in the dual M-Theory from wrapping an M2-brane along the curve 
\be \label{CMShet}
C_{\rm M2} = l \, C_{\rm b} + (k -  l \, E_0) \, T^2 \,.
\ee 
Here $T^2$ is the fiber of $Y$, and
the offset, $E_0$, reflects the quantized Casimir energy along the single-wrapped string on $S^1_F$ \cite{Lee:2018urn}.
In the present context, we specialise to the solitonic heterotic string in F-Theory obtained by wrapping a D3-brane along the curve $C_{\rm b} = {\mathbb P}^1_f$, with \cite{Lee:2018urn}
\be
E_0 =  - \frac{1}{2}    C_{\rm b} \cdot_{B_2} \bar K = - \frac{1}{2} \mathbb P^1_f  \cdot_{B_2} \bar K = -1 \,.
\ee
Thus the curve becomes
\be
C_{\rm M2} = l \, (\mathbb P^1_f + T^2 ) + k T^2  \equiv  l \, C_U + k \,  C_T  = C_{k,l} \,,
\ee
which establishes the claim above.

This identification is readily checked at the level of masses.
The mass of an M2-brane wrapping the curve $C_{k,l}$ on $Y$ in units of the 5d Planck mass $M_{\rm Pl}$, as computed in the M-Theory frame, is given by
\be \label{M2klM2side}
\frac{M^2_{k,l}}{M^2_{\rm Pl}}  = (2 \pi)^2 \,  {\cal V}^2_{C_{k,l}}  \,  \frac{M^2_{11}}{M^2_{\rm Pl}}  =   (2 \pi)^2   \,     \left(k {T}  + l {U}  \right)^2   \,  \frac{M^2_{11}}{M^2_{\rm Pl}} \,.   
\ee
Here the curve volumes (\ref{STUdef}) are measured in units of $\ell_{11} = M^{-1}_{11}$.
A heterotic string with wrapping number $l$ and KK momentum $k$, on the other hand, has mass:
\be \label{M2klhetside}
\frac{M^2_{k,l}}{M^2_{\rm Pl,\rm het}} = (2 \pi)^2 \left( l R_A    +  \frac{k}{R_A}  \,   \right)^2 \frac{M^2_{\rm het}}{M^2_{\rm Pl,het}}   \,.
\ee
Here $R_A$ measures the radius of $S^1_A$ in units of $\ell_{\rm het} = M^{-1}_{\rm het}$, and
the Planck mass as computed in the heterotic duality frame is given by
 \be \label{Mplhetframe}
 \frac{M^3_{\rm Pl, {\rm het}}}{M^3_{\rm het}} = 4 \pi \frac{{\cal V}_{\widehat K3}  R_A}{g^2_{\rm het}} \,.
 \ee
In the limit, we identify the solitonic heterotic string that arises from an M5-brane wrapping the K3-fiber of $Y$,
 with the fundamental heterotic string of the dual frame.
This amounts to matching their tensions as measured in units of the Planck mass via 
\be
\frac{T_{\rm het}}{M^2_{\rm Pl}}  =  \frac{2 \pi M^2_{\rm het}}{M^2_{\rm Pl, {\rm het}}} \,,
\ee
and hence
\be
\frac{{\cal V}_{\rm K3}}{{\cal V}^{2/3}_Y} = \frac{g^{4/3}_{\rm het}}{({\cal V}_{\widehat K3}  R_A)^{2/3}} \,.
\ee
 Comparing now the mass formulae (\ref{M2klM2side}) and (\ref{M2klhetside}), we find 
in particular that
\be
\frac{T}{U} = \frac{1}{R_A^2} \,.
\ee

For later purposes, let us note that we can identify the $U(1)_{\rm KK}$ Kaluza-Klein gauge symmetry associated with $S^1_A$, as well as the 
`winding $U(1)_{\rm w}$' (under which an $l$-times wrapped string has charge $q_{\rm w} = l$), with the abelian gauge groups on the M-Theory side as follows:
\bea
U(1)_{\rm KK} &=& U(1)_T \,, \\
U(1)_{\rm w} &=& U(1)_U \,.\nn
\eea
As an important check, note also that at the self-dual radius $R_A =1$, the left-moving $U(1)_L$ of the heterotic string enhances to $SU(2)_L$.
On the M-Theory side we therefore expect a massless BPS particle that represents the W-boson. Indeed, for $T=U$ the holomorphic curve $C_{(1,-1)} = U - T = \widehat{\mathbb P}^1_f$ approaches zero volume, and the expected massless W-boson is furnished by
an M2-brane wrapping this curve.

The final task is to understand the counting of the BPS states, and in particular to prove that the states associated with $C_{k,l}$ generally give rise to an infinite tower of states.
From M-Theory/F-Theory duality, we know that the Gopakumar-Vafa invariants counting the BPS states along the curve $C_{k,l}$ are encoded in the elliptic genus of the wrapped heterotic string. 
This is reflected by the fact that the meromorphic modular form, which also characterizes the 1-loop quantum corrections to the prepotential for the heterotic compacitifcation on $\widehat K3 \times S^1_A$, has the following infinite expansion \cite{Harvey:1995fq,Kawai:1996te}
\be
\Theta(q) = 2 \frac{E_4(q) E_6(q)}{\eta^{24}(q)} = -\frac{2}{q} + 480 + \ldots =: \sum_{n \geq -1, n \in \mathbb Z} c(n) q^n \,.
\ee
This coincides precisely with the elliptic genus   \cite{Schellekens:1986xh}
of the heterotic string on any $K3$ surface (and in particular on the dual surface we denoted by  $\widehat K3$).
In particular the BPS invariants for the curve $C_{k,l}$ are given by 
\be
N_{C_{k,l}} = c\left(\frac{1}{2} C^2_{k,l}\right) = c( k l) \,.
\ee
It has been verified, e.g. in \cite{Kawai:1996te,Klemm:2004km}, for explicit realizations of K3-fibrations $Y$ of the type under consideration, that these coefficients indeed correctly reproduce the genus-zero Gopakumar-Vafa invariants associated with the curve classes $C_{k,l}$.

To summarize, we have established the interpretation of the lattice $L_2$ in (\ref{L2sublattice}) in terms of BPS excitations due to the wrapped heterotic string with equal quanta of winding and momentum: $k=l$. As noted, this is only a sublattice of the tower of BPS states which become asymptotically massless as the K3-fiber shrinks: For each $C_{k, l}$ with $k \,  l \geq -1$, there exists such a BPS state.
Specifically there is an infinite tower of states with $l=0$ for arbitrary $k$. These states represent the $k$-th Kaluza-Klein states associated with the massless spectrum of the heterotic string on $\widehat K3$.
In agreement with this, the BPS index $N_{C_{k,0}} = c(0)$ coincides with the Euler characteristic of the Calabi-Yau $Y$.

\subsubsection{One-parameter K3}

For the previously discussed class of K3-fibrations, which have an extra, compatible genus-one fibration,
 the infinite tower of BPS states that arise
from wrapped M2-branes is encoded in the hyperbolic sublattice (\ref{HinLambda}) of the full BPS lattice, $\Lambda$. 
As discussed around (\ref{Lkdef1}), 
for more generic K3-fibrations the lattice responsible for the BPS tower can be even smaller, and in the extreme case
is solely given solely by the one-dimensional sublattice
\be
\Lambda = L_k\,,
\ee
which is generated by some element $f$ with $f^2 = k$. 
Our claim is that the resulting tower of BPS states is still related to a wrapped heterotic string.

To understand this, we start from an elliptically fibered K3-fibration as discussed above, and go to the point of self-dual radius $R_A = 1$ of the heterotic dual, where $U(1)_L$ enhances to $SU(2)_L$.
As explained in \cite{Kachru:1995wm} in the context of Type IIA - heterotic duality in four dimensions, one can now switch on an instanton background described by a vector bundle on $\widehat K3$ with structure group $SU(2)_L$.
This breaks the $SU(2)_L$ gauge group and reduces the rank of the gauge group by one. In the present 5d context it furthermore freezes $R_A=1$ so that we cannot take a 6d limit any longer.

Our point is that there still exists a tower of 5d BPS particles obtained
by wrapping the fundamental heterotic string on $S^1_A$, however with mutually locked momentum and winding numbers.  Indeed, the Cartan $U(1)_L$ of $SU(2)_L$ is given by the linear combination\
\be
U(1)_L = U(1)_{\rm KK} - U(1)_{\rm w} \equiv U(1)_{T} - U(1)_{U}  \,.
\ee
This follows, for instance, from our identification of the $SU(2)_L$  W-boson, which carries $U(1)_L$ charge $q_L = 2$, with the BPS state with charges $q_U= 1$ and $q_T = -1$ on the M-Theory side.
Breaking $SU(2)_L$ and thus $U(1)_L$ by a gauge background therefore breaks
\be
U(1)_{\rm KK} \times U(1)_{\rm w}   \longrightarrow U(1)_+  = \frac{1}{2} (U(1)_{\rm KK} + U(1)_{\rm w}) \,.
\ee
This leaves only states with charge $q_{\rm w} = q_{\rm KK}$ in the spectrum, i.e. states with equal momentum and winding numbers
(we normalize here $U(1)_+$ charge lattice
such that the minimal charge, corresponding to $q_{\rm KK} = q_{\rm w}=1$, is $q_+ =1$). 

As for the counterpart of the gauge symmetry breaking on the M-Theory side, note that only formal multiples of the curve $\frac{1}{2} C_{1,1}$ can be realized geometrically. The important  factor of $\frac{1}{2}$ reflects the correct normalization of $U(1)_+$ charge lattice, 
which translates into integrality conditions on the curve classes.
One can think of this as the result of a topology change from the elliptically fibered Calabi-Yau $Y$ to a new K3-fibration, denoted by $\hat Y$,
 due to an extra monodromy action on the fibral curve classes $C_U$ and $C_T$,
\be
Y        \xrightarrow[\text{background}]{\ {SU(2)_L} \ }        \hat Y \,.
\ee

 At the self-dual point $T=U$ on $Y$, the volumes of $C_U$ and $C_T$ are equal and we can therefore envisage a monodromy action exchanging $C_U$ and $C_T$ along closed paths on $\mathbb P^1_b$. The result of such a monodromy would be to project out all curve classes $C_{k,l}$ with $k \neq l$, keeping only integer multiples of the properly normalised class $\frac{1}{2} C_{1,1}$ with $\frac{1}{2} C_{1,1} \cdot \frac{1}{2} C_{1,1} = \frac{1}{2}$.
Let us denote the curve corresponding to $\frac{1}{2} C_{1,1}$ by $C_0$.
This relates to a BPS sublattice 
\be
L_{{1}/{2}} \subseteq \Lambda
\ee
pertaining to the new Calabi-Yau, $\hat Y$. 

The tower of M2-branes wrapping the fibral curves $k C_0$, for $k \in \mathbb Z$, within $\hat Y$, therefore corresponds to a wrapped heterotic string with wrapping number and Kaluza-Klein momentum $k$ along the circle $S^1_A$,  at fixed self-dual radius $R_A = 1$, after switching on an $SU(2)_L$ bundle on the dual heterotic $\widehat K3$.
The BPS numbers associated with this tower can again be read off from the quantum corrected prepotential of the heterotic theory. Even without resorting to an explicit realisation of the geometry, modularity guarantees the existence of a tower of BPS states because the BPS invariants for $k \geq 1$ are non-vanishing.

As a concrete example, we consider 
the Calabi-Yau 3-fold $\hat Y = \mathbb P^5_{1,1,2,2,6}[12]$, which has been studied in great detail in the literature beginning with \cite{Candelas:1993dm,Hosono:1993qy}.
The 3-fold is a K3-fibration over base $\mathbb P^1_b$, which spans the Mori cone along with a rational curve $C_0 :=\mathbb P^1_f$ in the K3-fiber,~ie.,
\be
{\mathbb M}(\hat Y) =  \langle \mathbb P^1_b, \mathbb P^1_f \rangle \,.
\ee
The dual K\"ahler cone generators $J_S$ and $J_T$, defined as
\bea
&J_S \cdot \mathbb P^1_b = 1 \qquad & J_S \cdot \mathbb P^1_f = 0 \\
&J_T \cdot \mathbb P^1_b = 0 \qquad & J_T \cdot \mathbb P^1_f = 1 \,,
\eea
have the following triple intersection numbers
\be
J_T^3 = 4 \,, \quad J_T^2 J_S= 2 \,, \quad J_S^2 J_T = 0 \,, \quad J_S^3 = 0 \,.
\ee
Note in particular that $J_S$ is the class of the K3 fiber of $\hat Y$.
If we parametrise the classical K\"ahler form as
\be \label{JforexK31}
J = T J_T + S J_S \,,\ee
the curve volumes in units of the string length are given by
\be
{\cal V}_{\mathbb P^1_b} = S \,,\quad \quad {\cal V}_{\mathbb P^1_f} = T \,.
\ee
Classically the limit
\be \label{classlimitB}
S  = \lambda \, , \qquad T = \frac{a}{\sqrt{\lambda}} \,, \qquad \lambda \to \infty\,, \qquad a \, \,  {\rm finite}\,,
\ee
realizes a limit of $J$-class B with the property that asymptotically
\be \label{limitclassical}
{\cal V}_{Y} \to \frac{1}{3} a^2 \,, \quad {\cal V}_{K3} = \frac{a^2}{\lambda}  \to 0 \,,\quad \, {\cal V}_{\mathbb P^1_b} = \lambda \to \infty \,, \quad {\cal V}_{\mathbb P^1_f} = \frac{a}{\sqrt{\lambda}} \to 0 \,.
\ee
This is indeed the characteristic behaviour for a limit of Type K3.

Now, the lattice $\Lambda$ as defined in (\ref{Lambdalatticedef}), coincides with the lattice $L_{1/2}$.
The associated heterotic prepotential was computed in \cite{Kawai:1996te}. The BPS numbers are encoded in a meromorphic modular form with respect to the modular subgroup $\Gamma_0(4)$, whose expansion reads
\bea
\Theta(q) &=& 2 \frac{E_4(q) \hat G_6(\tau)}{\eta^{24}(q)} = \frac{2}{q} - 252 - 2496 q^{1/4} - 223752 q - 725504 q^{5/4} + \ldots \\
&=& \sum_{n \in \mathbb Z \cup \frac{1}{4}\mathbb Z} c(n) q^n \,.
\eea
For the definition of the   meromorphic modular form $\hat G_6(\tau)$ we refer to  \cite{Kawai:1996te}.
The BPS numbers for the curve class $k \, C_0$, as predicted by the duality with the wrapped heterotic string, then follow as
\be
N_{k C_0} = c\left( k^2/4 \right) \,.
\ee
These indeed match \cite{Kawai:1996te} the Gopakumar-Vafa invariants on $\hat Y = \mathbb P^5_{1,1,2,2,6}[12]$ as computed in \cite{Candelas:1993dm,Hosono:1993qy,Henningson:1996jf}.  Further examples of K3-fibrations with a rank one lattice $\Lambda$ and their BPS numbers have been studied in \cite{Klemm:2004km}.

\subsection{Limits of Type $T^4$  and Type IIB theory on D-manifolds} \label{sec_Dmanifolds}

For the remaining type of infinite distance limits, as classified in Section \ref{classification-result}, the shrinking fiber is an Abelian surface, i.e. an algebraic torus of complex dimension 2.  Topologically this is the same as a real $4$-torus, $T^4 \simeq S^1 \times S^1 \times S^1 \times S^1$. 
Any product of two elliptic curves is an obvious example. However, a generic Abelian surface is not of this simple kind and indeed
Calabi-Yau three-folds can admit more general Abelian surface fibrations than direct products of elliptic curves.

Infinite distance limits of Type $T^4$  in M-Theory are again equi-dimensional and
form in some sense the middle ground between limits of Type $T^2$ and Type K3, by
sharing properties with each of them. In particular they follow the pattern of the limits of Type K3 studied in the previous section. That is, they exhibit asymptotically massless BPS towers whose structure and 
mass scales exactly parallel the three types of states and mass scaling behaviour listed in (\ref{hetstringgen1}), (\ref{hetstringgen2}) and (\ref{hetstringgen3}).   
The main difference is in the nature of the string described by the wrapped, asymptotically tensionless M5-brane, and the theory to which the M-Theory is dual in this limit.
In fact, an M5-brane wrapping a shrinking Abelian surface fiber gives rise to a weakly coupled, asymptotically tensionless Type II string, rather than to a heterotic string as in (\ref{hetstringgen2}).
A first hint comes from the duality 
\begin{center}
\begin{minipage}{4cm}
Heterotic on $T^4$ 
\end{minipage}
\begin{minipage}{1cm}
$\longleftrightarrow$
\end{minipage}
\quad \quad \begin{minipage}{4cm}
Type IIA on K3\,,
\end{minipage}
\end{center}
by which the heterotic NS5-brane wrapped on $T^4$ leads to a solitonic IIA string.
In the Horava-Witten picture of the heterotic string, the heterotic NS5-brane turns into an M5-brane, which leads to our claim. 

In the limit of vanishing fiber volume, M-Theory on an Abelian surface fibration $Y$   therefore reduces to Type II string theory compactified on
some dual background. This raises the question as to what the nature of this background is.
In order to preserve the same amount of 8 real supercharges, the dual fundamental string cannot probe a completely flat
geometry, rather than a geometry equipped with suitable extra defects.
As we will argue, the duality is of the form
\begin{center}
\begin{minipage}{6cm}
M-Theory on \\
Abelian surface fibration $Y$
\end{minipage}
\begin{minipage}{1cm}
$\longleftrightarrow$
\end{minipage}
\quad \quad \begin{minipage}{4cm}
Type IIB theory \\
$Z \times S^1_A$\,,
\end{minipage}
\end{center}
where $Z$ is a D-manifold in the sense of \cite{Bershadsky:1995sp}. 
The Type IIB background is in general a non-geometric background as will be explained at the end of this section.
The scalings (\ref{VT4sec2}) and (\ref{VCT4Type}) in the infinite distance limit of M-Theory translate into
\be  \label{VZscaling}
{\cal V}_{Z} \sim \lambda^2 \,, \qquad  \gB \sim {\lambda} \,,  \qquad R_A \sim  1
\ee
of the dual, five-dimensional Type IIB theory.

As a consequence of this duality, we can understand the origin of the BPS tower arising from wrapped M2 branes
as the analogue of the tower for K3 as characterized in (\ref{hetstringgen1}):
Now this tower is due to wrapped Type IIB strings with some extra KK momenta along $S^1_A$. 
To describe this tower more quantitatively, we must determine the lattice 
\be \label{Lambdalatticedef-b}
\Lambda = [\iota^* H^2(Y,\mathbb Z)]^\vee \,,
\ee
where $\iota: {\cal S} \to Y$ denotes the embedding of
 the Abelian surface fiber into $Y$, and again identify curves of non-negative self-intersection within $\Lambda$.

In the sequel we will first assume that the Abelian surface is a product of two elliptic curves.
In this case, the lattice $\Lambda$ is the hyperbolic lattice $\Lambda = {\bf H}$, and hence there exists a 2-dimensional lattice of BPS states giving rise to a tower of light particles from wrapped M2 branes.
This is very similar in spirit to a elliptically fibered K3-surface, as discussed in the previous section.
For more generic Abelian surface fibrations, the lattice of BPS states is only one-dimensional, and we will briefly discuss these
at the end of the next section.

\subsubsection{$T^4 = \cE_1 \times \cE_2$: Schoen manifold}\label{Schoen}

As laid out above, let us assume for now that the Abelian surface fiber is a product of two elliptic curves, i.e. there exists a fibration
of the form
\bea \label{E1E2} \nn
\pi : \;\; \cS \simeq \cE_1 \times \cE_2 \;\; \rightarrow \;\;Y \\ 
\downarrow \\ \nn
\mathbb P^1_b \,
\eea
In order for the three-fold $Y$ to exhibit genuine $SU(3)$ holonomy, both elliptic curves must be non-trivially fibered over 
$\mathbb P^1_b$. Here, as in the rest of this article, we are only interested in smooth such three-folds.

Note that these can also be viewed as a fiber product,
\be
\tilde Y = \cB_1 \times_{\mathbb P^1_b} \cB_2 \,,
\ee
of two smooth ``relatively minimal'' rational elliptic surfaces $\cB_1$ and $\cB_2$, or, in the presence of singularities, as a suitable blowup thereof.
These go by the name of Schoen manifolds \cite{Schoen}. 
Recall that a ``relatively minimal'' rational elliptic surface can be obtained from $\mathbb P^2$ by blowing up nine points in suitably non-generic positions, and that it is called ``relatively minimal'' because the elliptic fiber contains no $(-1)$ curves.
We will refer to such surfaces informally as dP$_9$ (``del Pezzo-nine") surfaces.

One can equivalently view the fibration
 $Y$, as defined in (\ref{E1E2}), as an elliptic fibration  over the rational elliptic surfaces ${\cal B}_i$ in two different ways,
\bea \label{E1} \nn
\pi_i : \;\; \cE_i \;\; \rightarrow \;\;Y  \nn \\  
\downarrow \nn \\  
\rho_i : \;\; \cE_{j} \;\; \rightarrow \;\;\cB_i \\ 
\downarrow \nn \\ \nn 
\mathbb P^1_b \,
\eea
for $(i,j)=(1,2)$ or $(2,1)$.
For example, the discriminant locus of the elliptic fibration $\pi_1$ is then in the class
\be
\Delta_1  = 12  \, \bar K_{\cB_1}  = 12 \,  \cE_2 \,,
\ee
where we used that 
 the  anti-canonical divisor of a dP$_9$ surface equals its elliptic fiber, and similarly for the fibration $\pi_2$.
 As is manifest in the representation given in (\ref{E1}), each $\cB_i$, $i = 1,2$, is an elliptic fibration over $\mathbb P^1_b$, with generic fibers $\cE_j$, $j=2,1$, respectively. The generic fibers of $\cB_i$ degenerate into  Kodaira fibers over a set of points $S^{(i)} = \cup_i \, p^{(i)}_\alpha$ on $\mathbb P^1_b$.
 For Kodaira fibers different from a nodal (Kodaira type $I_1$) or cuspidal (Kodaira type $II$) curve, this means that the fiber splits into a chain of intersecting $(-2)$ curves.

As described in detail in \cite{Schoen}, if $\Delta_1$ is a union of non-degenerate fibers of  ${\cB_1}$, the total Euler characteristic vanishes, $\chi(Y)=0$.
In this case
\be \label{delta1charact}
\Delta_1 = \sum_{a=1}^{12}  \rho_1^{-1} (Q_a)\,, \qquad Q_a \notin S^{(1)} \,,
\ee
with the understanding that some points $Q_a$ can be the same.
If $\Delta_1$ contains also some of the degenerate fibers of ${\cB_1}$, then $\chi(Y) \neq 0$.

After this preparation, we now turn to our goal of
 analyzing the infinite distance limit of Type $T^4$ for the direct product fibration $Y$ given by (\ref{E1E2}). The limit corresponds to taking
\be \label{SchoenLimit1}
{\cal V}_{\mathbb P^1_b} \sim \lambda \,, \, \qquad {\cal V}_{\cal S} \sim \frac{1}{\lambda} \, \qquad \lambda \to \infty\,,
\ee
and we impose in addition that both elliptic curves $\cE_1$ and $\cE_2$ {\it scale in the same way} as
\be \label{SchoenLimit2}
{\cal V}_{\cE_1} \sim \lambda^{-1/2} \,,\qquad {\cal V}_{\cE_2} \sim \lambda^{-1/2} \,.
\ee
The latter condition is required from the classification (\ref{VCT4Type}),
 and ensures that the limit is genuinely different from the limit of Type $T^2$.

Our idea is to understand the physics of this limit by first considering the more generic, asymmetric
 limit (\ref{SchoenLimit1})  in such a way that
\be \label{SchoenLimit3}
{\cal V}_{\cE_1} \sim \lambda^{-\frac12 - x} \,,\qquad {\cal V}_{\cE_2} \sim \lambda^{-\frac12 +x}  \,,\qquad x >0 \,,
\ee
rather than as in (\ref{SchoenLimit2}), and then taking $x \to 0$ in a second step.
Due to the temporary asymmetric scaling, (\ref{SchoenLimit3}) then realizes  a limit of Type $T^2$, which takes us to F-Theory on an elliptic fibration with projection $\pi_1$ and base $\cB_1$ as in (\ref{E1}). As discussed around (\ref{VolFandM}), to compute the volumes in the 6d F-Theory frame,
we rescale the K\"ahler form on $\cB_1$ by scaling out the factor of ${\cal V}^{-1/2}_{\cE_1} = \lambda^{1/4 + x/2 }$ from the M-Theory frame volumes. The volume of the base curve $C$ in the F-Theory frame then becomes
\be
{\cal V}_{C,\rm F} = \lambda^{-\frac{1}{4} - \frac{x}{2} } \,{\cal V}_{C}  \,,
\ee
i.e.
\be \label{dP9limit}
{\cal V}_{\cE_2,\rm F} = \frac{1}{\mu}  \to 0 \,,\qquad \quad {\cal V}_{\mathbb P^1_b,\rm F} = \mu \to \infty \,, \qquad \quad \mu = \lambda^{\frac34 - \frac{x}{2} } \,.
\ee

For sufficiently small $x >0$, we thus have arrived 
at a six-dimensional compactification of F-Theory on a base $\cB_1$, which by itself is elliptically fibered;  the volume of the generic fiber $\cE_2$, as measured in the six-dimensional frame, vanishes  as ${\cal V}_{\cE_2,\rm F} = \frac{1}{\mu} \to 0$, such that the total volume of
 $\cB_1$ stays finite.
 As $\mu \to \infty$ a tensionless string emerges from a D3-brane wrapping a generic elliptic fiber ${\cE_2}$.
This situation is very similar to the kind of infinite distance limits studied for six-dimensional F-Theory in \cite{Lee:2018urn}, the difference being that here an elliptic fiber rather than a rational fiber of the F-Theory base asymptotes to zero size 
at finite base volume (as measured in the F-Theory frame).\footnote{See Appendix \ref{App_Surface-Limits} for a summary of infinite distance limits for six-dimensional F-Theory.}

As we will argue, the asymptotically tensionless string is a Type II string probing a rather non-trivial background. 
We first carefully analyze the consequences of the limit (\ref{dP9limit}) in the six-dimensional F-Theory regime. In particular, we will discuss how certain BPS invariants on $Y$ reflect the elliptic genus of a Type II string on a highly non-perturbative and in general non-geometric background. We then consider the implications for the symmetric limit of the form (\ref{SchoenLimit2}), which is of our actual main interest.

\paragraph{Limits of Type $T^2$} For the six-dimensional F-Theory, it is well-known (see e.g. \cite{Weigand:2018rez} for a review) that the discriminant locus on $\cB_1$ is wrapped by 7-branes.
For $\chi(Y) = 0$, we have characterized the discriminant in (\ref{delta1charact}).
This means that the 7-branes wrap 12 copies of the generic elliptic fiber $\cE_2$ of $\cB_1$. 
For  $\chi(Y) \neq 0$, on the other hand, 
there are now 7-branes also wrapping some of the degenerate fibers of $\cB_1$, i.e. that there are 7-branes on curves which either self-intersect (as in the case of Kodaira fibers of Type $II$ or Type $I_1$), or on chains of mutually intersecting (-2) curves.

We focus now on the solitonic string that arises from wrapping a D3-brane on $\cE_2$.
By the same arguments as in \cite{Lee:2019xtm}, this string is a Type IIB string which in the limit (\ref{dP9limit}) becomes tensionless and weakly coupled.
Asymptotically it takes the role of a fundamental Type IIB string probing a highly non-perturbative, dual six-dimensional background, which we now describe.

The non-perturbative nature of this dual background can be inferred already from the fact that 
 the gauge theory on the 7-branes becomes strongly coupled in the limit (\ref{dP9limit})  because the inverse gauge coupling is proportional to ${\cal V}_{\cE_2,\rm F}$.
Its physics is studied easiest by performing two T-dualities along the elliptic fiber $\cE_2$ such as to enter the large volume regime.
Taking into account the effect of T-duality on the Type IIB dilaton, we arrive at Type IIB string theory, on a background where there are certain defects localised on the base $\cB_1$. This applies to the regime 
\be \label{dP9limitTdual}
\hat {\cal V}_{\cE_2,\rm F} = {\mu}  \to \infty \,,\qquad \quad {\cal V}_{\mathbb P^1_b,\rm F} = \mu \to \infty \,, \qquad \gB = \mu  \,  \gB^{(0)}    \,,
\ee
where $\gB^{(0)}$ is a reference value for the string coupling prior to T-duality.\footnote{Due to the variation of the axio-dilaton along $\cB_1$ in F-Theory this can be thought of as an asymptotic value away from the 7-brane positions, which can be taken to be small.}   
Note that the six-dimensional Planck scale stays invariant under T-duality,  as a consequence of the co-scaling of $\gB$. 
T-duality furthermore maps 
\begin{center}
\begin{minipage}{7cm}
D3 on ${\cE_2}$ \vspace{1mm}   \\
(p,q) 7-brane on   $\rho_1^{-1}(Q_a)$  ($Q_a \notin S^{(1)}$)
\end{minipage}
\begin{minipage}{1cm}
$\longrightarrow$
\end{minipage}
\quad \quad \begin{minipage}{6cm}
unwrapped D1 string  \vspace{1mm}  \\
(p,q) 5-branes on $\sum_{a=1}^{12} Q_a$
\end{minipage}
\end{center}
The requirement that $Q_a \notin S^{(1)}$ in the second line indicates that this  straightforward T-duality rule can a priori only be applied to the 7-branes along copies of the generic, non-degenerate fiber, while the fate of 7-branes wrapping degenerate fibers is less obvious.
We will argue that the degenerate fibers map to certain defects on $Z$ after (\ref{MMII}).
We hence end up with a 6d $N=(1,0)$ supersymmetric compactification of Type IIB string theory on a D-manifold, in the sense of \cite{Bershadsky:1995sp}, obtained by taking into account the backreaction of 5-branes (or more general defects) at $12$ points on $\cB_1$. For clarity we refer to this D-manifold as $Z$, to distinguish it from the dP$_9$ surface $\cB_1$. 
The tensionless solitonic string from the D3-brane along $\cE_2$ has turned into the D1-string probing this dual background.

There are two types of asymptotically massless BPS towers in the limit (\ref{dP9limitTdual}):
As the D1-string becomes light, it takes over the role of the fundamental string of tension
\be
\frac{T_{\rm D1}}{M_s^2} = \frac{2\pi}{\gB} \sim \frac{1}{\mu} \,,
\ee
and hence gives rise to a tower of light particle excitations of mass $M^2/M^2_s\sim 1/\mu$. Alternatively we can perform an S-duality 
transformation, taking us to weakly coupled string theory with an asymptotically tensionless, fundamental F1-string.
This F1-string probes the D-manifold obtained from $Z$ by dualising the $(p,q)$-5-branes into $(q,p)$-5-branes.
Either way, the string excitation scale
 coincides with the mass scale of the Kaluza-Klein tower,
\be
\frac{M^2}{M^2_s} \sim \frac{1}{{\cal V}_{\mathbb P^1_b,\rm F}} \sim \frac{1}{\mu} \,.
\ee
Therefore the theory remains effectively six-dimensional, in the sense that the tower of KK states is locked to the tower of string excitations
and so cannot become dense independently.
 This is precisely like for the limits considered in \cite{Lee:2018urn} and \cite{Lee:2019xtm}.

 Before turning to the five-dimensional limit (\ref{SchoenLimit2}) of genuine Type $T^4$ we are actually interested in, we like to point out that 
 non-trivial information about the nature of the light D1-string probing the D-manifold $Z$ is encoded in its elliptic genus. 
 More precisely, we can relate the elliptic genus of the solitonic string from a D3-brane on $\cE_2$ to suitable BPS invariants on the Schoen three-fold $Y$.
This works analogously as in the discussion of the heterotic string in Section \ref{sec_TypeBK3M}:
According to the general logic of \cite{Klemm:1996hh}, the modes of the solitonic string with KK momentum $k$ and winding number $l$ along the M-Theory circle map to M2-branes in M-Theory wrapped $C_{\rm M2}$ on $Y$ given by 
\be \label{CMSSchone1}
C_{\rm M2} =  l \, C_{\rm b} + (k -  l \, E_0) \, T^2 = l \cE_2 + k \cE_1 \,.
\ee
Recall also the discussion around (\ref{CMShet}).
Indeed, $C_{\rm b} = \cE_2$ is the curve on the base of the elliptic fibration $Y$ wrapped by the D3-brane that produces the string, and the role of the elliptic fiber $T^2$ of $Y$ is played by $\cE_1$. 
Note that the elliptic curves $\cE_1$ and $\cE_2$ have intersection numbers
\be
\cE_1 \cdot_{\cal S} \cE_1 =0 \,, \quad \cE_2 \cdot_{\cal S} \cE_2 =0 \,, \qquad \cE_1 \cdot_{\cal S} \cE_2 =1\,,
\ee
and hence 
generate a 2-dimensional hyperbolic lattice ${\bf H}$, in analogy with what we found for the  elliptic K3-fibration.
Furthermore, the vacuum energy of the string is
\be
E_0 = \frac{1}{2} C_{\rm b} \cdot \bar K_{\cB_1} =  \frac{1}{2} \cE_2 \cdot \cE_2 = 0 \,.
\ee
 This is as expected for a Type II string. 
 Applying the same reasoning as in \cite{Haghighat:2013gba,Huang:2013yta,Haghighat:2014vxa,Huang:2015sta,Haghighat:2015ega,Gu:2017ccq,DelZotto:2017mee,Duan:2018sqe,Lee:2018urn}  to the single-wrapped Type II string, 
 the genus-zero Gopakumar-Vafa invariants of $C_{\rm M2}$ are to be identified with the
 Fourier coefficients of the elliptic genus of the solitonic string, ie., of
 \be \label{Zelldef}
 {\mathbf Z}_{\rm ell}(\tau) =  {\rm Tr}_{\rm RR} (-1)^F F^2 q^{H_L}  \bar q^{H_R}   = \sum_{n=0}^\infty c_k \, q^k \,.
 \ee
 Here $q\equiv e^{2\pi i \tau}$, where $\tau$ corresponds to
  the modular parameter of the torus which is now interpreted as the world-sheet of the string.

It is well-known that for the perturbative heterotic string, ${\bf Z}_{\rm ell}$ transforms as a modular form of weight -2 \cite{Schellekens:1986xh}, while for the Type II string its weight vanishes.
 Its precise relation to the genus-zero Gopkumar-Vafa invariants of $Y$ is 
 \be \label{NE1E2ck}
 c_k =  N_{\cE_1 + k \cE_2} \,.
 \ee
This is rooted \cite{Lee:2018urn} in the more general correspondence between the full topological string partition function on $Y$ and the elliptic genus of solitonic strings in F-Theory, as studied extensively in the literature \cite{Haghighat:2013gba,Huang:2013yta,Haghighat:2014vxa,Huang:2015sta,Haghighat:2015ega,Gu:2017ccq,DelZotto:2017mee,Duan:2018sqe}.
On the other hand, it is clear that in absence of further refinements, the elliptic genus of the Type II string can only be a constant as this is the only modular form of weight zero, i.e.
\be \label{ZelltauNE1}
{\mathbf Z}_{\rm ell}(\tau) = c_0 = N_{{\cal E}_2}  \,.
\ee
Consistency with  (\ref{NE1E2ck}) then requires that 
\be \label{NE1kE2chiY}
N_{{\cal E}_2 + k \cE_1} = 0 \,,  \qquad \quad \forall \, \,  k \neq 0 \,.
\ee
By symmetry, we must have that also $N_{{\cal E}_1 + k \cE_2} = 0$ for $k \neq 0$. Since $\cE_2$ can be viewed as the elliptic fiber of the elliptic fibration  $\pi_2: Y \to \cB_2$ in (\ref{E1}),
we conclude that
\be
{\mathbf Z}_{\rm ell}(\tau)  = \chi(Y) \,.
\ee
Here we used the fact, invoked already in (\ref{NT2}), that the BPS invariants of a generic elliptic fiber equal the Euler characteristic of the Calabi-Yau 3-fold. More precisely
\be \label{NE1E20}
 N_{n \cE_1} = N_{n \cE_2} = \chi(Y)  \,.
 \ee

For $\chi(Y) = 0$, these assertions are in agreement with the result of \cite{1997alg.geom..9027H} that the BPS invariants for the curves in the lattice ${\bf H}$ spanned by $\cE_1$ and $\cE_2$ vanish.
It would be interesting to verify the validity of eq.~(\ref{NE1kE2chiY}) also for explicit examples of Schoen three-folds with $\chi(Y) \neq 0$.

The vanishing of the elliptic genus for $\chi(Y) = 0$ traces back to the absence of field theoretic chiral anomalies \cite{Schellekens:1986xh} of the six-dimensional $N=(1,0)$ supergravity theory.
From the space-time perspective, the vanishing of all BPS invariants in ${\bf H}$  signals that the prepotential of Type IIA string theory on a Schoen threefold, with $\chi(Y)=0$, only receives instanton corrections from sectional curves or curves related to these by a flop, which lie in an $E_8$ lattice \cite{1997alg.geom..9027H}. This in turn reflects the underlying structure of spontaneously broken $N=4$ spacetime supersymmetry for compactifications on 3-folds with $\chi(Y)=0$   \cite{KashaniPoor:2013en}.

The degeneration of the fiber $\cE_2$ into chains of intersecting (-2) curves (whenever $\cB_1$ contains singular Kodaira fibers other than type $I_1$ or $II)$ invites an intriguing observation:
A D3-brane wrapping a (-2) curve is known to lead to a non-critical string, often referred to as an M-string \cite{Haghighat:2013gba}. 
Thus, as a consequence of the splitting in geometry,
the fundamental Type II string, when approaching such a degenerate fiber, must split into a pair of interacting M-strings;
schematically
\be\label{MMII}
{\rm M} + {\rm M}  \to  {\rm Type\ II} \,.
\ee
If the degenerate fiber is not wrapped by 7-branes (as can happen already if $\chi(Y)=0$), the non-critical string is literally an M-string, while more generally it couples to the enhanced gauge theory on the 7-branes.
Evidently, these considerations parallel the known relation \cite{Haghighat:2014pva} between the non-critical 
E-string and the heterotic string: Geometrically, an analogous splitting of the form ${\rm E} + {\rm E}  \to  {\rm H}$ occurs when the generic  fiber of a blowup of a Hirzebruch surface (giving rise to a heterotic string) degenerates into two (-1) curves, each associated with an E-string, in F-Theory. 
 From the perspective of the heterotic dual, the degeneration of the Hirzebruch fiber is equivalent to the presence of NS5-branes on the dual K3-surface probed by the heterotic string. This reflects the appearance of E-strings spanned between M5-branes and the $E_8$-walls
in Horava-Witten theory. 

 It is tempting to speculate that a similar mechanism is at work for the Type II strings probing the background $Z$. In particular, the reducible degenerate fibers of $\cB_1$  should be dual to pointlike defects 
 on $Z$, at which the Type II string splits into non-critical M-strings (or their generalisation in presence of 7-branes along the degenerate fibers). It would be interesting to make this more concrete in future work.

\paragraph{Limits of Type $T^4$}
So far we have focused on M-Theory on 
an $T^2\times T^2$-fibered threefold, $Y$, in the limit  (\ref{SchoenLimit3}) of Type $T^2$. If we insist instead on the 
restricted, symmetric scaling as defined in (\ref{SchoenLimit2}), which
exibits a  more genuine limit of a shrinking Abelian surface, the Kaluza-Klein tower from the circle reduction of F- to M-Theory is no longer dominant over the tower of string excitations from the M5-brane. Rather it sits exactly at the same mass scale
\be \label{MoverM11scale}
\frac{M}{M_{11}} \sim \frac{1}{\sqrt{\lambda}} \,,
\ee
as measured with respect to the fundamental scale of M-Theory. 
Indeed, recall that the KK excitations associated with the F-Theory limit arise from M2-branes wrapping the elliptic fiber~$\cE_1$.
This means that in this degenerate limit, the theory stays effectively five-dimensional, similarly to what happens in limits of K3-fibered type.
M-Theory on $Y$ in the limit (\ref{SchoenLimit1}), (\ref{SchoenLimit2}) is therefore dually described by an effectively five-dimensional compactification of Type IIB string theory on the D-manifold $Z \times S_A^1$, where
\be \label{RAsaclings}
{R_A} \sim 1 \,, \qquad {\cal V}_{\cE_2} \sim {\lambda} \,,\qquad \quad {\cal V}_{\mathbb P^1_b} \sim {\lambda} \to \infty \,, \qquad \gB = {\lambda} \, \gB^{(0)}  \,.
\ee
Here all volumes are measured with respect to $\ell_{11}$, but since the Planck scale is finite, 
\be
\frac{M^3_{\rm Pl}}{M^3_{11}} \sim \frac{{\cal V}_Z}{\gB^2} \sim 1 \,,
\ee
 this is parametrically the same as measuring all volumes with respect to the five-dimensional Planck length. This leads to our claim (\ref{VZscaling}).

In the five-dimensional theory there now appear three towers of BPS states at the same scale $1/\sqrt{\lambda}$ (with respect to the Planck mass): In addition to the tensionless string excitations and the Kaluza-Klein states, these include the wrapping modes of the {light Type IIB D1-string around $S^1_A$}. These are precisely the modes associated with M2-branes on the family of curves (\ref{CMSSchone1}).

For $\chi(Y) \neq 0$, the BPS index for M2-branes along multiples of $ \cE_1$ and of $\cE_2$ is manifestly non-vanishing, while for $\chi(Y) = 0$ it is  zero.
Note, however, that in general the vanishing of a BPS index for a curve class need not mean that there are no BPS states from wrapped M2-branes, but rather that only the 5d {\it index} computed by the Gopakumar-Vafa invariants vanishes.
After all, from the dual Type IIB picture it is clear  that a tower of 5d BPS states arises from wrapped Type IIB strings,
 with increasing quanta of KK momenta along $S^1_A$.  What (\ref{NE1E20}) shows is merely that the 5d BPS index as such is too coarse to resolve their degeneracy due to the bose-fermi cancellation
 alluded to above.

\subsubsection{General Abelian surface fibration}

We conclude this section with some more tentative remarks concerning the Type II string theory duals of
general Abelian surface fibrations, whose  generic fiber is not of the product form $\cE_1 \times \cE_2$. To unravel the physics of such configurations, we will follow the same heuristic reasoning as in Section \ref{sec_TypeBK3M},  where we interpolated between K3-fibrations with a compatible elliptic fibration and more general K3-fibrations:  Now we will interpolate between fibers of direct product type and more generic Abelian surface fibers.

Starting from an $\cE_1 \times \cE_2$-fibration with $\chi(Y) = 0$, as 
investigated in the previous section, we approach the point in moduli space where
\be
{\cal V}_{\cE_1} = {\cal V}_{\cE_2} \,.
\ee
In  analogy with the discussion in Section \ref{sec_TypeBK3M}, where we considered K3-fbrations and their dual heterotic geometries,
this corresponds in the dual Type IIB theory on $Z \times S^1_A$
to a circle at self-dual radius
\be \label{sdpointIIB}
R_A = 1\,.
\ee
Recall that for the heterotic string on $\hat K3 \times S^1_A$, this point exhibits a non-abelian gauge enhancement $U(1)_L \to SU(2)_L$, for one of the two abelian gauge symmetries associated with the heterotic string on $S^1_A$.
On the M-Theory side this enhancement is reflected by the shrinking of a rational curve $\mathbb P^1_f$ inside the K3-fiber of $Y$.

In the present situation, no such gauge enhancement occurs: On the M-Theory side, the Abelian surface fiber does not contain a rational curve that shrinks at the self-dual point (\ref{sdpointIIB}). Correspondingly, for a Type II fundamental string on $Z \times S^1_A$, there cannot occur any 
non-abelian gauge group enhancement at $R_A =1$. Instead, the symmetry of the theory enhances by a $\mathbb Z_2$-valued involution, ${\cal I}_2$, which exchanges the KK and winding modes of the Type IIB string along $S^1_A$, or more precisely the left and right-moving sectors of the string world-sheet. We can therefore mod out the theory by this stringy symmetry, which leads to Type IIB theory on a non-geometric background of the form $Z \times S^1_A/{\cal I}_2$.
The effect is to break the abelian gauge symmetries realized in the left- and right-moving sectors to $U(1)_L \times U(1)_R \to U(1)_+ = \frac{1}{2} (U(1)_L + U(1)_R)$, and hence to reduce the number of vector multiplets by one.

In the M-Theory frame, the $\mathbb Z_2$ symmetry corresponds to an exchange symmetry between $\cE_1$ and $\cE_2$.
We can therefore envisage a monodromy fibration where the two algebraic curves $\cE_1$ and $\cE_2$ are exchanged along a closed path on the base $\mathbb P^1_b$.
The expected effect of such a monodromy would be to pass from a $\cE_1 \times \cE_2$-fibration to a more genuine $T^4$-fibration. 

Even without specifying the precise nature of the most general, dual D-manifold background on the Type IIB side, it is clear that there still exists a tower of BPS particles arising from the wrapped Type II string. The fact that the M-Theory geometry contains only a single fibral curve class of non-negative self-intersection, reflects the correlation between the wrapping and winding numbers of the string along $S^1_A$ after modding out by ${\cal I}_2$.

We leave a more detailed study of such Type IIB non-geometric backgrounds for future investigations.

\subsection{Infinite distance limits at infinite volume} \label{sec_infvolume}

In the previous sections we have described the behaviour of M-Theory in various infinite distance limits in which the total Calabi-Yau volume stays finite.
Let us now drop this extra requirement. While the general expectation is that such limits correspond to decompactification, one might 
nevertheless wonder whether tensionless strings can arise at a scale comparable with the KK scale.
As we will now show, this is not the case.

As explained already in Section \ref{subsec_Kahlercone}, we parametrise the limit by writing ${\cal V}_Y = \mu {\cal V}'_Y$ and
define a rescaled K\"ahler form
\be \label{Jy'scaling}
J_Y' = \mu^{-1/3} \, J_Y = \sum_i {T'}^i J_i \,, \qquad \quad {\rm as }\ {\mu} \to \infty \,.
\ee
If all rescaled K\"ahler parameters  ${T'}$ are finite, there cannot appear any towers of states at a scale comparable with the KK scale
\be \label{MKKoverall}
\frac{M_{\rm KK}}{M_{11}} \sim {\cal V}_Y^{-1/6} \sim \mu^{-1/6} \,.
\ee
As explained in Section \ref{subsec_Kahlercone}, this is because all curves shrinking in the metric $J'_Y$ are contractible. The physics is thus a straightforward decompactification limit with KK scale given in~(\ref{MKKoverall}).

If some ${T'}^i$ become large, the rescaled K\"ahler form $J_Y'$ undergoes a nested finite volume infinite distance limit. Every such limit must correspond to one of the limits described in Theorem \ref{classify}.
The remaining question is now how the mass scales associated with these limits compare to the scale~(\ref{MKKoverall}). 

Assume first that $J'_Y$ undergoes a limit of Type $T^2$, with scaling parameter $\lambda$ as in Theorem~\ref{classify}. 
This means that, taking into account the overall scaling (\ref{Jy'scaling}), 
\be
{\cal V}_{T^2}    \sim \frac{\mu^{1/3}}{\lambda^2} \,, \qquad \quad {\cal V}_{B_2} \sim   \mu^{2/3} \lambda^2 \,.
\ee
Due to the anisotropy of the metric,
 the overall  scale (\ref{MKKoverall}) is always higher than the KK scale set by the volume of the large base $B_2$,
\be
\frac{M_{{\rm KK}, B_2}}{M_{11}} \sim {\cal V}^{-1/4}_{B_2} \sim \mu^{-1/6} \lambda^{-1/2} \,.
\ee
This scale competes with the mass scale associated with the tower of M2-branes wrapping the fiber,
\be
\frac{M_{\rm M2}}{M_{11}} \sim {\cal V}_{T^2} \sim \mu^{1/3} \lambda^{-2} \,.
\ee
Since the two scaling parameters $\mu$ and $\lambda$ are independent, we must distinguish the following three regimes: \\

\paragraph{1) {$\mathbf \mu^{-1/6} \lambda^{-1/2} \prec \frac{\mu^{1/3}}{\lambda^2}$}:}
The theory undergoes decompactification before the tower of wrapped M2-branes along $T^2$ becomes relevant.  \\

\paragraph{2) {$\mathbf \mu^{-1/6}  \lambda^{-1/2}  \succ \frac{\mu^{1/3}}{\lambda^2}$}:}
The theory first probes the mass scale associated with the tower of M2-branes along $T^2$. At this mass scale, the theory partially decompactifies from five to six dimensions, corresponding to taking an F-Theory limit.
We now match the (diverging) Planck scale of the five-dimensional theory and the (diverging) Planck scale in the F-Theory frame by equating
\be
\mu {\cal V}'_Y  \sim \frac{M^3_{\rm Pl,5d}}{M^3_{11}} = \frac{M^4_{\rm Pl,6d}}{M^4_s}   \sim {\cal V}_{B2, {\rm F}} \,.
\ee
This is accomplished by measuring all volumes within $B_2$ in the F-Theory frame via the K\"ahler form 
\be
J_{B_2,\rm F} = \frac{\mu^{1/6}}{\lambda} J_{B_2}. 
\ee
The KK scale of the six-dimensional theory is now set by the largest curve volume on $B_2$ as measured by $J_{B_2,\rm F}$. 
If all curves on $B_2$ scale as ${\cal V}_{C,{\rm F}} \precsim \mu^{1/2}$, this implies that the decompactification scale sits at
\be
\frac{M_{\rm KK}}{M_{s}} \sim \mu^{-1/4}  \,.
\ee
If there exists a curve $C$ on $B_2$ with ${\cal V}_{C,{\rm F}} \succ \mu^{1/2}$, we know from Lemma~\ref{lemmaBlimit} in Appendix~\ref{App_Surface-Limits} that $B_2$ must be fibered with generic fiber $T^2_f$ or $\mathbb P^1_f$, and $C$ must be the base of this fibration, $\mathbb P^1_b$.
In particular this means that
\be
{\cal V}_{T^2_f/\mathbb P^1_f, {\rm F}} \sim \mu^{1/2} \rho^{-1} \,, \qquad \quad {\cal V}_{\mathbb P^1_b, {\rm F}} \sim \mu^{1/2} \rho\,,
\ee
for some parameter $\rho \to \infty$. A D3-brane wrapping a curve, $T^2_f$ or $\mathbb P^1_f$, gives rise to a Type II or heterotic string that becomes asymptotically tensionless.
However, since $\rho \to \infty$,
 the KK scale associated with the large curve ${\mathbb P^1_b}$ is always lower than the mass scale set by the tower of string excitations. This means that the string tower does not lead to an equi-dimensional limit, unlike for pure limits of Type $T^2$ followed by a finite volume limit on the base; rather, the six-dimensional theory decompactifies further with KK scale
\be
\frac{M_{\rm KK}}{M_{s}} \sim {\cal V}^{-1/2}_{\mathbb P^1_b,{\rm F}}  \sim \mu^{-1/4} \, \rho^{-1/2} \,.
\ee

\paragraph{3) {$\mathbf \mu^{-1/6} \sim \frac{\mu^{1/3}}{\lambda^2}$}:}
We have a decompactification to six-dimensional F-Theory, whose scale coincides with
the scale associated with the supergravity modes on $B_2$. The appearance of two  towers of KK states at the same scale clearly signals decompactification. \\

Finally, consider the situation where the rescaled metric $J'_Y$ undergoes a limit of Type K3 or Type $T^4$ as defined in Theorem \ref{classify}. This means that with respect to $J_Y$ the fiber and base volumes scale as
\be
{\cal V}_{K3/T^4} \sim \frac{\mu^{2/3}}{\lambda} \, , \qquad \qquad {\cal V}_{\mathbb P^1_b} \sim \mu^{1/3} \, {\lambda} \,.
\ee
An M5-brane along the K3/$T^4$ fiber gives rise to a heterotic/Type II string, respectively,
whose associated mass scale compares to the KK scale from the large base curve as follows:
\be
\frac{M_{\rm string}}{M_{11}}  \sim \frac{\mu^{1/3}}{\lambda^{1/2}}   \,, \qquad \quad  \frac{M_{\rm KK, \mathbb P^1_b}}{M_{11}} \sim \mu^{-1/6} \lambda^{-1/2} \,.
\ee
Again, this signals decompactification, rather than an equi-dimensional limit, with KK scale $\mu^{-1/6} \lambda^{-1/2}$.

\section{Large distance limits in quantum K\"ahler geometry} \label{sec}
\subsection{General Considerations on Quantum Volumes and Mirror Symmetry \label{sec_gencons1}}

So far our discussion has been restricted to limits at infinite distance in the {\it classical} K\"ahler geometry of $Y$, as probed by M-Theory. Such limits are distinguished by whether or not the classical volume of $Y$ remains finite.
In compactified Type IIA string theory, on the other hand, we must differentiate between the behaviour of the total Calabi-Yau volume ${\cal V}_Y$ and the behaviour of the Planck scale 
\be\label{MPL}
\frac{M^2_{\rm Pl}}{M^2_s} = \frac{4 \pi}{\gA^2} {\cal V}_Y  \,.
\ee
The first question we need to address is whether there exist infinite distance limits in quantum K\"ahler moduli space in  which ${\cal V}_Y$ remains finite.
From the discussion in Section \ref{sec_classical} we know that these are necessarily limits in which a classically non-contractible cycle becomes small, and the only candidates are limits in which a $T^2$, a K3 or a $T^4$ fiber shrinks while the base expands.
However, for Type IIA compactifications it is well-known (see eg.~\cite{Aspinwall:1993xz,Greene:1996tx,Greene:2000ci}) that in the regime of small volumes stringy quantum geometry takes over and the classical notion of volume can be drastically modified. We therefore need to properly define what we mean by volumes, and carefully re-evaluate the physics of the limits. 
The outcome of this analysis will be that in all three cases the quantum volume of $Y$ tends to infinity as the non-contractible fiber becomes small and the base expands. Hence no finite volume infinite distance limits exist in the quantum K\"ahler moduli space.

Next, to answer the question of whether or not there exists an equi-dimensional limit at infinite distance, we must consider two different situations:

\begin{enumerate}
\item {\it Limits without co-scaling} \\ 
If we keep the ten-dimensional dilaton $g_{\rm IIA}$ fixed when taking the limit, the divergence of the total volume ${\cal V}_Y$ implies that the four-dimensional Planck scale, in units of $M_s$, always tends to infinity. 
\item {\it Limits with co-scaling} \\ 
Alternatively, we can co-scale $g_{\rm IIA}$ in such a way that the four-dimensional Planck scale stays finite in units of $M_s$.
If the volume diverges as ${\cal V}_Y = \mu \, \,  {\cal V}'_Y$, we must scale
 \be \label{gIIAcosaling}
 g_{\rm IIA} = \mu^{1/2} \, \, g^{(0)}_{\rm IIA} 
 \ee
 with $g^{(0)}_{\rm IIA}$ finite.
 For a discussion of such co-scaling limits in Type II theory see also \cite{Marchesano:2019ifh,Font:2019cxq}.
 The significance of co-scaling limits in light of mirror symmetry will be explained around (\ref{MplIIB}) below.
\end{enumerate}

In both types of limits, there is a universal candidate 
 for a KK tower signalling decompactification. In limits without co-scaling, this is the supergravity tower at mass scale
 \be \label{Mkkuniver}
 \frac{M_{\rm KK}}{M_s} \sim {\cal V}_Y^{-1/6} \sim \mu^{-1/6} \,.
 \ee
In limits with co-scaling, the candidate tower is given by the tower of D0 branes at mass scale 
 \be \label{MD0}
  \frac{M_{\rm D0}}{M_s} = \frac{2\pi}{g_{\rm IIA}} \sim \mu^{-1/2} \,,
  \ee
which is clearly leading compared to (\ref{Mkkuniver}). 
Unless there exists a tower of stringy modes at a comparable scale, the first tower would indicate conventional decompactification, while the second tower
indicates decompactification to five-dimensional M-Theory on $Y$.
 In the five-dimensional M-Theory frame, all volumes (in units of the 11-dimensional Planck scale $M_{11})$ are measured by the rescaled K\"ahler form \cite{Witten:1995ex}
 \be \label{JMJV-a}
 J_{\rm M} = \frac{J_Y}{\gA{}^{2/3}} \,.
 \ee 
Hence M-Theory now probes the geometry of $Y$ at finite volume ${\cal V}_{Y,\rm M} = {\cal V}'_Y$. This justifies our claim, made at the beginning of Section \ref{sec_Mtheorylimits}, that finite volume infinite distance limits of M-Theory capture also the physics of certain infinite distance limits in Type IIA string theory at finite Planck scale.

However, to establish decompactification we must exclude that there can exist a competing tower of stringy states, associated with a shrinking non-contractible divisor, at the same scale as (\ref{Mkkuniver}) or (\ref{MD0}).
Such strings, if any at all, can occur only in the quantum versions of the three types of limits we are analyzing. Hence, it suffices to scrutinize these limits in turn, with and without co-scaling $g_{\rm IIA}$. All other infinite distance limits are automatically decompactification limits.
 
Let us now explain in more detail the strategy for investigating the quantum geometry of the three types of infinite distance limits.
As is well-known and briefly reviewed in Appendix~\ref{app_mirror},  
a suitable definition of quantum volumes can be given in terms of globally defined period integrals that are associated with the complex structure moduli space, $\MCS(X)$,  of the mirror manifold, $X $\cite{Aspinwall:1993xz,Greene:1996tx,Greene:2000ci}. In terms of the periods, 
one can define flat coordinates by taking ratios. A priori there is no canonical way of doing this, and in order to make
contact with classical geometry, one needs to fix an integral symplectic frame in the large complex structure limit which is canonically mirror dual to
the classical large volume limit.  Specifically, at such a point of ``unipotent monodromy''
there is a unique symplectic $3$-cycle, ${\gamma^0}$, 
whose period integral yields a pure power series in a suitable coordinate patch and which can be normalized as follows:
$$
X^0(z) = \int_{\gamma^0} \Omega(z)\ =\ 1+\CO(z)\,, \qquad z\rightarrow 0 \,.
$$
Here $z$ stands for the collection of coordinates of $\MCS(X)$.
This serves as a reference with respect to which the flat coordinates of $\MCS$ near the large complex structure limit can be defined as
$$
t^a(z) =\frac{X^a(z)}{X^0(z)}\equiv \frac{\int_{\gamma^a}\Omega(z)}{ \int_{\gamma^0}  \Omega(z)}\,.
$$
Here $a$ runs over a basis of symplectic $A$-cycles in a suitable polarization of $H_3(X,\IZ)$. 
Near $z=0$ these coordinates match, via mirror symmetry, the (complexified) classical K\"ahler parameters of $Y$
(see eq.~(\ref{tadefinition})). Correspondingly, the integral $i\int_X \Omega\wedge\bar\Omega$ 
coincides with the K\"ahler volume of $Y$ in the large volume limit,
\be \label{quantumvolume}
\frac18\frac{i\int_X \Omega\wedge\bar\Omega(z)}{|X^0(z)|^2}\ \
\stackrel{z(t)\simeq0}{=}\ \ \frac{1}{3!} \int_Y J^3\ 
\equiv\ {\cal V}_Y(t)\,.
\ee
Away from $z\simeq 0$, the left-hand side of (\ref{quantumvolume}) defines the quantum corrected, 
globally defined analytical continuation of ${\cal V}_Y(t)$ over all of the quantum K\"ahler moduli space of $Y$.
Note that the requirement to fix a geometrical basis near a given
large complex structure point is not just a minor point of fixing some normalization or gauge.
Since periods, including $X^0$, can  undergo non-trivial monodromy when transported globally over
 $\MCS$, this is necessarily a local choice; 
in particular there is no intrinsic choice for $X^0$ 
deep inside the bulk of the moduli space, where generically all periods are power series.
In this sense there is no {\it a piori} defined quantum volume, and, in turn, Planck scale (\ref{MPL}),
\be\label{MPLB}
\frac{M^2_{\rm Pl}}{M^2_s}(z)= \frac{i\pi }{2\gA^2}  \frac{\int_X\Omega\wedge\bar\Omega}{|X^0|^2}(z)\,,
\ee
which would be independent of a choice of initial frame at infinity.

Before we continue, let us comment on the relation between the string couplings, $g_{\rm IIA}$ and $g_{\rm IIB}$,
on both sides of mirror symmetry. Near the large complex structure limit,
the distinguished 3-cycle $\gamma^0$ that defines the period $X^0$ is a special Lagrangian 3-torus, and the three-fold $X$ is locally a fibration of this 3-torus over the dual 3-cycle \cite{Strominger:1996it}. 
Mirror symmetry amounts to performing three T-dualities along this 3-cycle in an adiabatic way, which basically maps the 3-cycle of $X$
into a 0-cycle on $Y$. This relates the string couplings of both theories as 
\be \label{gAinGb}
\gA = \frac{g_{\rm IIB}}{{\cal V}_{\gamma^{0}}} =  g_{\rm IIB}  \frac{(i \int_X \Omega_X \wedge \bar\Omega_X)^{1/2}}{ |X^0|\,  (8{\cal V}_X)^{1/2}} \,,   \, 
\ee
where we used the calibration condition (\ref{calibration}). Clearly this relationship is sensitive to the
specific analytic continuation along a given path in the complex structure moduli space.
For illustration, let us consider the monodromy along a path in $\MCS(X)$  that starts and ends at the same large 
complex structure point.
This monodromy acts as a symplectic duality
transformation on the period vector which leaves $\int_X \Omega_X \wedge \bar\Omega_X$ invariant, while $X^0$ may change as
\be
X^0 \to {\hat X^0} \,.
\ee
Thus, in order for the physics to remain invariant,  the Type IIA coupling must co-transform\footnote{This reflects Buscher's rules for T-duality \cite{Buscher:1987sk,Buscher:1987qj} in the present setting.}
as
\be \label{gAtrafopath}
\gA \to {\hat g}_{\rm IIA} = \gA \frac{{|X^0|}}{{|\hat X^0|}} \,,
\ee
while on the IIB side the coupling stays invariant (since there the volume does not change). 
Depending on whether ${\hat X^0}$ diverges or not, such a rescaling can be very significant.

Importantly, we note here that the limits with co-scaling (as introduced around (\ref{gIIAcosaling})) represent the mirror duals of infinite distance limits in Type IIB complex structure moduli space, at fixed Type IIB dilaton $g_{\rm IIB}$.\footnote{We thank Eran Palti for discussions on this point.} Limits of this type have been analyzed in \cite{Grimm:2018ohb,Grimm:2018cpv}. Indeed, in Type IIB theory compactified on a Calabi-Yau three-fold $X$, the four-dimensional Planck scale is given by 
\be \label{MplIIB}
\frac{M^2_{\rm Pl}}{M^2_s} = \frac{4 \pi}{g^2_{\rm IIB}} {\cal V}_X \,,
\ee
where the volume of $X$ depends only on the K\"ahler moduli of $X$. Limits in the complex structure moduli space of $X$, at fixed $g^2_{\rm IIB}$, leave both ${\cal V}_X$ and the ratio (\ref{MplIIB}) invariant, but change the periods as computed from $\Omega_X$.
The direct mirror dual of such limits is Type IIA string theory on the mirror three-fold $Y$, with ${\cal V}_Y$ computed as in (\ref{quantumvolume}), and with $g_{\rm IIA}$ as given in (\ref{gAinGb}).
By construction, the four-dimensional Planck scale is the same on both sides.
From  (\ref{quantumvolume}) and  (\ref{gAinGb}) it is evident that if on the Type IIA side ${\cal V}_Y$ diverges, $g_{\rm IIA}$ automatically compensates for this.
We will comment more on the interpretation of large distance limits in $\MCS(X)$ in Section \ref{sec_Conclusions}.

\begin{figure}[t!]
\centering
\includegraphics[width=10cm] {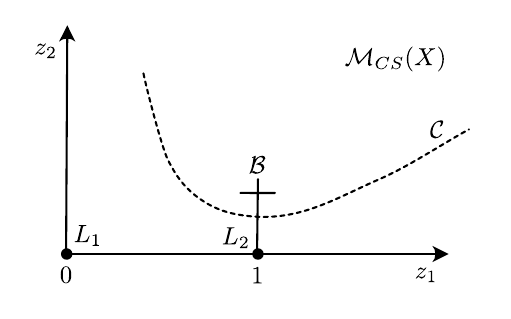}
\caption{Schematic representation of (part of) 
the complex structure moduli space of the three-fold $X$, where the coordinates $z_1$ and $z_2$ govern the
quantum volumes of fiber and base, respectively, of the mirror fibration, $Y$. The locus $z_2=0$ corresponds to a partial long distance limit where the base volume is large, and the $z_1$-axis coincides with the moduli space of the fiber.
The large complex structure point, $L_1$, corresponds to large total volume
${\cal V}_Y$, and $L_2$ to the limit where, in classical geometry, the fiber would have zero volume. For more details see the text.}
\label{f:Fig1MCS}
\end{figure}

In the main part of the present Section \ref{sec},
we will re-consider the three types of limits we have discussed for M-Theory, but now in the context of Type IIA string theory,
where the quantum geometry of volumes and monodromies become relevant. That is, we consider
limits where the local geometry of $Y$ at large volume is given by a fibration of either an elliptic curve, a K3 surface, or an abelian surface,
respectively. In each of these cases, the (relevant part of the) moduli space of the mirror, $\MCS(X)$, is schematically of the form shown in Figure \ref{f:Fig1MCS}. For illustration we depict both fiber and base directions as one-dimensional.

More concretely, the $z_2$-axis is related to the size of the base such that the locus $z_2=0$ corresponds, via mirror symmetry,
 to the locus of infinite base volume, and analogously for $z_1$.  Thus the point of large complex structure is located at the origin $L_1:\,(z_1,z_2)=(0,0)$.  The horizontal axis corresponds to the moduli space of the fiber $\cal F$ in question, with $z_1=1$ denoting the singular locus of the fiber where its mirror, naively, has zero volume.
Furthermore $\cal C$ denotes the conifold locus of $Y$, whose point of tangency with the divisor $z_2=0$ has been resolved by a blowup divisor denoted by $\cal B$. We will be mostly interested in the region near $L_2:\,(z_1,z_2)=(1,0)$,
as well as in the path ${\cal P}_{12}=\overline{L_1L_2}$ connecting both points.
This is because, classically,  we have at $L_2$ a coincidence of large base and vanishing fiber. This means that, at least classically,
the total volume ${\cal V}_Y$ could remain finite, so gravity need not decouple.

However, in contrast to the classical picture, we will find that the total volume in units of $\ell_s$ always tends to infinity, due to quantum corrections. Essentially, our results can be summarized as follows:

\begin{itemize}
 \item For the elliptic fibration, $L_2$ turns out to be a large distance limit with vanishing elliptic fiber volume. As is typical for a large distance limit, the vanishing is relative and arises via dividing out a diverging $X^0$;  this just reflects that the cycle is non-contractible.  The  point $L_2$
corresponds to a $T$-dual copy of $L_1$ subject to a transformation of $\gA$, as indicated in (\ref{gAtrafopath}). In particular, the theory at $L_2$ with $\gA$ in the perturbative regime is equivalent to strongly coupled Type IIA theory at $L_1$, and thus is best described in terms of M-Theory. For limits without co-scaling, the intermediate five-dimensional M-Theory  undergoes additional decompactification, while in the co-scaled situation the physics depends on possible further, nested infinite distance scalings.

 \item For the K3 fibration, the fiber volume remains non-zero at $L_2$ due to quantum effects, while $X^0$ stays finite. 
 That is, the classically infinite distance limit turns into a finite distance limit as a consequence of quantum effects.
 For limits without co-scaling, the Planck scale diverges and gravity decouples in the sense that the theory undergoes decompactification.
 This is in agreement with the fact that $L_2$ 
 describes the weak coupling limit of $SU(2)$ Seiberg-Witten theory \cite{Kachru:1995fv}
(whose moduli space coincides with the blowup divisor $\cal B$). What we will add to this here is to
disentangle in detail the relative mass scales of the various excitations. 
In particular, as a consequence of the non-vanishing quantum volume of the 4-cycle of the K3 fiber, 
the heterotic string (arising from the wrapped NS5 brane) does not become tensionless in terms of the gauge theory scale, even though its tension does vanish in units of the (diverging) Planck scale. 

As before, the Planck scale can be kept finite by suitably co-scaling $\gA$. Interestingly, for the K3 fibration
such a limit is found to be equi-dimensional, in the sense that a tower of KK states sits parametrically at the same scale as the excitations of the heterotic string. Thus this limit is the four-dimensional version of the equi-dimensional limit of Type K3 in M-Theory that we discussed previously.

 \item  For limits of Type $T^4$, T-duality maps the small fiber regime at $L_2$ to the large fiber point $L_1$ at strong coupling, similarly as for limits of Type $T^2$. The difference is, however, that after decompactification to M-Theory the theory probes a finite volume three-fold $Y$, as measured in the five-dimensional M-Theory frame. In other words, there arises no need for co-scaling $g_{\rm IIA}$. In this sense the decompactification is less severe as compared to limits of Type $T^2$.

\end{itemize}

\subsection{Quantum large distance limit for elliptic fibrations} \label{sec_QuantumT2}

We begin by analyzing Type IIA string theory probing an infinite distance limit of fibration Type $T^2$.
Here the three-fold $Y$ is a $T^2$-fibration over some base space, $B_2$.
Let us parametrise its  {\it classical} K\"ahler volume as
\be \label{classicalvolumeelliptic}
{\cal V}_{Y,{\rm cl.}} = \frac{1}{6} \int_Y J_Y^3 = a \,  T^3 + \, T^2 S + c\,  T \, {\cal V}_{B_2,\rm cl.} + \Delta\,,
\ee
with 
\be
T:= \int_{T^2} J_Y\,, \qquad S:= \int_{C_S} J_Y \, , \qquad {\cal V}_{B_2,\rm cl.} = \frac{1}{2} \int_{B_2} J_Y^2  \,.
\ee
The numerical coefficients $a$ and $c$ depend on the intersection numbers of $Y$, and $C_S$ is a distinguished curve class on the base, $B_2$.
For a smooth Weierstrass model, this coincides with the anti-canonical divisor class of $B_2$.\footnote{In this case, the K\"ahler cone is generated by the pullback of the K\"ahler cone generators of $B_2$ together with $S_0 + \pi^\ast \bar K$, where $\bar K$ is the anti-canonical divisor on $B_2$.
If we expand the K\"ahler form as
$ J_Y = T (S_0 + \pi^\ast \bar K) + \pi^\ast J_{B_2}$ 
we identify
$a= \frac{1}{6} \bar K \cdot_{B_2} \bar K$, $c=1$, $S= \bar K \cdot_{B_2} J_{B_2}$. \label{aSparameters}
}
Possible extra non-negative terms subsumed in $\Delta$ 
can occur if the fibration contains curves different from the generic fiber $T^2$, or fibral surfaces. To remain inside the K\"ahler cone, their contributions must vanish in the limit where the classical fiber volume $T^2$ is taken to zero. 

Our question is whether it is still possible to take an infinite distance limit in such a way that the total volume stays finite,
while the base volume tends to infinity. There are two potential obstructions to taking this limit in quantum geometry:
\begin{enumerate}
\item
It is a priori not guaranteed that the fiber volume can be taken to be arbitrarily small, i.e. that the corresponding point is contained in the quantum K\"ahler moduli space of $Y$.
\item
The expression for the quantum volume of $Y$ differs in general  from its classical analogue. Taking the fiber volume to zero, 
if at all possible, need not guarantee that the total quantum volume of $Y$
remains constant, even if we allow for a co-scaling of the base volume.
\end{enumerate}

Both problems can be addressed by starting near large K\"ahler volume of $Y$, switching to the mirror three-fold $X$, and
analytically continuing the periods  (\ref{Pivector}) and  total volume (\ref{VYqvolume}) over the complex structure moduli space
to the point where the volume of $T^2$ would vanish.  That is, referring to Figure \ref{f:Fig1MCS}, we consider an analytic continuation
from $L_1$ to $L_2$.

For the presently discussed elliptic fiber, the analysis is very simple: The only singularities in its moduli space are cusps, which are $T$-dual to the
large volume/large complex structure limit.  This means that $L_2$ must be equivalent to $L_1$ up to $T$-duality.
 An equivalent description of the analytic continuation is thus to mod out the complex structure moduli space by identifying the cusp points, and to consider a closed loop starting and ending at $L_1$. Then, essentially, the effect of
analytical continuation turns into monodromy, which is generated by encircling the 
singularity that arises as the fixed locus under this modding.\footnote{Note however, that this monodromy need not act nicely as a symplectic transformation on the periods, rather what generically happens is that it acts as a symplectic transformation up to an overall rescaling of
the periods, which is physically irrelevant as only their ratios matter. We will come back to this later.}

The above arguments characterize the situation in full generality. 
For further illustration, we consider a three-fold for which the elliptic fiber is given by the curve in $\mathbb P^2_{(1,2,3)}[6]$
defined by the vanishing of
\be\label{sextic}
W(x,z)=\frac16{x_1}^6+\frac13{x_2}^3+\frac12{x_3}^2-z^{-1/6}x_1 x_2 x_3\,.
\ee
A concrete example is the well-studied fibration over $\mathbb P^2$ in $\mathbb P^4_{1,1,1,6,9}[18]$ \cite{Hosono:1993qy,Candelas:1994hw}.
In a context closely related to ours, this and other 
elliptically fibered three-folds have recently been discussed in refs.~\cite{Blumenhagen:2018nts,Grimm:2018cpv,Corvilain:2018lgw};
further related works include \cite{Alim:2012ss,Klemm:2012sx,Huang:2015sta,Schimannek:2019ijf}.
Concretely, in terms of its periods, the mirror map of the fibral curve is given by 
 \bea\label{F21}
t(z)&=& \frac{X^1(z)}{X^0(z)}\,, \qquad{\rm where}\nn\\
X^0(z) &=& \ {}_2F_1(1/6,5/6,1;z)\,,\\
X^1(z) &=& i{}_2F_1(1/6,5/6,1;1-z)\,.\nn
\eea
This is equivalent to writing the mirror map in terms of the modular $J$-function as \cite{Lian:1994zv}
\be\label{ellJ}
J(q)=  {432\over z(1-z)}\,,\qquad  q\equiv e^{2\pi i t(z)}\,,
\ee
which makes it manifest that the singularities at $z=\{0,1\}$ map to the two cusp points at $t=\{i \infty,0\}$. 
This reflects the fact that $W(x,1-z)=\frac z{z-1} W({x}',z)$ and thus describes the same world-sheet CFT as $W(x,z)$.  Obviously the involution
 $\CI: z\leftrightarrow 1-z$ exchanges the periods (modulo the factor $i$ in (\ref{F21})) and acts as modular transformation
 $$
\CI:t(1-z)= -\frac1{ t(z)}\,.
 $$
As remarked above, this transformation can also be interpreted as monodromy upon modding out $\CI$ and encircling the singularity at $z=1/2$.
It is easy to see that the volume of the elliptic fiber scales like ${\cal V}_{T^2}(z\simeq 1)\sim 1/T$ and thus vanishes 
at $L_2$, if it is initially defined in the frame where ${\cal V}_{T^2}(z\simeq 0)\sim T$ is large ($T=\Im t$).

An analogous structure persists for the three-fold fibration $Y$ of the curve \cite{Hosono:1993qy,Candelas:1994hw,Huang:2015sta}, and in fact also for
general elliptic three-fold fibrations \cite{Klemm:2012sx}. Thus 
the point $L_2:~(z_1,z_2)=(1,0)$ of arbitrarily small fiber volume {\it is} indeed
part of the quantum K\"ahler moduli space. The associated analytic continuation
acts as a symplectic transformation
on the period vector (\ref{Pivector}) (up to an overall prefactor), 
in particular by exchanging a D2-brane along $T^2$ with a D0-brane at a point on $Y$. A more formal statement concerning the full
period vector of $Y$ can be made in terms of a Fourier-Mukai transform acting as a symplectic automorphism in the derived category of coherent sheaves, as explained in \cite{Andreas:2000sj,Andreas:2001ve,Aspinwall:2002nw,Klemm:2012sx,Schimannek:2019ijf}. 
In particular, \cite{Schimannek:2019ijf} shows that T-duality along the fiber is well-defined, in this sense, in the presence of degenerate fibers. 
What is important for us is 
that this transformation does not change the  inner product implicit
in $\int_{X}\Omega \wedge \bar \Omega$. Therefore the quantum volume near $L_2$, 
\be
{\cal V}_{Y}|_{L_2} = \frac18 \frac{i \int_{X} \Omega \wedge \bar\Omega|_{L_2}}{ |X^0|^2|_{L_2}} \,,
\ee
is closely related to the one near $L_1$. 
The difference is that $T$ is replaced by $\hT=1/T$, while the periods in (\ref{F21}) exchange. 
The relevant  feature is that the reference period, $X^0$, as canonically defined at $L_1$, blows up at $L_2$:  
\be
{X^0|_{L_2}}\sim \hT \,, \qquad \quad \text{as} \quad \hT\to \infty \,.
\ee
Thus, as compared to (\ref{classicalvolumeelliptic}), the total quantum volume at $L_2$ takes the asymptotic form
\be\label{VL2}
{\cal V}_{Y}|_{L_2} = a \,  {\hT} + \,  S + c\,  {\hT}^{-1} \, {\cal V}_{B_2,\rm cl.} + {\hT}^{-2} \, \Delta + \ldots \, \ee
and diverges for\footnote{The case $a \neq 0$ covers the generic situation; for $a=0$ see below.} $a\not=0$, even though the elliptic
fiber shrinks to zero size.

Nevertheless the theory at $L_2$ must equivalent to $L_1$ because it is $T$-dual to it and is described by the same world-sheet CFT.
To see this explicitly, recall that the path $L_2 \to L_1$ is equivalent to a closed monodromy loop and
hence the discussion around (\ref{gAtrafopath}) applies.
Thus by re-adjusting $\gA$, in accordance with Buscher's rules,
we recover equivalent physics at $L_1$ and $L_2$. Indeed it is evident that
the tower of D2-branes wrapped around the
shrinking elliptic fiber at $L_2$ is equivalent to the tower of D0-branes 
at $L_1$, which becomes dense at strong coupling. 
What is slightly more subtle is the way in which the Kaluza-Klein spectra at $L_1$ and $L_2$  map to one another.

To understand this in more detail, let the fiber and base volumes scale as follows
\be \label{L2volscaling}
{\rm at} \ \ L_2: \qquad  {\cal V}_{T^2} = \frac{1}{\lambda^2} \,,\qquad {\cal V}_{B_2} = {\cal V}'_{B_2} \lambda^2 \,, \qquad  \gA = \gA^{(0)} \,, \qquad \lambda \to \infty \,.
\ee 
By T-duality this is equivalent to the large volume limit at $L_1$, for which
\be \label{L1volscaling}
{\rm at} \ \ L_1: \qquad {\cal V}_{T^2} = \lambda^2 \,,\qquad {\cal V}_{B_2} = {\cal V}'_{B_2}\lambda^2 \,, \qquad  \gA = \gA^{(0)} \lambda^2 \,, \qquad \lambda \to \infty \,.
\ee
As required by T-duality, the Planck scale manifestly behaves in the same way at $L_1$ and $L_2$,
\be\label{planckT2}
\frac{M_{\rm Pl}^2}{M^2_s} = \frac{4 \pi}{\gA{}^2} {\cal V}_Y \sim  \frac1{\gA^{(0)}{}^2}\left( a\,  \lambda^2 + S + c \,  {\cal V}'_{B2,\rm cl.} + \ldots \right)\,,
\ee
{as is readily checked by computing the left-hand side either at $L_1$ using (\ref{classicalvolumeelliptic}) and (\ref{L1volscaling}) or at $L_2$ using (\ref{VL2}) and (\ref{L2volscaling}).}

\subsubsection{Limit without co-scaling}

Consider first the theory at $L_1$ and choose, for definiteness, $\gA^{(0)}$ being small and fixed. This is a {\it limit without co-scaling} in the language of Section~\ref{sec_gencons1}. 
From (\ref{L1volscaling}) we see that we enter the strong coupling regime when  approaching $L_1$.
There are two types of light BPS towers, the D0-brane tower and the supergravity Kaluza-Klein tower, as measured in the Type IIA frame. Their respective mass scales are as follows: 
\be \label{D2KKatL1}
\begin{aligned}
&\text{D0 at $L_1$}:\qquad &&\frac{M}{M_s} \sim   \frac{1}{\gA}  \sim      \frac{1}{\gA^{(0)} \lambda^2} \\
&\text{Type IIA KK at $L_1$}: \qquad &&\frac{M_{\rm KK}}{M_s} \sim  \frac{1}{{\cal V}^{1/6}_Y} \sim  \frac{1}{ \lambda} \,.
\end{aligned}
\ee
The second line, however, does not correspond to a meaningful Kaluza-Klein scale in the effective field theory, for the following reason.
Before the naive Kaluza-Klein tower becomes relevant, the
parametrically lower scale of the D0-brane tower will be dominant. This signals decompactification to five-dimensional M-Theory
before we enter the geometrical decompactification regime. Thus,  once in M-Theory,
 the supergravity Kaluza-Klein spectrum must be re-evaluated in the M-Theory frame.
This is done via the relation \cite{Witten:1995ex}
\be \label{JMJV}
J_{\rm M} = \frac{J_Y}{\gA{}^{2/3}} \,.
\ee 
Here the K\"ahler form $J_{\rm M}$ measures volumes  in the M-Theory frame in units of $\ell_{11}$, and the K\"ahler form $J_Y$ refers to the Type IIA string frame.
Together with (\ref{L1volscaling}) this gives 
\be
\begin{split} \label{VMtheory1}
& {\cal V}_{T^2,\rm M} = \lambda^{2/3} \,, \qquad  {\cal V}_{B_2, \rm M} =\lambda^{-2/3} \, {\cal V}'_{B_2, \rm M}  \,, \qquad    
 {\cal V}_{Y,\rm M} = a\,  \lambda^2 + S + c \,  {\cal V}'_{B2,\rm M}  + \ldots 
\end{split}
\ee
with ${\cal V}'_{B_2, \rm M}$ finite as $\lambda \to \infty$.
For $a \neq 0$, the diverging Calabi-Yau volume then implies that the five-dimensional M-Theory undergoes further decompactification.\footnote{The special case $a=0$ will be treated below.} Indeed
the five-dimensional Kaluza-Klein scale for this second decompactification can be estimated as 
\be \label{KKMtheory}
\frac{M_{\rm KK}}{M_{11}} \leq \frac{1}{{\cal V}^{1/6}_{Y,\rm M} } \precsim \lambda^{-1/3} \,,
\ee
rather than the naive value given in (\ref{D2KKatL1}).
{The precise value depends on the scaling of 
$S = {\cal V}_{C_2}$.}

The same conclusions are reached by studying the system at the physically equivalent point $L_2$.
In the Type IIA frame, we find the following relevant mass scales:
\be \label{D2KKatL2}
\begin{aligned}
&\text{D2 at $L_2$}:\qquad &&\frac{M}{M_s} \sim   \frac{{\cal V}_{T^2}}{\gA}  \sim      \frac{1}{\gA^{(0)} \lambda^2} \\
&\text{Type IIA KK at $L_2$}: \qquad &&\frac{M_{\rm KK}}{M_s} \sim  \frac{1}{{\cal V}^{1/6}_Y} \sim  \frac{1}{\lambda^{1/3}} \,.
\end{aligned}
\ee
The D2 and D0-spectra at $L_2$ and $L_1$ obviously agree in (\ref{D2KKatL2}) and (\ref{D2KKatL1}).
At $L_2$, it is the parametrically leading spectrum of D2-branes on $T^2$ which acts as the dominant tower of Kaluza-Klein states, and thus signals decompactification to five dimensions. 
The Kaluza-Klein spectrum in the five-dimensional M-Theory is then computed via (\ref{JMJV}), but using the value of $\gA = \gA^{(0)}$ which is appropriate at $L_2$.
This leads to the same expression (\ref{KKMtheory}). In this sense the Kaluza-Klein spectra as computed at $L_1$ and $L_2$ agree, despite the apparent mismatch of the scales computed in the second lines of (\ref{D2KKatL2}) and (\ref{D2KKatL1}).

To conclude this discussion, consider now the special case $a=0$ in (\ref{classicalvolumeelliptic}) and (\ref{VMtheory1}).
We will show that the M-Theory decompactifies  further,
even if the limit is taken such that  the total M-Theory volume, ${\cal V}_{Y,\rm M}$, remains finite. 
To understand this, note first from (\ref{VMtheory1}) that ${\cal V}_{Y,\rm M}$   can remain finite only if 
\be \label{SvolumeCS}
S \equiv {\cal V}_{C_S} = \lambda^{4/3} \, {\cal V}_{C_S,{\rm M}}
\ee
stays  finite in the limit as well.
For instance, for a smooth Weierstrass model over $B_2$, footnote (\ref{aSparameters}) implies that
$a=0$ is possible for $\bar K \cdot_{B_2} \bar K= 10 - h^{1,1}(B_2) = 0$, where we recall that the curve $C_S = \bar K$ for a smooth Weierstrass model.
Following the discussion in Section \ref{sec_Dmanifolds}, the base $B_2$ is a dP$_9$ surface, and $C_S = \bar K$ is its elliptic fiber.
Now, in order for (\ref{SvolumeCS}) to stay finite,  ${\cal V}_{C_S,{\rm M}}$ must scale as $\lambda^{-4/3}$. At the same time, ${\cal V}_{B2,\rm M}$ must scale as $\lambda^{-2/3}$ if we insist on finite ${\cal V}_{Y,\rm M}$.
This means that the base $\mathbb P^1_b$ of the elliptically fibered dP$_9$ surface must scale as $\lambda^{2/3}$ and hence altogether
\be
{\cal V}_{\mathbb P^1_b,\rm M} \sim \lambda^{2/3} \,,\qquad {\cal V}_{T^2,\rm M} \sim \lambda^{2/3} \,,\qquad {\cal V}_{C_S,\rm M} \sim \lambda^{-4/3} \,.
\ee
As discussed in Section \ref{sec_Dmanifolds}, the elliptic fibration over dP$_9$ is a double elliptic fibration over $\mathbb P^1_b$, with fiber ${\cal E}_1 \times {\cal E}_2 =T^2 \times C_S$, see (\ref{E1}).
The fact that ${\cal E}_2=C_S$ shrinks suggests that we should view it as elliptic fiber of $Y$, whose base ${\cal B}_2$ is 
a rational elliptic surface with fiber ${\cal E}_1 = T^2$.
Indeed, the scaling of ${\cal E}_2=C_S$  is  exactly as for an M-Theory infinite distance limit of Type $T^2$.
The theory therefore decompactifies further to a six-dimensional F-Theory compactification
on the dP$_9$ base ${\cal B}_2$.  Thus, for $a=0$ we obtain altogether the following chain of decompactification limits:
\be \label{decomp-chain}
\text{4d Type IIA on $Y$} \quad  \stackrel{\text{D0/D2 on $T^2$}}{\longrightarrow} \quad \text{5d M-Theory on $Y$} \quad \stackrel{\text{M2 on $C_S$}}{\longrightarrow} \quad \text{6d F-Theory with base dP$_9$}
\ee
where we indicate the tower of BPS states responsible for each decompactification step.

In summary, we see that Type IIA strings probe the elliptic fibration geometries and their infinite distance limits in a very different manner 
as compared to M-Theory: It is not possible to perform an infinite distance limit where the elliptic fiber shrinks 
in such a way that the total volume would stay finite, without  effectively running
into a decompactification to a higher dimensional theory, in one way or another.

\subsubsection{Limit with co-scaling}

Let us go back to (\ref{L1volscaling}) at the point $L_1$, which is the T-dual to the point $L_2$ of vanishing fiber volume.
According to the general discussion in Section \ref{sec_gencons1}, we can co-scale $g^{(0)}_{\rm IIA}$ in such a way as to keep the volume finite in units of $M_s$.
For $a \neq 0$ and $S \precsim \lambda^2$ this requires taking
\be\label{rescale-g0IIA}
g^{(0)}_{\rm IIA} \sim \lambda \,.
\ee
The mass scale of the D0-branes becomes
\be \label{D2KKatL1coscale}
\begin{aligned}
 &&\frac{M}{M_s} \sim   \frac{1}{\gA}  \sim      \frac{1}{\lambda^3}  \,.
\end{aligned}
\ee
As a result, the theory decompactifies once more to M-Theory on $Y$, with all volumes in the five-dimensional frame scaling as
\be
\begin{split} \label{VMtheory1coscale}
& {\cal V}_{T^2,\rm M}  \sim 1 \,, \qquad  {\cal V}_{B_2, \rm M} =\lambda^{-2} \, {\cal V}'_{B_2, \rm M}  \,, \qquad    
 {\cal V}_{Y,\rm M} = a\,  + \lambda^{-2} \, S + c \, \lambda^{-2} \, {\cal V}'_{B2,\rm M}  + \ldots  \,.
\end{split}
\ee
By construction, ${\cal V}_{Y,\rm M}$ is finite and any residual, subsequent limit must be one of the three infinite distance limits at finite volume of five-dimensional M-Theory, for which we refer to Section \ref{sec_Mtheorylimits}.
The same conclusion is reached if $S \succ \lambda^2$, in which case the rescaling~\eqref{rescale-g0IIA} of $g^{(0)}_{\rm IIA}$ to keep the total volume finite has to change correspondingly.
For $a = 0$ and $S$ finite, on the other hand, no additional rescaling of $g^{(0)}_{\rm IIA}$ is necessary as the volume is already finite. As discussed above, this leads to the decompactification chain (\ref{decomp-chain}).

\subsection{Quantum small fiber limit for K3 fibrations}\label{sec_QK3IIA}

As discussed in Section \ref{sec_TypeBK3M}, M-Theory on a K3-fibered Calabi-Yau three-fold $Y$ admits an equi-dimensional infinite distance limit,
 for which the classical volume of the K3-fiber goes to zero such that the total volume of $Y$ stays finite.
In this section we will analyze analogous  limits in Type IIA string theory, where we will find that they are obstructed by quantum effects. 
As we will argue, any such limit implies decompactification 
and a diverging Planck scale if we keep the string coupling, $g_{\rm IIA}$, finite.
A finite Planck scale can only be obtained by performing a simultaneous rescaling of the string coupling: $g_{\rm IIA} \to \infty$. 
Interestingly, such a co-scaled limit {\it is} an
 equi-dimensional limit because a tensionless heterotic string sits at the same scale as the D0 tower which normally would take us to five-dimensional M-Theory.

We will first give a physical argument why a quantum obstruction against taking a finite volume K3-fibered limit in Type IIA string theory must occur. This discussion is completely general without reference to a specific realisation of the fibration.
This will be then exemplified for a particular K3-fibration, by studying the regime of small fiber volume via mirror symmetry.

\subsubsection{Implications for quantum geometry from M-Theory}

Our first task is to determine if the limit of vanishing K3-fiber volume is part of the quantum K\"ahler moduli space of $Y$.
Since the stringy quantum volume of a cycle is determined by the BPS tension of a D-brane wrapping it, 
an equivalent question is whether a D4-brane wrapping the K3 fiber can become massless.
Typically obstructions would be due to worldsheet instantons in the stringy K\"ahler geometry of $Y$, 
which can also be dually interpreted as perturbative 1-loop corrections to the classical M-Theory effective action 
compactified on a circle \cite{Lawrence:1997jr,Gopakumar:1998ii,Gopakumar:1998jq}.
This motivates us to carefully investigate the circle reduction linking  M-Theory and Type IIA string theory.

In flat space, the D4- and NS5-branes in the Type IIA theory originate from M5-branes wrapping or not wrapping the M-Theory circle, $S^1$,
which we take to be of some radius $R$.
Equating the respective expressions for the tension identifies  \cite{Witten:1995ex}
\be \label{MIIArelations}
\frac{1}{\ell_s} \equiv M_s = \frac{\gA}{R} \,, \qquad \quad \gA = (M_{11} R)^{3/2} \,.
\ee

If we consider M-Theory on the Calabi-Yau $Y$, 
an M5-brane along the K3-fiber gives rise to a solitonic heterotic string in five dimensions of tension
\be
\frac{T_{\rm het}}{M^2_{11}} = 4 \pi {\cal V}_{\rm K3,M}  = 2\pi \int_{\rm K3} J_M^2 \,.
\ee
The subscript of $J_M$ and ${\cal V}_{\rm K3,M}$ reminds us that these quantities refer to the M-Theory frame. 
Essentially, the heterotic string wrapped on the additional $S^1$ corresponds in Type IIA theory
to a  D4-brane wrapped on the K3 fiber, but we have to be careful what exactly we mean by a D4-brane in this context due to curvature effects.
Recall that the Chern-Simons action of a D(p+1)-brane along a $p$-cycle $\Gamma$  is
\be \label{SCS}
S_{\rm CS} = {2\pi} \int_{\mathbb R \times \Gamma} \sum_k  \RRC_{k} \, \wedge  \sqrt{\hat A(\Gamma)} \, \wedge \, e^{F}\,,
\ee
where $\hat A(\Gamma)$ is the A-roof genus and where we set $\ell_s\equiv 1$. 
In addition to its $\RRC_5$ charge, a D4-brane wrapping the whole
 K3 therefore carries one unit of induced  $\RRC_1$ charge,
  because the A-roof genus contributes the coupling 
  $S_{\rm CS} \supset 2\pi \int_{\mathbb R} \RRC_1 \times \frac{1}{24} \int_{\rm K3} c_2({\rm K3}) =2\pi \int_{\mathbb R} \RRC_1$.
If we are interested in the object carrying only $\RRC_5$ charge and no lower-form Ramond-Ramond charge,
 we must rather consider  a bound state of one D4-brane on K3 with (-1) D0-brane (interpreted as an anti-D0-brane). 
 We will call this object for brevity a D4$_{-1}$ brane.

In the limit of small ${\cal V}_{\rm K3,M}$, we identify the BPS particle that is
obtained by wrapping a D4$_{-1}$-brane along the K3 fiber with the dual solitonic heterotic string
 wrapped once around the extra $S^1$. 
The important observation is that this wrapped string has some non-vanishing Casimir energy, $E_0$, which produces an offset for the BPS mass of the particle for any finite value of the circle radius $R$ (and thus also for any finite value of $\gA$). 
In the present context we have $E_0 = -1$, as computed from the left-moving sector of the heterotic string.
More generally, $E_0$ relates to the 1-loop gravitational anomaly on the stringy world-sheet and is
given by 
\be\label{vacuumshift}
E_0 =-\frac\chi{24}\,, 
\ee
where $\chi$ is the Euler number of the shrinking fiber ($\chi=24$ for K3 under present consideration).

The total mass of the wrapped heterotic string obtained in this way is thus
\bea
\frac{M}{M_{11}} &=& \left|  w \frac{R}{M_{11}}  T_{\rm het}  + \frac{E_0}{R M_{11}} \right|   \,,
\eea
with winding number $w=1$ and Casimir energy $E_0=-1$.
The limit of vanishing {\it classical} K3-volume corresponds to $T_{\rm het}  \to 0$, and using (\ref{MIIArelations}) this yields
\be
\frac{M}{M_{s}} =  \frac{1}{g_{\rm IIA}}  \,.
\ee
This is the minimal mass of a single D4$_{-1}$-brane wrapping the K3-fiber in Type IIA string theory.  Our reasoning therefore predicts that the quantum volume of the K3 cannot become smaller than $\ell_s^4$ anywhere in moduli space. Indeed, since the object D4$_{-1}$ carries only $\RRC_5$ charge, its entire mass
is due to the (quantum) volume of K3, and if this mass cannot reach zero,
 this means that the quantum volume of the K3 cannot vanish anywhere in moduli space.

Moreover, the heterotic string with winding number $w=1$, vacuum energy $E_0 = -1$ plus $n_{\rm KK}$ extra quanta of Kaluza-Klein momentum,
gives rise to a BPS state of mass
\be
\frac{M}{M_{11}} = \left|   \frac{R}{M_{11}}  T_{\rm het}  + \frac{E_0 + n_{\rm KK}}{R M_{11}} \right|  \,. 
\ee
We interpret this as a bound state of one D4$_{-1}$-brane with $(-n_{\rm KK})$ extra D0-branes (or rather $n_{\rm KK}$ anti-D0-branes).
 In the limit where $T_{\rm het} \to 0$, we obtain for its mass:
\be \label{cancellationMtheory}
 \frac{M}{M_{s}} =  \frac{1}{g_{\rm IIA}} \left|  n_{\rm KK} -1 \right| \,.
\ee
Hence the bound state of one D4$_{-1}$-brane on K3 with $(-n_{\rm KK}) = -1$ additional D0-brane  - i.e. altogether a D4$_{-2}$ brane - {\it can} become massless 
exactly at the point in the quantum moduli space that is the analogue of the vanishing fiber volume in M-Theory.
We will confirm this general prediction later by a non-trivial computation in mirror symmetry for an example.

What about the tension of the NS5-brane wrapped on K3 at the same point in moduli space where the D4$_{-2}$ state becomes massless? 
This could potentially give rise to a tensionless heterotic string, which would be against all expectations (e.g. from experience with realisation of Seiberg-Witten theory in Type IIA theory \cite{Kachru:1995fv}).
However we can argue that the NS5-brane on K3 probes the same quantum volume as the D4$_{-1}$ brane. This is because both objects carry vanishing D0-brane charge and hence qualify as the objects obtained from the M5-brane on K3
 - depending on whether or not they wrap the M-Theory circle. Combined with our above findings, this means that an 
asymptotically tensionless heterotic string from the K3-fiber will not appear, as measured in units of $M_s$, as long as $g_{\rm IIA}$ is kept small and finite.

More subtle is the fate of a D2-brane wrapped on some curve inside the K3-fiber. 
As explained in Section \ref{sec_TypeBK3M}, the M2-brane along the same curve is dual to
the fundamental  heterotic string wrapped on the circle of $\hat K3 \times S^1_A$. 
The question is whether these particles acquire quantum corrections to their mass in going from M-Theory to Type IIA theory. 
The naive expectation is that this should be generically the case, and we will verify this explicitly for an example discussed below.

There is another important conclusion we can draw from this analysis:
There exists no infinite tower of (single-particle) BPS bound states of the form 
\be \label{D4-2bound}
n \times \text{(D4$_{-2}$ on K3)} \,.
\ee
As just explained, the bound state of a  D4-brane on K3  with (-2) D0-branes is dual to the heterotic string wrapped once on
 the $S^1$ with a net number of $E_0 + n_{\rm KK}=-1 +1 =0$ Kaluza-Klein quanta.
An immediate question might be whether the heterotic string winding $n$ times around $S^1$ can form non-trivial BPS bound states, 
but this is known not to be the case \cite{Minahan:1998vr}.
 Hence at the point in moduli space where the {\it classical} volume of K3 vanishes,
 only one BPS state becomes massless (with respect to the string scale) from a single D4$_{-2}$ bound state; as we will discuss later, this state is
 nothing but the $W$-boson of $N=2$ supersymmetric Seiberg-Witten gauge theory. 
 
To summarize, we predict a quantum obstruction against the vanishing of the K3-fiber in the Type IIA moduli space. Combined with the fact that 
the volume of the base scales to infinity in the given limit, we expect that the total Calabi-Yau volume should become infinite as well. We will
demonstrate this further below for the example.
We therefore posit that in {\it weakly coupled} 4d N=2 compactifications of Type IIA string theory, it is not possible to take a long distance limit in which the Planck scale stays finite in terms of the 10d string scale.
As we will see, such a limit is possible only by suitably co-scaling $g_{\rm IIA} \to \infty$, which will turn out to be an equi-dimensional limit.

In the remainder of this section we will verify these predictions by explicitly analyzing the quantum geometry of a prototypical K3-fibration. 
In view of the preceding discussion, we expect the properties encountered in this example to apply very generally.

\subsubsection{Quantum geometry of the small K3 fiber limit}\label{sec_QK3}

For the purposes of this section the details of the embedding of the K3-fiber into the full Calabi-Yau $Y$ are inessential. It
suffices to consider a simple K3 fibration for which the image (\ref{Lambdalatticedef}) of the K3-lattice embedded into $Y$ is one-dimensional. This
captures the quantum geometric properties we are interested in, namely those which affect the classical zero-volume limit of the K3 fiber.
As a prototype we choose the Calabi-Yau three-fold $Y = \mathbb P_{1,1,2,2,6}[12]$, which has been well studied starting from 
refs.~\cite{Candelas:1993dm,Hosono:1993qy} and whose Mori cone and intersection numbers we have introduced at the end of Section~\ref{sec_TypeBK3M}. 
From the point of view of the swampland distance conjecture it has recently been discussed in ref.~\cite{Blumenhagen:2018nts}.  
The classical limit in question can be parametrized as in (\ref{JforexK31}) and (\ref{classlimitB}).
The real curve volumes $S$ and $T$ appearing in (\ref{JforexK31}) are now the imaginary parts of the complexified volumes
\be
t^1 = \int_{\mathbb P^1_f} {\bf J}_Y     \,, \qquad  t^2 = \int_{\mathbb P^1_b} {\bf J}_Y \,,
\ee
as defined in (\ref{tadefinition}).
In terms of the variables
\be
z_a = e^{2 \pi i t^a}\,,
\ee
the classical limit (\ref{classlimitB}) corresponds to the regime where
\be \label{xyparametrisation}
z_1 = e^{-a/\sqrt{\la}} \to 1 \,, \qquad \quad \quad z_2 = e^{-\la} \to 0   \,.
\ee
To analyze whether a corresponding limit can be taken in the quantum corrected K\"ahler moduli space of $Y$, 
we interpret $z_1$ and $z_2$ as the usual coordinates of the complex structure moduli space, $\MCS(X)$, of the mirror manifold $X$
as studied in detail in~\cite{Candelas:1993dm,Hosono:1993qy}. Their dependence on $t^a$ will be strongly modified away from $z^a\simeq0$.  As recalled in Appendix \ref{App_K3quantum},
the moduli space becomes singular along the discriminant locus, which contains as a factor the conifold locus
\be
 \Delta_c \equiv(1728z_1-1)^2-4(1728z_1)^2z_2\ =\ 0\,.
\ee
The relevant part of the moduli space is hence precisely as described at the end of Section \ref{sec_gencons1}:
The double intersection point $\{z_2=0\} \cap \Delta_c$ is resolved by introducing a resolution divisor ${\cal B}$ and the 
regime (\ref{xyparametrisation}) corresponds to 
the region near the intersection point 
\be
L_2 = {\cal B} \cap  \{z_2=0\} \,.
\ee
Of course a classical interpretation of the K\"ahler geometry of $Y$ is a priori  possible only near the large volume point $L_1$,
where $t_1,t_2 \to \infty$. Its mirror in $\MCS$ corresponds to $z_i \to 0$ for $i=1,2$, which
is denoted by $L_1$ in Figure~\ref{f:Fig1MCS}. Near $L_1$ we fix an integral symplectic basis of $H_{2p}(Y,\mathbb Z)$ as
\be \label{hatgammabasis}
 C_0\,,  \quad   C^1_{2} =  \mathbb P^1_f\,, \quad    C^2_{2} =  \mathbb P^1_b \,, \quad   C_{4,2} = J_S \,, \quad   C_{4,1} = J_T \,, \quad    C_6 \,,
\ee
where $ C_0$ and $ C_6$ represent the unique classes in $H_0(Y,\mathbb Z)$ and $H_6(Y,\mathbb Z)$, respectively, and we recall that $J_S$ 
denotes the K3-fiber.
On the mirror three-fold $X$, these cycles map to an integral symplectic basis $\{\gamma_A, \gamma^B\}$ of $H_3(X,\mathbb Z)$ 
according to the following scheme:
\be
\begin{aligned}
C_0        &\longleftrightarrow& \gamma^0   \,,  &  \qquad   C_6  &\longleftrightarrow&\ \ \  \gamma_0 & \ ,\cr
C_{2}^a  &\longleftrightarrow& \gamma^a   \,, & \qquad C_{4,a}  &\longleftrightarrow& \ \ \ \gamma_a &\ .
\end{aligned}
\ee

To study the quantum geometry in the regime of vanishing fiber volume,
we need to analytically continue the periods from $L_1$ to the regime near $L_2$.
For this, one first computes a vector of solutions, $\hat \Pi$, to the Picard-Fuchs equations valid at $L_2$, see eq. (\ref{hatPisolution}).
These solutions can be interpreted as the periods with respect to some ad hoc basis of 3-cycles $\{ \hat \gamma_{A}, \hat \gamma^B\}$ on $X$.
The cumbersome step then consists of finding how this generally non-integral basis
 relates to the integral symplectic basis  $\{\gamma_A, \gamma^B\}$, 
 which is mirror, at large complex structures, to the $2p$-cycle basis (\ref{hatgammabasis}) on $Y$.
The analytic continuation of the periods from $L_1$ to $L_2$ has been performed in \cite{LMunpubl} (see also \cite{Curio:2000sc}), 
and we present the result in Appendix \ref{App_K3quantum}. The upshot is that
the analytic continuation relating those two bases acts as a matrix, ${\cal N}$, via
\be
\gamma = {\cal N}^{-1} \cdot  \hat\gamma \,,
\ee
which is given in equation (\ref{calNmatrix}).
This then yields  the following expressions for the  periods (\ref{periods})
with respect to the integral basis  $\{\gamma_A, \gamma^B\}$,  near the regime $L_2$:
\be
\begin{aligned} \label{K3periodscont}
X^0&:= \int_{\gamma^0} \Omega_X = \frac{1}{2 \pi X} + \ldots     \qquad \quad  &F_0&:= \int_{\gamma_0} \Omega_X = \frac{2 i }{\pi^2}  \sqrt{\hat z_1} {\rm log}(\hat z_2) + \ldots \cr
X^1&:= \int_{\gamma^1} \Omega_X =  \frac{i}{2 \pi X}  +\ldots    \qquad \quad &F_1&:= \int_{\gamma_1} \Omega_X = - \frac{1}{2\pi^2 X}   {\rm log}(\hat z_2 \hat z_1^2)  + \ldots  \cr
X^2&:= \int_{\gamma^2} \Omega_X =-\frac{i}{4 \pi^2 X}  {\rm \log}(\hat z_2 \hat z_1^2) + \ldots          \qquad \quad    &F_2&:= \int_{\gamma_2} \Omega_X = \frac{1}{\pi X} + \ldots\,.
\end{aligned}
\ee
Here we have omitted all subleading terms, and $X= { \Gamma \left(\frac{3}{4}\right)^4}/{\sqrt{3} \pi ^2}$.
The local blowup coordinates vanish by design near the point $L_2: \hat z_i \to 0$ and are given by
\bea
{\hat z}_1 = 1-1728 z_1 \,, \qquad 
{\hat z}_2 = \frac{4z_2{1728 z_1}^2 }{(1-1728 z_1)^2} \,. 
\eea

The periods (\ref{K3periodscont}) then determine the exact BPS masses (\ref{BPStension}) 
of D-branes wrapping  the mirror dual cycles (\ref{hatgammabasis}) on $Y$, in the regime near $L_2$.
For the D2-branes along 2-cycles, the tension is directly proportional to the quantum volume of the wrapped curve,
 because such a brane carries no lower-dimensional brane charge.
In stark contrast to the classical limit (\ref{limitclassical}), we see that the quantum volume of $\mathbb P^1_f$ does not vanish at $\hat z_i \to 0$, but rather tends to a constant value of order one in string units because of:
\be
\frac{M}{M_s} =\frac{1}{g_s} \left|\frac{X^2}{X^0}\right|  = \frac{1}{g_s} = \frac{{\cal V}_{\mathbb P^1_f}}{g_s} \qquad \text{for a D2-brane on} \, \,  \mathbb P^1_f \,.
\ee
For D4-branes wrapped on 4-cycles, we must subtract the contribution from the induced D0-charge to isolate the quantum volume.
As explained below eq.~(\ref{SCS}),  a D4-brane along the K3-fiber carries one unit of induced D0-charge, and the object whose tension is proportional to the quantum volume of K3 is the bound state we had denoted by D4$_{-1}$. Its BPS mass is
\be
\frac{M}{M_s} =\frac{1}{g_s} \left|\frac{F_2 - X^0}{X^0}\right|  = \frac{1}{g_s}= \frac{{\cal V}_{\rm K3}}{g_s}\,.
\ee
As advertised before, 
this non-zero value for the minimal quantum volume perfectly agrees with the Casimir energy of the heterotic string when wrapped
on the M-Theory circle, $S^1$.

Since the quantum volume of the K3 does not vanish near $L_2$, the volume of the total Calabi-Yau $Y$ should diverge because the base becomes large.
We can confirm this expectation
by recalling that the total volume of $Y$ in general differs from the period $F_0$ associated with the quantum 6-cycle, but rather is given by
\be \label{volumeVY}
{\cal V}_{Y} = \frac{{ i \int_X \Omega_X \wedge \bar\Omega_X}}{8 |X^0|^2} =  \frac{ i (\bar X^i F_i - X^i \bar F_i)  }{ 8 |X^0|^2}     =- \frac{1 }{2\pi} {\rm log}(\hat z_2 \hat z_1^2) + \ldots \,,
\ee
and so indeed diverges for $\hat z_2 \to 0 \,,\hat z_1 \to 0$.

On the other hand, from our M-Theory considerations, where we argued for the existence of a single massless BPS particle, D4$_{-2}$,  
we expect that there should exist some {\it integral} linear combination 
of cycles whose quantum volume does vanish at $L_2$. 
By inspection of the matrix ${\cal N}$ displayed in (\ref{calNmatrix}), the only integral linear combination for which this is the case is
\be \label{gammaWdef}
\gamma_W =   \gamma_2- 2\gamma^0 \,.
\ee
By definition the quantum volume of this cycle is given by the tension of the bound state of a D4-brane wrapped
on K3 with (-2) D0-branes, and it scales near $L_2$ as
\be \label{gammaWperiod}
\int_{\gamma_W} \Omega_X =  F_2- 2 X^0=  \frac{1}{\pi} \sqrt{\hat z_1} + \ldots  \to 0   \qquad {\rm at} \, \ {L_2} \,.
\ee
This matches the expected behaviour for the D4$_{-2}$ bound state on K3, which we anticipated from arguing via M-Theory
and which reflects the cancellation of the Casimir energy on $S^1$ against one unit of D0-brane charge. 

To wrap up this section, let us recapitulate and list the asymptotic
BPS masses of the bound states denoted by D4$_{-n}$, in the regime of large base where $t_2\to \infty$.
For the pure $D4$ brane, the central charge is defined by $F_2=\partial_2 F(t_1,t_2)$, where  $F(t_1,t_2)$ is given in (\ref{FK3}) and
whose relevant part reads
$$
F(t_1,t_2) = -{t_1}^2t_2+ b_2 t_2+\dots\,, \qquad\  {\rm where}\ \ \ \ b_2=\frac1{24}\int  c_2\wedge J_2=\frac1{24}{\chi(K3)}=1\,.
$$
Thus we have in this limit\footnote{Note that for $t_2\to \infty$  there are no instanton corrections to these volumes.} 
\be
\begin{aligned} \label{D42n}
M(D4)\ &\sim\  |F_2|&= &\ |{t_1}^2-1| \,,\ \cr
M(D4_{-1})\ &\sim\  |F_2-X_0|&= &\ |{t_1}^2| \,,\ \cr
M(D4_{-2})\ &\sim\  |F_2-2X_0|&= &\ |{t_1}^2+1| \,,\ \cr
\end{aligned}
\ee
which exhibits the interplay of curvature-induced $D0$ charge, Casimir energy $E_0=-b_2$ as discussed
in (\ref{vacuumshift}), and mass shifts. From the mirror map of the sextic K3 surface \cite{Lian:1994zv} we know that
$$
J(q_1)= \frac{1728}{z_1}\,,  \qquad  q_1\equiv e^{2\pi it_1}\,,
$$
so that at the conifold point $L_2: z_1=1$ we have $t_1=i$. Thus the quantum volume of the K3 fiber, $M(D4_{-1})$, which also gives the tension of the heterotic string arising from a wrapped NS5 brane, is of order one. On the other hand, the mass of the single BPS state $D4_{-2}$ vanishes, and this
is entirely consistent with the expectation that at the conifold point, which corresonds to a finite distance limit in the moduli space, only a finite number of states become massless.

We also see a crucial difference as compared to the elliptic fibration discussed above, and to the abelian fibration to be discussed later: For the K3 fiber the
quantum shift traces back to a non-zero $b_2=\frac1{24}{\chi(K3)}$.  For toroidal fibers,
$\chi$ vanishes, and there is no such shift. Correspondingly, the quantum volumes of the elliptic or abelian fibers do vanish at $L_2$.

\subsubsection{Weak coupling as gravity decoupling limit}

To summarize our findings so far, we have shown that the quantum version of the limit (\ref{limitclassical}) corresponds to a decompactification limit in which the Calabi-Yau volume (\ref{volumeVY}) in string units diverges. This traces back to a non-vanishing K3 fiber volume at the quantum level, which perfectly
ties in with our previous arguments in M-Theory.

Let us first analyze the physical consequences for limits where we keep $g_{\rm IIA}$ fixed.
In view of the highly
anisotropic scaling of $Y$, we identify the Kaluza-Klein scale, $M_{\rm KK}$, as the scale dominated by the volume of the 
base 2-cycle $\mathbb P^1_b$. This leads to
\be \label{MKKquantumK3}
\frac{M^2_{\rm KK}}{M^2_{s}} \sim \frac{1}{{\cal V}_{\mathbb P^1_b}}  = \left| \frac{X^0}{X^2} \right| = \frac{2\pi}{ |- {\rm log}(\hat z_2 \hat z_1^2)|} + \ldots  \to 0 \,.
\ee
An asymptotically low Kaluza-Klein scale is indeed a hallmark of decompactification. Correspondingly,
for finite string coupling $g_{\rm IIA}$, the Planck mass diverges in terms of the string scale as follows:
\be \label{MplK3QGa}
\frac{M^2_{\rm Pl}}{M^2_s} = \frac{{4 \pi}}{g^2_{\rm IIA}} {{\cal V}_{Y} } =   \frac{2}{g^2_{\rm IIA}} { \left|-  {\rm log}(\hat z_2 \hat z_1^2)\right|} \to \infty \,, \qquad {\rm for\  finite} \, \, g_{\rm IIA}  \,.
\ee
The decompactification and the divergence of the Planck scale can be avoided only by simultaneously rescaling $g_{\rm IIA}$, 
 hence taking us back to the classical M-Theory description. This will be discussed below in the next section.

For now we would like to study the mass scales of the  degrees of freedom associated with branes wrapping the various relevant
cycles, in the regime near $L_2$. It is instructive to list the parametric behavior of the BPS masses,
as determined by the general mass formulae (\ref{M2p1}) and (\ref{M2p2}) in conjunction with the quantum periods (\ref{K3periodscont}):
\be
\begin{aligned} \label{towerscalingK3a}
{\rm KK \, \, scale}: \qquad  &\frac{M}{M_s} \sim (- {\rm log}(\hat z_2 \hat z_1^2))^{-1/2} &\qquad \frac{M}{M_{\rm Pl}} &\sim& \frac{g_{\rm IIA}}{-{{\rm log}(\hat z_2 \hat z_1^2)}}      \cr
{\rm D0}\, /\, {\rm D2 \, \, \, on \, \, \mathbb P^1_f} /\qquad  &&\vspace{-1mm}\cr
 {\rm NS5 \, \, \, on \, \, K3} \,  / \,  {\rm D4 \, \, \, on \, \, K3}: \qquad & \frac{M}{M_{s}} \sim \frac{1}{g_{\rm IIA}}  \qquad &\frac{M}{M_{\rm Pl}} &\sim& \frac{1}{\sqrt{-{\rm log}(\hat z_2 \hat z_1^2)}} \cr
{\rm -2 D0 + D4 \, \, \, on \, \, C_W}: \qquad &\frac{M}{M_s} \sim \frac{\sqrt{\hat z_1}}{g_{\rm IIA}}  \qquad &\frac{M}{M_{\rm Pl}} &\sim& \frac{\sqrt{\hat z_1}}{\sqrt{-{\rm log}(\hat z_2 \hat z_1^2)}} \cr
{\rm  D6 \, \, \, on \, \, C_6}: \qquad &\frac{M}{M_s} \sim \frac{- {\rm log}(\hat z_2) \sqrt{\hat z_1}}{g_{\rm IIA}}  \qquad &\frac{M}{M_{\rm Pl}} &\sim& \frac{- {\rm log}(\hat z_2) \sqrt{\hat z_1}}{\sqrt{-{\rm log}(\hat z_2 \hat z_1^2)}} 
\end{aligned}
\ee
where $C_W$ denotes the holomorphic cycle mirror to $\gamma_W$ in (\ref{gammaWdef}).
Note that the fate of the D6-brane that wraps the six-cycle depends on the precise 
but unspecified ratio of $\hat z_2$ and $\hat z_1$ in the limit $\hat z_i \to 0$. 

With regard to relevance for weak gravity conjectures, there are the following sources of potential\footnote{Focusing on scaling properties only, we do not consider issues of stability here. For related recent discussions, see ref.~\cite{Grimm:2018ohb}.} infinite towers of BPS particles:
\begin{enumerate}
\item
The towers of BPS particles present in M-Theory compactified on $Y$ continue to exist as light states:
These are bound states of $n$ D2-branes along the fibral curve $\mathbb P^1_f$ on the one hand,
and particle excitations of the effective heterotic string associated with a (single wrapped) NS5-brane along the K3-fiber on the other. 
Also for the Type IIA string, both these towers are naturally associated with the dual heterotic string that emerges in the infinite distance limit.
The difference to the M-Theory compactification
 is that the mass scale {\it in string units} of both BPS towers does not vanish in the limit. This is a consequence of the quantum corrections discussed above. Nonetheless they become arbitrarily light as measured in Planck units, because the Planck scale diverges itself. 
\item
BPS bound states of $n$ D0-branes on $Y$ are well-known to realize the conventional 5d Kaluza-Klein states associated with the circle reduction from M-Theory to Type IIA string theory. They sit at the same parametric mass scale as the BPS towers that are present already in M-Theory.
\item
For finite string coupling, $g_{\rm IIA}$, the Kaluza-Klein tower (associated with the large base volume) is parametrically lighter than all other BPS towers. In this sense we reach the decompactification scale {\it before} the infinite towers of states associated with the wrapped D0, D2 or NS5-brane excitations become relevant. 
\end{enumerate}

Apart from particles and strings, wrapped branes can also give rise BPS objects in four dimensions, i.e. 2+1 dimensional domain walls. Their mass scales have been studied in various large volume limits in \cite{Font:2019cxq} (prior to possible quantum corrections).
Unlike for light weakly coupled strings, their quantization is not expected to give rise to a tower of particles which would compete with a Kaluza-Klein tower such as to avert effective decompactification. Apart from this general fact, for dimensional reasons their associated energy scales lie above the mass scale of the solitonic heterotic string in the type of limit considered here. In this sense, they are subleading.

In addition to the light BPS {\it towers}, there exist in the Type IIA theory also individual light BPS {\it states} which do not have direct analogues in M-Theory: For example,
the BPS particle that arises from a single D4-brane wrapping the K3 fiber, and furthermore
 the bound state D4$_{-2}$ associated with $\gamma_W$ that we have discussed before. 
Moreover, a single D6-brane along the six-cycle mirror-dual to $\gamma_5$ can be interpreted as Kaluza-Klein monopole from the M-Theory perspective, 
but to what extent this state becomes light near $L_2$ depends on the ratio of $\hat z_2$ and $\hat z_1$ in the limit $\hat z_i \to 0$.

In fact, the last two types of states have an interpretation in the well-known realisation of Seiberg-Witten theory for $SU(2)$
via Type IIA string theory on the K3-fibration $Y$ \cite{Kachru:1995fv}:
The bound state D4$_{-2}$  represents the W-boson of charge $(q_{m}, q_e) = (0,2)$ and the wrapped D6-brane the monopole of charges $(q_m, q_e) = (1,0)$
(which becomes massless on the conifold locus, which however does not intersect $L_2$ after the blowup). 
From the perspective of Seiberg-Witten theory, approaching $L_2$ corresponds to taking the weak coupling limit. It is known that in this regime 
the stable BPS states besides the $W$-boson are given by dyons of charges $(1,2n)$ with $n \in \mathbb Z$. In the present situation these
are realized as bound states of one D6-brane on $C_6$  and $n$  D4$_{-2}$ bound states on the mirror-dual, $C_W$, of  $\gamma_W$. 
This fits nicely to our argumentation around (\ref{D4-2bound}), in that for
$n \neq \pm1$ no bound states of $n$ D4$_{-2}$ branes with zero ``magnetic'' D6-charge exist; there the reasoning was based on
the fact that multiple wrappings of heterotic strings do not form new bound states.

It is interesting to compare our findings with the BPS mass spectrum of Seiberg-Witten theory in the weak coupling limit.
From the perspective of quantum field theory, rather than of quantum gravity, the relevant scale is not the Planck mass but the  
gauge theory scale, $\Lambda$. The latter is defined via the running of the gauge coupling from a reference scale in the ultra-violet to the strongly coupled infra-red regime. 
In string compactifications, the reference scale at which the gauge coupling is defined, can be identified with the scale, $M_{\rm KK}$, of Kaluza-Klein excitations. 
As is well known, the complexified gauge coupling in Seiberg-Witten theory can be written as
\be
\frac{\partial a_D}{\partial a} = \frac{\theta}{\pi} + \frac{8 \pi i }{g^2_{\rm YM}} \,.
\ee
Here $a$ and $a_D$ control the central charges and hence the masses of states with electric-magnetic charges $(q_m, q_e)$ as
\be
Z = q_m a_D + q_e a \,.
\ee
From the specific form of the quantum periods and the above identification of the states with charges $(0,2)$ and $(1,0)$ we deduce, for the specific example,
\be
\frac{8 \pi^2}{g^2_{\rm YM}} = - 4 \,  {\rm log}(\hat z_2).
\ee
Identifying this with the value of the gauge coupling at the KK scale gives
\be
\frac{{\Lambda}}{M_s} = \frac{{M_{\rm KK}}}{M_s} \, {\rm exp} \left({- \left. \frac{8 \pi^2}{\beta_0}   \frac{1}{  g^2_{\rm YM}}\right |_{M_{\rm KK}}}\right)   = \frac{M_{\rm KK}}{M_s} \,  \hat z_2^{\alpha}\,.
\ee
Here the power $\alpha$ is determined in terms of  the beta-function coefficient, $\beta_0$, of the gauge theory as  follows:
\be
\alpha = \frac{4}{\beta_0}\,,
\ee
which for pure $SU(2)$ Seiberg-Witten theory is given by $\beta_0= 4$.

Together with (\ref{MplK3QGa}) this implies that 
\be
\frac{\Lambda}{M_{\rm Pl}} \sim \, \frac{   g_{\rm IIA}   \hat z_2^{\alpha } }{ {- \,  {\rm log}(\hat z_2 \hat z_1^2)  } }\to 0  \,.
\ee

The difference between analyzing the theory `in the presence of gravity' versus `in the field theory limit' corresponds to whether we measure the
masses with respect to $M_{\rm Pl}$ or $\Lambda$. In terms of the latter, the relevant BPS masses scale as follows:
\be
\begin{aligned}
{\rm KK}: \qquad  \frac{M}{\Lambda} &&\sim&   \, \,   \hat z_2^{-\alpha}    \to \infty       \\
{\rm D0}\, /\, {\rm D2 \, \, \, on \, \, \mathbb P^1_f}\, /\, {\rm NS5 \, \, \, on \, \, K3}\, /\, {\rm D4 \, \, \, on \, \, K3} : \qquad  \frac{M}{\Lambda} &&\sim& \, \, g_{\rm IIA}^{-1} \sqrt{- \,  {\rm log}(\hat z_2 \hat z_1^2)  }   \,  \hat z_2^{-\alpha}    \to \infty   \\
{\rm -2 D0 +D4 \, \, \, on \, \, C_W}:  \qquad  \frac{M}{\Lambda} &&\sim&  \, \,  g_{\rm IIA}^{-1} \sqrt{- \,  {\rm log}(\hat z_2 \hat z_1^2)  }      \hat z_2^{-\alpha} \hat z_1^{1/2}              \\
{\rm   D6 \, \, \, on \, \, C_6}:  \qquad  \frac{M}{\Lambda} &&\sim& \, \,   g_{\rm IIA}^{-1} \, \sqrt{- \,  {\rm log}(\hat z_2 \hat z_1^2)  } \,  (-{\rm log}(\hat z_2)) \,    \hat z_2^{-\alpha} \, \hat z_1^{1/2}    \,.    \nonumber  
\end{aligned}
\ee
Note that the behavior of the $W$ and the D6-brane mass with respect to $\Lambda$ depends on the relative rate at which $\hat z_1$ and $\hat z_2$ vanish in the limit. 
For instance, if the limit is taken in such a way that the mass of the monopole, i.e. of the D6-brane, goes to infinity in terms of $M_s$, both the $W$-boson and the monopole mass scale to infinity with respect to 
$\Lambda$.

Importantly, the first two lines include towers of particles which become light with respect to $M_{\rm Pl}$. These towers decouple from the gauge theory by becoming infinitely heavy with respect to $\Lambda$.
This explains why in Seiberg-Witten theory no towers of asymptotically light states are observed in the weak coupling limit; massless towers arise only with respect to $M_{\rm Pl}$. 
This is in agreement with the general intuition underlying the Swampland Distance conjecture, namely
that infinite distance limits for moduli in quantum field theory do not necessarily lead to towers of massless states, 
while infinite distance limits in the presence of gravity do.

\subsubsection{Equi-dimensional limit via co-scaling of $g_{\rm IIA}$}

In view of (\ref{MplK3QGa}) it is clear that we can keep the Planck scale finite at the point $L_2$ if we choose to
co-scale the coupling, $g_{\rm IIA}$, as
\be \label{gIIAdoublescaling}
g_{\rm IIA} = g^{(0)}_{\rm IIA} \,  \mu^{1/2} \,, \qquad      \mu \sim -{\rm log}(\hat z_2 \hat z_1^2) \to \infty \,.
\ee
As a result, all towers of states now appear at the same mass scale, as follows from (\ref{towerscalingK3a}):
\be
\begin{aligned}
{\rm KK \, \, scale}: \qquad  &\frac{M}{M_s} \sim (- {\rm log}(\hat z_2 \hat z_1^2))^{-1/2} \sim \mu^{-1/2}    \cr
{\rm D0}\, /\, {\rm D2 \, \, \, on \, \, \mathbb P^1_f} /    
 {\rm NS5 \, \, \, on \, \, K3} \,  / \,  {\rm D4 \, \, \, on \, \, K3}: \qquad & \frac{M}{M_{s}} \sim \frac{1}{g_{\rm IIA}} \sim  \mu^{-1/2}     
\end{aligned}
\ee
We are therefore in the situation where a tower of string excitations - given by the excitations of the NS5-brane along the fiber - sits at the same parametric scale as the various particle towers, namely those associated with the supergravity KK modes arising
from the large base $\mathbb P^1_b$, plus the towers of D0 and D2-branes, respectively. 
According to the logic applied many times in this work, the appearance of a dense tower of stringy excitations at the same parametric mass scale
as of the particle excitations,  implies that the system does not undergo  decompactification, i.e. the limit effectively stays equi-dimensional. 

Note that obviously the statements we made here rest upon the assumption that the heterotic string and its particle excitations remain stable in the limit $g_{\rm IIA} \to \infty$.

\subsection{Abelian variety fibration: Schoen manifold in Type IIA String Theory}

We  briefly discuss the last of the three fibration types, namely Abelian fibrations, now for Type IIA strings where quantum phenomena 
can come into play. As for the classical case, these geometries form in some sense the middle ground between genus-one and K3 fibrations, by sharing properties of both of them.

Indeed, even without resorting to any specific realisation of an Abelian surface fibration, the general behaviour can be deduced using a combination of arguments that we invoked also for the quantum geometry of the $T^2$- and K3-fibrations.
Recall first the discussion at the beginning of Section \ref{sec_QK3IIA}. Analogously, the quantum volume of the $T^4$ fiber is proportional to the mass of a D4-brane wrapping the fiber $T^4$, which is the same as an M-Theory M5-brane on $T^4 \times S^1$.
The M5-brane on $T^4$ gives rise to a Type II string, whose vacuum energy vanishes because 
\be
E_0 = - \frac{\chi(T^4)}{24} = 0 \,.
\ee
From the reasoning after (\ref{vacuumshift}) one infers  that there cannot be any non-vanishing offset for the quantum volume, and hence, unlike for K3-fibers, the point of vanishing $T^4$ volume is part of the quantum moduli space. 
This is the regime
 \be \label{L2volscalingT4}
L_2: \qquad  {\cal V}_{T^4} \sim \frac{1}{\lambda} \,,\qquad {\cal V}_{\mathbb P^1_b} \sim  \lambda \,, \qquad  \gA = \gA^{(0)} \,, \qquad \lambda \to \infty \,.
\ee 
The mass scales of states becoming light at this point are then as follows:
\be
\begin{aligned}
{\rm KK \, \, scale}: \qquad  &\frac{M}{M_s} \sim   \frac{1}{\sqrt{\lambda}} \,,   \cr
{\rm D2 \subset T^4/ NS5 \, \, on \, \,  T^4}: \qquad  &\frac{M}{M_s} \sim   \frac{1}{\gA \sqrt{\lambda}}\,,    \cr
{\rm D4 \, \, on \, \,  T^4}: \qquad  &\frac{M}{M_s} \sim   \frac{1}{\gA {\lambda}}  \,.  \cr
\end{aligned}
\ee
Note that, contrary to a K3-fiber,  D4-branes on a $T^4$ fiber generate a tower of particles because a $T^4$ admits arbitrary smooth multiple covers, i.e. a D4-brane along $n T^4$ gives rise to BPS particles for every value of $n$. The masses of these particle states are parametrically smaller than those of the towers of string excitations from the NS5-brane on $T^4$, or of any other towers. They indicate decompactification along a circle to five-dimensional M-Theory.

This claim is readily understood by noting that, as for $T^2$-fibers, the limit of a vanishing $T^4$-fiber is T-dual to the large volume regime near 
\be \label{L1volscalingT4}
L_1: \qquad {\cal V}_{T^4} \sim \lambda \,,\qquad {\cal V}_{\mathbb P^1_b} \sim \lambda \,, \qquad  \gA = \gA^{(0)} \lambda \,, \qquad \lambda \to \infty \,.
\ee
 At the large volume point $L_1$, the classical volume of the $T^4$-fibration is reliably computed as
 \be
 {\cal V}_{Y, \rm cl.} \ \sim\  \, {\cal V}_{T^4} \, {\cal V}_{\mathbb P^1_b} \,,
 \ee
up to   subleading terms independent of ${\cal V}_{\mathbb P^1_b}$.
 The Planck mass scales as
\be\label{planckT2}
\frac{M_{\rm Pl}^2}{M^2_s} = \frac{4 \pi}{\gA^2} {\cal V}_Y \sim 1 
\ee
if we take $\gA^{(0)}$ finite. 
The tower of D4-branes wrapping $T^4$ at $L_2$ maps to a tower of D0-branes at $L_1$. As is familiar by now, this tower indicates decompactification to five-dimensional M-Theory before any of the other towers at subleading mass scales become relevant. Similarly to the discussion around (\ref{JMJV}), invoking
\be
J_M = \frac{J_Y}{\gA^{2/3}} =  \frac{J_Y}{(\gA^{(0)})^{2/3} \, \lambda^{2/3}} \,,
\ee 
the M-Theory volumes scale as
\be
{\cal V}_{T^4,\rm M} \sim \lambda^{-1/3} \,, \qquad \quad {\cal V}_{\mathbb P^1_b,\rm M} \sim \lambda^{1/3} \,, \qquad \quad {\cal V}_{Y,\rm M} \sim 1 \,.
\ee
Hence, we end up precisely with a limit of Type $T^4$ for M-Theory on $Y$, with no further decompactification beyond five dimensions, and the asymptotic physics is as discussed in Section \ref{sec_Dmanifolds}.
In particular, for limits of Type $T^4$, there is no need for a co-scaling of $\gA^{(0)}$ in order to keep the Planck scale finite. 
This is a significant difference as compared to limits of Type $T^2$.
Note that this conclusion is fully consistent with the analysis at the end of Section~\ref{sec_QuantumT2}.

As for a concrete example, let us re-visit for a case study the
 Schoen manifold, whose abelian fiber is given by a product of two
cubic elliptic curves.  
The mirror, $X$, of this Schoen manifold can be represented as the following complete intersection in $\IP^1\times \IP^2\times \IP^2$:
\bea\label{schoenCS}
W_1(x,p,z)&=&\left(\frac 13\sum {x_i}^3- \frac1{z_1} x_1x_2x_3\right)p_0 - \left(\frac{z_0}{z_1z_2}\right)^{1/2}\!\!\!\!x_1x_2x_3\, p_1 \ =\ 0\,,               \\
W_2(y,p,z)&=&\left( \frac13\sum {y_i}^3\,-\, \frac1{z_2} y_1y_2y_3\right)p_1\, -  \left(\frac{z_0}{z_1z_2}\right)^{1/2}\!\!\!\! y_1y_2y_3\, p_0\ =\ 0\,.\nn
\eea
It exhibits a 3-dimensional sub-space of the 19 dimensional complex structure moduli space, in a certain patch,
where the coordinates $z_0$ and $z_{1,2}$ correspond, via mirror symmetry, to the volumes of the base $\IP^1_b$ and to the two elliptic fibers, respectively.

The quantum mirror geometry of this three-fold has been discussed in~\cite{1997alg.geom..9027H}, to which we refer for details. 
Suffice it here to mention that in the large base limit, the prepotential is given by
\be\label{schoenprep}
F(t_0,t_1,t_2)\ =\ -9t_0 t_1 t_2 -\frac32 t_1^2 t_2 - \frac32 t_1 t_2^2 +\frac32 t_1 + \frac32  t_2+\CO(e^{-t})\,,
\ee
where $t_0$ is the K\"ahler parameter of the base, and $t_1$, $t_2$ those of the two fibral tori.  As in Section \ref{Schoen}, in order to capture a limit that is 
genuinely different from a limit of type $T^2$, we consider the 
situation where we scale the two curves in the same way, ie., set $t_1=t_2=t$ and analogously for the complex structure moduli.
In the large base limit, the periods of the fiber become the products of the periods of the cubic curve, i.e.,
\be\label{T4periods}
X^0_{\CE\times\CE}\cdot\{1,t,t^2\}. 
\ee
where $X^0_{\CE\times\CE}=(X^0)^2$ and moreover
 \bea\label{F21schown}
t(z)&=& \frac{X^1(z)}{X^0(z)}\,, \qquad{\rm where}\nn\\
X^0(z) &=& {}_2F_1(1/3,2/3,1;z^3)\,,\\
X^1(z) &=& \frac i{\sqrt3}{}_2F_1(1/3,2/3,1;1-z^3)\,.\nn
\eea

From (\ref{T4periods}) it is evident that in the large base limit, the quantum volumes of
$0$-, $2$- and $4$-cycles, as determined by the periods, are given by their classical versions without any instanton corrections, quite as expected. 
The point $L_2$ hence corresponds to the regime
\be
t_0 \sim \lambda \,, \qquad \quad t\sim \lambda^{-1/2} \,, \qquad \lambda \to \infty \,.
\ee
At $L_2$ the fundamental period  of the fiber diverges as $X^0_{\CE\times\CE}\sim \hat T^2$, 
where $\hat T=1/|t|$.
The transformation (\ref{gAtrafopath}), applied to the value of $g_{\rm IIA}$ at $L_1$ and $L_2$, is precisely as dictated by T-duality, see (\ref{L1volscalingT4}) versus (\ref{L2volscalingT4}).
Furthermore, while the fiber volume tends to zero at $L_2$,
 the Calabi-Yau volume (\ref{quantumvolume}) remains finite at $L_2$.
 Technically this is due to the fact that $\int_{X} \Omega \wedge \bar \Omega$ does not change between $L_1$ and $L_2$ except for the replacement $|t| \to |\hat T| = 1/|t|$, while $X^0_{\CE\times\CE}\sim \hat T^2$ at $L_2$ but 
 $X^0_{\CE\times\CE}\sim 1$ at $L_1$, i.e.
 \be
 {\cal V}_Y =    \frac18\frac{i\int_X \Omega\wedge\bar\Omega(z)}{|X^0(z)|^2}\ \
 \sim       \frac{|t_0 \hat T^2 + {\cal O}(\hat T^3)|}{|\hat T^4|} = \frac{|t_0|}{|\hat T^2|} \sim 1 \,.
 \ee

Compared to the situation for K3-fibrations, there is an important difference.
While the mass of a wrapped D4-brane is again just given by the classical expression
\be
M(D4)\ \sim\  |\partial_{t_0}F| \ = \ |t^2| \,,\ 
\ee
there is no constant shift as compared to eq.~(\ref{D42n}), which applies to a K3 fiber. 
This is due to the absence of a term $\frac1{24}t_0\chi(T^4)=t_0\frac1{24}\int c_2\wedge J_0$ in the
prepotential (\ref{schoenprep}) since  $\chi(T^4)=0$.  As mentioned at the beginning of this section, this reflects the absence of a vaccum energy, $E_0$,
for the Type II string.  As a consequence, and in contrast to a K3 fibration, the regime of vanishing fiber volume does lie in the moduli space, 
and the asymptotically massless, multiply wrapped D4-branes at $L_2$ play the T-dual role of the D0-branes at $L_1$.  This signals decompactification to M-Theory.

\section{Summary and Discussion} \label{sec_Conclusions}

The present work has studied the physics and geometry of  infinite distance limits in the vector multiplet
 moduli space of M-Theory and Type IIA string theory compactified on Calabi-Yau three-folds.
Our findings lend support to a conjecture with a more general scope formulated in the introduction, according to which
 any equi-dimensional infinite distance limit of a gravitational theory should give rise to an asymptotically weakly coupled and tensionless string theory. 
The appearance of a unique asymptotically tensionless fundamental string could be avoided if the theory undergoes (generally partial) decompactification, or alternatively, if quantum corrections obstruct an equi-dimensional limit at infinite distance in the first place (for example by turning a classically infinite distance into a finite distance limit).

The framework of our explicit analysis is sufficiently rich in order to address a variety of these and other related phenomena.
First, in the realm of M-Theory compactified on a Calabi-Yau three-fold, we have classified the possible infinite distance limits in the K\"ahler moduli space, which is classically exact. Due to the absence of quantum corrections, the only way to avoid the appearance of a light weakly coupled string and its excitation spectrum would be by partial decompactification, according to the conjecture. 
This perfectly matches the results of our analysis: As we have shown, infinite distance limits with finite classical volume in the K\"ahler moduli space admit a complete classification, where either a $T^2$ fiber, a $K3$ fiber or a $T^4$ fiber of the Calabi-Yau three-fold shrinks in a unique way,
 while the base expands such as to keep the volume finite. 
Limits of Type $T^2$ realize the phenomenon of partial decompactification in that a tower of M2-branes wrapped on the torus fiber signals decompactification to six dimensions, which leads to the standard F-Theory limit (see also 
\cite{Corvilain:2018lgw,Lee:2019xtm}).
For limits in which a K3 or $T^4$ fiber shrinks, by contrast, we  have observed the appearance of a unique asymptotically tensionless heterotic or Type II string, respectively.
The resulting string excitation spectrum sits parametrically at roughly the same mass scale as the KK spectrum.
Since the string spectrum is much denser than the KK spectrum, the
appearance of the light string modes prevents the physics from running into a purely field theoretic decompactification limit. 

 The emergence of a {\it unique} asymptotically light heterotic or Type II string rests on a beautiful interplay of various aspects of K\"ahler geometry and is a highly non-trivial consistency check of the compactifications under scrutiny.
 The existence of towers of 2-brane particle states corresponding to wrapping modes of this string in a dual compactification is guaranteed by the connection between BPS invariants and modular forms in the relevant geometries.
 Our analysis also points to what we believe is a new realisation of string dualities in five dimensions, namely a duality
  between M-Theory on an Abelian surface fibration, and a non-geometric D-manifold background for Type II string theory.
 These backgrounds deserve further investigation in the future.

Once we pass on to compactifications of Type IIA string theory on the same kinds of fibered three-folds, 
we need to watch out for potential obstructions against taking equi-dimensional limits. 
Indeed it is well known that quantum geometry can drastically modify naive classical geometry. 
In limits of Type $T^2$ and $T^4$, the shrinking of the fiber is not prohibited by quantum geometric effects; nonetheless the theory undergoes a partial decompactification because T-duality relates the regime of small toroidal fiber to the large fiber regime, however at large value of the ten-dimensional string coupling $g_{\rm IIA}$. As a result, the theory decompactifies to M-Theory.
In limits of Type $T^2$, whether or not the five-dimensional theory decompactifies further (possibly in a nested series of limits), depends on whfether or not $g_{\rm IIA}$ is co-scaled such as to keep the Planck scale finite, while for limits of Type $T^4$ no such co-scaling is necessary and there occurs no further decompactification  beyond five dimensions.

For limits where classically a K3-fiber would shrink, on the other hand, quantum corrections obstruct a vanishing fiber volume. This phenomenon can also be understood in terms of the vacuum energy of a wrapped M5-brane on $K3 \times S^1$, where $S^1$ is 
the familiar M-Theory circle. As a consequence of the non-vanishing quantum volume of the K3 fiber,
 the total Calabi-Yau volume necessarily diverges.  If $g_{\rm IIA}$ remains fixed and finite,  this leads to straightforward decompactification (or decoupling of gravity in four dimensions). 
 Indeed this limit corresponds to the regime where Seiberg-Witten gauge theory decoupled from gravity emerges 
 \cite{Kachru:1995fv}, and it occurs
 at finite distance from the perspective of the K3-surface.
 Hence in a sense quantum corrections have turned the infinite distance into a finite distance limit, where only a finite number of states become massless. This is similar in spirit to the observations of \cite{Marchesano:2019ifh} where a classical strong coupling/large distance limit is drastically modified by quantum corrections.\footnote{Such quantum effects become even more drastic in 4d $N=1$ supersymmetric compactifications, as exemplified for K3-fibrations in \cite{Cicoli:2018tcq}.}
 The most interesting limit is obtained when the string coupling is co-scaled such as to keep the four-dimensional Planck scale finite. The tower of D0-branes at strong coupling now sits parametrically at the same scale as the tower of heterotic string excitations (from the M5-brane along the K3-fiber). Hence, the limit remains equi-dimensional. This represents, in fact, the only possible realisation of an equi-dimensional infinite distance limit in the vector moduli space of Type IIA theory on Calabi-Yau three-folds.

Our findings and the logic of our ``Emergent String Conjecture'' also shed interesting light on other frameworks of compactifications.
Mirror symmetry relates Type IIA string theory on $Y$ to Type IIB string theory on its mirror three-fold $X$. The corresponding
infinite distance limits in the complex structure moduli space of $X$ correspond to certain degeneration limits of 3-cycles. Wrapped D3-branes on  such vanishing 3-cycles give rise to mirror towers of asymptotically massless particle states, and this has been studied in the context of infinite distance limits in \cite{Grimm:2018ohb,Grimm:2018cpv}. This analysis makes use of the theory of limiting mixed Hodge structures and results in a classification of possible boundaries of complex structure moduli space.

As stressed in section \ref{sec_gencons1}, infinite distance limits in  complex structure moduli space on $X$ map to limits with co-scaled $g_{\rm IIA}$ (and thus finite Planck scale) on the Type IIA side. In such limits, a tower of D0 branes appears which is parametrically leading and signals decompactification to five dimensions. The exception is for limits of Type K3, where the D0 tower sits at the same mass scale as the excitations of an emerging heterotic string. The question arises how this spectrum manifests itself in the Type IIB mirror picture. Evidently
a natural interpretation would be in terms of a tower of D3-branes wrapping some special Lagrangian 3-torus fiber on $X$ \cite{2000math8018G,2012arXiv1212.4220G,Morrison:2010vf}. Indeed it is well known from the SYZ picture of mirror symmetry \cite{Strominger:1996it,Gross:1999hc}, that at the large complex structure point, which corresponds to maximal unipotent monodromy, 
 a special Lagrangian 3-torus shrinks to zero volume.  Since we are observing a tower of D0-branes in every infinite distance limit on the Type IIA side,  the interesting question arises which precise cycles degenerate on  the mirror $X$ in more general limits than those of maximal unipotent monodromy. 
Furthermore, for limits of Type K3, the appearance of a heterotic string at the same scale as the D0-brane tower implies that 
an emerging heterotic string should play an analogous role on the Type IIB mirror side. 

Another important open question is to what extent our conjecture applies to infinite distance limits in the hypermultiplet moduli space, which
is much more challenging than the vector multiplet moduli space we have considered in this paper.
 While there exist well-known candidates for emergent strings, quantum corrections that are difficult to control complicate the analysis.
For the sake of argument, let us first ignore quantum effects for a moment.
The candidates for light strings are, in Type IIA language,
 D4-branes wrapping shrinking 3-torus  fibers (near complex structure degenerations), or  in Type IIB language,
D3/D5-branes wrapping 2- or 4-torus fibers (near K\"ahler degenerations).
For the latter, the strings correspond to light Type IIB strings in the vanishing fiber limit. This parallels the emergence of tensionless Type IIB strings in six dimensions, which arise  \cite{Lee:2019xtm} for Type IIB string theory on K3 at infinite distance of the moduli space.

The emergence of light strings is also implied by T-duality:
After an even number of T-duality transformations, we can translate the small fiber regime to the large fiber regime, however at large value of the 10d coupling, $\gB$. In this description, the D1-string becomes asymptotically light.
For Type IIA theory, it would take an odd number of T-dualities to arrive at large fiber volume, which maps the theory to Type IIB on the mirror
dual three-fold. Still, mirror symmetry suggests that also the D4-brane on a $T^3$-fiber would become a fundamental string in the limit of vanishing fiber volume.
However, as mentioned above, what impedes a straightforward analysis of the infinite distance limits in hypermutliplet space, is the appearance of potentially drastic quantum corrections. This phenomenon has already been addressed in a combined large volume - strong coupling limit in Type IIB theory in \cite{Marchesano:2019ifh}. At any rate, it seems plausible that this corner of string compactifications is in agreement with our conjecture, either directly or due to quantum obstructions against taking the limit.

As a particularly interesting testing ground of our conjecture, we finally turn to infinite distance limits in the complex structure moduli space of a Calabi-Yau three-fold, as probed by M-Theory.
Unlike for Type IIA theory, there is no natural candidate for a tensionless string from wrapping some 4-brane around a degenerating 3-cycle. The only source of light particles is the supergravity KK tower associated with large 3-cycles in the limit (at finite total volume). However
at the same time, M2-brane instantons become unsuppressed (and membranes from wrapped M5-branes become parametrically light), and these could potentially obstruct an infinite distance limit.
As far as the membranes are concerned, their energy scale sits indeed at the same scale as the KK tower arising from large 3-cycles. 

Hence, from a parametric scaling perspective, one might view this as a potential counter-example to the conjecture that every denser but parametrically competing tower of particles, if any, should come from the excitations of a string.
Note that for the Type IIA theory in the same limit, no such potential counter-example arises, because the membrane (domain wall) from an NS5-brane along a three-cycle is always accompanied by a string from a D4-brane along the same cycle \cite{Font:2019cxq}.
However, even in M-Theory it is generally not expected that tensionless higher-dimensional objects give rise to a tower of particle-like excitations, in the same way as a critical string does. 
If this line of reasoning is correct, the lack of an emergent tensionless string would conform with our expectations, as the theory either undergoes partial decompactification or the limit is prohibited by quantum effects in the first place. It would be interesting to support this picture by a quantitative analysis.

\subsubsection*{Acknowledgements}

We thank Markus Diriegl, Antonella Grassi, Thomas Grimm, Andy Katz, Dieter L\"ust, Fernando Marchesano, Luca Martucci, Dave Morrison, Eran Palti, Emanuel Scheidegger, Cumrun Vafa, Irene Valenzuela, and Max Wiesner for helpful discussions.

\newpage
\begin{appendix}

\section{Classification of Large Distance K\"ahler Limits}\label{pf_thm1}

In this Appendix we will deduce Theorem~\ref{classify} as stated in Section \ref{classification-result}, starting from the classification of finite volume limits in terms of what we call $J$-class A and B limits, see~\eqref{classA-sec2} and~\eqref{classB-sec2}, respectively.
See also Figure~\ref{f:FigMlimits} for an overview.
More precisely we will show the following:

\begin{enumerate}

\item
Every $J$-class A limit as specified in ~\eqref{classA-sec2}  gives rise to an infinite distance limit of Type $T^2$ as defined in Theorem~\ref{classify}.
If the genus-one fibered three-fold $Y$ admits a K3-fibration, the respective fibers scale as
\be \label{stement1B1}
{\cal V}_{T^2} \sim \lambda^{-2} \,, \qquad {\cal V}_{K3/T^4} \sim \lambda^{-1}   \quad {\rm or} \quad  \lambda^2 \,.
\ee 
This will be shown in Appendix \ref{AppA_subTypeT2}.

\item
Every $J$-class B limit as specified in ~\eqref{classB-sec2}  implies that the three-fold $Y$ admits a K3 or $T^4$-fibration over base $\mathbb P^1_b$.
Focusing  for definiteness on a K3-fibration, we will prove in Appendix \ref{SubsecTypeB} the following structure:
Either the limit is of Type K3 as specified in Theorem~\ref{classify}, or the K3-fibration admits a compatible $T^2$-fibration with base $B_2$ and the limit is of Type $T^2$ with scaling behaviour
\be \label{VT^2beginningApp}
{\cal V}_{T^2} \sim \mu^{-2} \,, \qquad {\cal V}_{K3} \sim \mu^{-\frac{4}{1 + 2x}} \,, \qquad  {\cal V}_{\mathbb P^1_b} \sim \mu^{\frac{4}{1 + 2x}} \,.
\ee
Here the scaling parameter $\mu$ is related to the scaling parameter $\lambda$ appearing in ~\eqref{classB-sec2} as
\be \label{lambdavsmu}
\lambda = \mu^{\frac{4}{1 + 2x}}\,,
\ee
and the range of possible values of $x$ depends on the topology of the base $B_2$ of the genus-one fibration:
\be
\begin{split} \label{boundAppB}
0 < x      & \qquad \quad  \text{if} \, \,  B_2 = \mathbb P^1_f \times \mathbb P^1_b  \cr
0 < x \leq \frac{3}{2}         &       \qquad \quad           \text{ otherwise\,.     }   
\end{split}
\ee
\end{enumerate}

\subsection{Type $T^2$-limits from $J$-class A} \label{AppA_subTypeT2}

In the infinite distance limits where the K\"ahler class is of the form~\eqref{classA-sec2}, the nef effective divisor $J_0$ of $Y$ obeys 
\beq \label{ellfibcond}
J_0^2 \neq 0 \,, \quad\quad J_0^3 =0 \,.
\eeq
As shown in~\cite{oguiso} and~\cite{PMHWilson}, a Calabi-Yau three-fold $Y$ which has a nef divisor satisfying (\ref{ellfibcond})  must be genus-one fibered:\footnote{There is a technical assumption that either $J_0$ is effective or $J_0 \cdot c_2(Y)>0$.} 
\be
\ba \label{EllfibrationY}
\text{$J$-class A:} \qquad  \pi :\quad T^2 \ \rightarrow & \  \ Y \cr 
& \ \ \downarrow \cr 
& \ \   B_2   \,.
\ea
\ee
The homology class of (in general a multiple of) the generic genus-one fiber  is given by 
\be
 n \, T^2 = J_0^2    \qquad (n \geq 1) \,.
\ee
In other words, $J_0$ is a vertical divisor on $Y$, {\it i.e.}, the pullback of a divisor on $B_2$ to $Y$.

Note furthermore that the fibration has either a $k$-section or a rational section $\sigma$, which defines the embedding of the base $B_2$ (or a multi-cover of $B_2$) into $Y$ with the property 
that 
\be
\sigma(B_2) \cdot T^2 = k\,,\qquad k \geq 1 \,.
\ee
In the special case of an elliptic fibration there exists a rational section, for which $k=1$.
The fiber and the base volumes scale with $\la\to \infty$ as follows:
\bea\label{VT2}
{\cal V}_{T^2} &=& \frac{1}{n} \int_Y J_0^2 \wedge J = \sum_{\alpha\in\cI_1} \frac{\hat a_\alpha n_\alpha}{ n}    \, \frac{1}{\la^2} ~\to~ 0  
\,, \label{VT2A} \\
{\cal V}_{B_2} &=& \frac{1}{2k} \int_Y \sigma(B_2)  \wedge J^2 = \frac{1}{2n} \la^2 + \cdots ~\to~ \infty\,,  \label{VB2A}        
\eea
where we have used Eqs~\eqref{classA-sec2} and~\eqref{classA-details}, and $n_\alpha :=\int_Y J_0^2 \wedge J_\alpha$ are positive integers. Here, the ellipsis contains terms that scale as $\la^2$ or slower. Note that terms of the former type may in principle be included here in case there exist some $J_r$ with $\gamma_r =1$ as defined in~\eqref{classA-details}; this is because the four-forms $J_0 \cdot J_r$ and $J_r \cdot J_s$ are proportional to $J_0^2$ (see Proposition 2 in Appendix D.1 of \cite{Lee:2019tst}). 
Note also that there exist divisors whose volumes scale to zero as well. For instance, as far as the divisors associated with the K\"ahler cone generators are concerned, we have
\be \label{volJrapp}
{\cal V}_{J_r}  = \frac{1}{2}\int_Y J^2 \wedge J_r = (J_0 \cdot J_r \cdot \sum_{\alpha \in \cI_1} J_\alpha \, \hat a_\alpha)  \frac{1}{\la}  \sim \frac{1}{\lambda} \,.
\ee 
The last assertion follows because $J_0 \cdot J_r = n_r J_0 \cdot J_0$ for $n_r > 0$ (see again Appendix D.1 of \cite{Lee:2019tst}), and furthermore we must have at least one $\hat a_\alpha \sim 1$ in order for the total volume to satisfy ${\cal V}_Y \sim 1$. 

In particular, this shows that $J$-class A limits can never give rise to limits of Type K3 or $T^4$, even if the three-fold $Y$ admits in addition a compatible  K3 or $T^4$-fibration.
This follows from the scaling behaviour
\be \label{VK3T4appB1}
{\cal V}_{K3/T^4} \succsim \frac{1}{\lambda}
\ee
for the volume of the respective fiber, if present. 
By contrast, a limit of Type K3 or $T^4$ would require a scaling behaviour like ${\cal V}_{T^2} \sim \lambda^{-2}$ and ${\cal V}_{K3/T^4} \sim \lambda^{-4}$, according to (\ref{VCK3Type}) and (\ref{VK3JB}) (after suitably rescaling $\lambda$ such as to compare (\ref{VCK3Type}) to the behaviour ${\cal V}_{T^2} \sim \lambda^{-2}$ considered in this section).
To show (\ref{VK3T4appB1}), first note that the class of such a surface fiber has to be proportional to one of the K\"ahler cone generators different from $J_0$. 
If the class is a generator $J_r$ for $r \in \cI_3$, then (\ref{volJrapp}) shows the claimed behaviour. If its class is $J_\alpha$ with $\alpha \in \cI_1$, its volume scales as ${\cal V}_{J_\alpha} \sim \lambda^2 J_0^2 \cdot J_\alpha + \ldots \sim \lambda^2$. 
This completes the proof of (\ref{stement1B1}).

\subsection{ Type $K3/T^4$ or Type $T^2$ limits from $J$-class B}  \label{SubsecTypeB}

In the infinite distance limits with the K\"ahler class of the form~\eqref{classB-sec2}, the nef effective divisor $J_0$ of $Y$ obeys 
\beq \label{J0condlassB}
J_0 \neq 0 \,, \quad\quad\quad J_0^2 =0 \,.
\eeq
Furthermore, one can easily show that
\beq \label{J0c2classB}
\int_Y J_0 \wedge c_2(Y) \geq 0 \,.
\eeq
In order to see this, let us first recall the adjuction formula,
\beq\label{adj}
0 \to TS_0 \to TY|_{S_0} \to \cO_Y (S_0) |_{S_0} \to 0 \,,
\eeq
where $S_0$ is the surface of class $[S_0]=J_0$. 
Note that the normal bundle ${\cal O}_Y(S_0)|_{S_0}$  in~\eqref{adj} is a trivial bundle,  ${\cal O}_Y(S_0)|_{S_0}  \simeq {\cal O}_{S_0}$ since $J_0^2 =0$. Then, it immediately follows that  
\beq
c_1(S_0) = c_1(TY|_{S_0}) - c_1(\cO_Y (S_0) |_{S_0}) =  0 \,,
\eeq
and hence that $S_0$ is either a K3 surface or an Abelian surface. 
We thus have 
\beq
c_2(Y) \cdot J_0 = c_2(Y)|_{S_0} = c_2(S_0) =
\begin{cases}
      24 & \text{if}\ S_0~\text{is a K3 surface}\,, \\
      0 &\text{if}\ S_0~\text{is an Abelian surface}\,.
    \end{cases}
\eeq
In the second case, $S_0$ is topologically a four-torus $T^4$.
According to the classification~\cite{oguiso} of fibration structures of Calabi-Yau three-folds, (\ref{J0condlassB}) and (\ref{J0c2classB}) guarantee that the Calabi-Yau $Y$ is a fibration with the class of the generic fiber 
given by $J_0$. This leads to the two following possibilities for $J$-class B limits, depending on whether
we have K3 or $T^4$ as fiber:
\be \label{TypeBfibrationstructure}
\ba
\text{$J$-class B ($K3$-/$T^4$-fibered):} \qquad  \pi :\quad \text{K3/$T^4$} \ \rightarrow & \  \ Y \cr 
& \ \ \downarrow \cr 
& \ \  \mathbb P^1_b    
\ea
\ee

In the remainder of this subsection we focus on the case of a K3-fibration.
We will need to understand the scaling behaviors of the volumes of base $\mathbb P^1_b$,  $K3$ fiber and of  curves embedded in the fiber, in terms of the large parameter $\la$ defined in (\ref{classB-sec2}).
First, note that $\mathbb P^1_b$ is the curve dual to the fiber, whose class is $J_0$. Hence it is clear that
\be \label{P1bvolume}
{\cal V}_{\mathbb P^1_b} =  \int_{\mathbb P^1_b} J \to \la \qquad \text{as}\quad \la\to \infty\,.
\ee
The volume of the $K3$-fiber in the limit $\la \to\infty$ can be computed by noting that
\be\label{VY-K3}
{\cal V}_{Y} = \frac{1}{6} \int_Y J^3 = \frac{1}{2} \la J_0  J^2 + \ldots = \la \, {\cal V}_{K3} + \ldots\,,
\ee
with the ellipsis denoting only parametrically subleading terms in $\la$. This will be proven in Appendix~\ref{K3unique}. In the parametric limit this implies
\be \label{VolK3T4}
{\cal V}_{K3}   \to  \frac{{\cal V}_{Y} }{\la}  \,, 
\ee
with ${\cal V}_{Y}$ finite. 
To be precise, this is only true when $Y$ admits a single $K3$ fibration. As we will make clear in Appendix~\ref{K3unique}, there may exist other order-one contributions in case the three-fold $Y$ is $K3$ fibered in multiple different ways. Even for such (more generic) situations, however, we can prove that there must exist at least one $K3$ fibration with the parametric behavior $\cV_{K3} \sim \frac{1}{\la}$, which is all we need in the rest of this section. 

Of special importance will be curves $C$ inside the $K3$ fiber of self-intersection $C \cdot_{K3} C \geq 0$ which also exist as curves inside the full Calabi-Yau three-fold $Y$.
The reason is that, as we will show in Section \ref{sec_TypeBK3M}, it is precisely these curves which support an infinite tower of BPS states (from wrapped 2-branes) that become massless in the limit $\la \to \infty$.
Our claim is that there can occur only the following two possibilities:
\begin{enumerate}
\item {\bf Type $K3$} \\
The first possibility is that the volume of the curves $C \subset K3$ embedding into $Y$ scales as
\be \label{VCrate}
{\cal V}_{C} \sim \frac{1}{\sqrt{\la}}  \qquad {\forall} \, \, C \subset K3 \quad {\text{with}} \quad C \cdot_{K3} C \geq 0 \,.
\ee
$K3$-fibrations with this property admit a finite volume limit of {\it Type K3}, as in Theorem~\ref{classify}.
\item {\bf Type $T^2$ (with a compatible $K3$ fibration)} \\
In every other case, the $K3$ surface is necessarily genus-one fibered
\be \label{K3ellfibration}
\ba
r :\quad T^2 \ \rightarrow & \  \ K3 \cr 
& \ \ \downarrow \cr 
& \ \  {\mathbb P}^1_f    
\ea
\ee
and 
\bea \label{Asymmlimitcurves}
{\cal V}_{T^2} \sim  \la^{-1/2-x} \,,  \qquad \quad 
{\cal V}_{\mathbb P^1_f} \sim  \la^{-1/2+x} \,,  \qquad   x > 0\,.
\eea
The only curves with non-negative self-intersection in the $K3$-fiber are given by 
\be \label{Ckldef1}
C_{k,l} := k T^2 + l (T^2 + \mathbb P^1_f)  \,, 
\ee
for $k \, l  \geq 0$. 
Such limits fall into Type $T^2$ as defined in Theorem~\ref{classify}: Indeed, the scaling laws (\ref{Asymmlimitcurves}) and  (\ref{Ellipticextracond}), (\ref{P1bvolume}) are the same upon a redefinition of the parameter $\la$.
\end{enumerate}

To show this, suppose that there exists a curve $C \subset K3$ with $C \cdot_{K3} C \geq 0$ and whose volume vanishes at a rate 
\be \label{Vctrial}
{\cal V}_{C, \rm trial} \sim \la^{-1/2 - \alpha} \qquad \text{with} \quad   \alpha  \neq 0 \,. 
\ee
In view of (\ref{VolK3T4}), it is convenient to uniformly scale out a factor of $1/\sqrt{\la}$  from the volumes of all curves $C_i \subset K3$, and define
\be
{\cal V}_{K3} = \frac{1}{\la}  {\cal V}'_{K3} \,, \qquad {\cal V}_{C_i} = \frac{1}{\sqrt{\la}}  {\cal V}'_{C_i} \,.
\ee
Our assumption therefore is that on the $K3$ with finite volume  ${\cal V}'_{K3}$, we have 
\be \label{Vtrial}
{\cal V}'_{C,{\rm trial}} \sim \la^{-\alpha} \,.
\ee

Assume first that $\alpha < 0$. 
This means that a curve of non-negative self-intersection on a finite volume K\"ahler surface $K3$ scales to infinity.
As recalled in Appendix \ref{App_Surface-Limits},
in this situation there must exist a unique curve $C_0$ with 
\be
 C_0 \cdot_{K3} C_0 = 0
 \ee
 whose volume vanishes at the inverse rate ${\cal V}'_{C_0} \sim \la^{\alpha}$ (with $\alpha <0$ by assumption). This curve is the fiber of a genus-one fibration on $K3$.\footnote{See Section 3.2 of~\cite{Lee:2019xtm} for a relevant discussion in the context of weak-coupling limits for two-form fields, in Type IIB compactification on $K3$.}
 The base of the fibration is a rational curve  $\mathbb P^{1}_f$ with ${\cal V}'_{\mathbb P^{1}_f} \sim \la^{-\alpha}$. 
On the other hand, if $\alpha >0$, this means that on the K\"ahler surface a curve of non-negative self-intersection shrinks.
By Lemma \ref{lemmak3limit} in Appendix \ref{App_Surface-Limits},
 this curve is the fiber of a genus-one fibration on $K3$.

Scaling the factor of $1/\sqrt{\la}$ back, we have established, for $\alpha \neq 0$, the existence of a genus-one  fibration structure for the $K3$, as indicated in 
(\ref{K3ellfibration}) together with (\ref{Asymmlimitcurves}). 
On such a genus-one fibered surface, all other elements of the Picard group are negative intersection curves $C_{\rm ex}$ with $C_{\rm ex} \cdot T^2 = 0 = C_{\rm ex} \cdot \mathbb P^1_f$.
This explains why the only curves of non-negative self-intersection are given by the combinations (\ref{Ckldef1}) for $k l \geq 0$.

Note that  (\ref{K3ellfibration}) implies that $Y$ is itself genus-one fibered, 
 \be \label{K3ellfibrationY}
 \ba
 p :\quad T^2 \ \rightarrow & \  \ Y \cr 
 & \ \ \downarrow \cr 
 & \ \  B_2
 \ea
 \ee
 where the base is a Hirzebruch surface (or its blowup) of the form,
 \be \label{Hirzebrucha}
 \ba
 s :\quad \mathbb P^1_f \ \rightarrow & \  \ B_2 \cr 
 & \ \ \downarrow \cr 
 & \ \  \mathbb P^1_b
 \ea
 \ee
 
We conclude that the $J$-class B limit, for the case in which $J_0$ corresponds to a K3-fibration, leads to one of the two possibilities (\ref{VCrate}) or (\ref{Asymmlimitcurves}), as stated in Theorem~\ref{classify}.
An analogous analysis applies if $J_0$ defines a $T^4$-fibration.

Finally, let us comment on the range of the parameter $x$ appearing in (\ref{Asymmlimitcurves}).
The scaling   (\ref{Asymmlimitcurves}) of the $T^2$-fiber volume of $Y$ implies that
\be \label{VB2scaling}
{\cal V}_{B_2} \sim \lambda^{\frac{1}{2} + x}\,,
\ee
because in the small fiber limit the leading contribution to ${\cal V}_Y$ is of the form ${\cal V}_{T^2} {\cal V}_{B_2}$, and ${\cal V}_{Y}$ is finite.
On the other hand, the volume of a surface $B_2$ which is fibered as in (\ref{Hirzebrucha}) is of the form
\be
{\cal V}_{B_2}  = a \, {\cal V}_{\mathbb P^1_f} \,  {\cal V}_{\mathbb P^1_b} + b   \, {\cal V}^2_{\mathbb P^1_f} + \ldots \sim a \, \lambda^{\frac{1}{2} +x} + b \, \lambda^{-1 + 2x} + \ldots \,,
\ee
where the omitted non-negative terms involve the blowup divisors (if present).
The topological intersection numbers $a$ and $b$ depend on the details of the fibration. In particular, $b = 0$ only if the fibration is trivial, i.e. if $B_2 = \mathbb P^1_f \times \mathbb P^1_b$.
If $b \neq 0$, the second term must not dominate over the first term as otherwise the scaling (\ref{VB2scaling}) would be compromised. This leads to the constraint
\be
x \leq \frac{3}{2}   \qquad \text{if} \qquad  b \neq 0 \,.
\ee
Altogether we have hence reproduced the claim (\ref{VT^2beginningApp}) together with the bound (\ref{boundAppB}).

\subsection{Example: Elliptic K3-fibration} \label{App_EllK3fibraEx}

Let us now exemplify the limits of $J$-classes A and B, as well as the fibration structures they imply more concretely. Consider the Calabi-Yau three-fold $Y=\mathbb P^5_{1,1,2,8,12}[24]$ as studied for instance in \cite{Scheidegger:2001cqa}. This Calabi-Yau three-fold admits both a K3 and a compatible genus-one fibration, and hence lends itself to illustrating the different finite volume limits that can be taken.

The K\"ahler cone is generated by three divisors, $J_A$, $J_B$, $J_C$ , with non-vanishing intersection numbers given by
\be
J_B^3 = 8 \,, \qquad J_B^2 J_A = 2 \,, \qquad J_B^2 J_C = 4 \,, \qquad J_B J_C^2 =2 \,,\qquad J_A J_B J_C =1 \,.
\ee
First of all, the Calabi-Yau admits a K3-fibration with
\be
\text{K3-fiber class} \qquad J_A 
\ee
over a base $\mathbb P^1_b$. The latter is the curve dual to the divisor $J_A$.
In addition, $Y$ admits a compatible genus-one fibration with
\be
\text{$T^2$-fiber class} \qquad \frac{1}{2} J_C^2 \,\,.
\ee
Compatibility means that the genus-one fiber $T^2$ of $Y$ coincides with the genus-one fiber of the K3 surface. Together with the base ${}{\mathbb P}^1_f$ of the K3-surface, viewed as a genus-one fibration, there are two independent 
curve classes in K3 which embed as linearly independent classes into $Y$:
\be
T^2 = J_A \cap J_C \,, \qquad \quad {}{\mathbb P}^1_f = J_A \cap (J_B - 2 J_C) \,.
\ee
Their intersection numbers inside the K3 fiber are
\be
T^2 \cdot_{\rm K3} T^2 = 0 \,, \qquad {}{\mathbb P}^1_f \cdot_{\rm K3} {}{\mathbb P}^1_f = -2\,, \qquad T^2 \cdot_{\rm K3} {}{\mathbb P}^1_f = 1 \,.
\ee
If we parametrise the K\"ahler form generically as
\be
J = T^A J_A + T^B J_B + T^C J_C \,, \qquad T^i \geq 0 \,,
\ee
we find the following volumes:
\bea
{\cal V}_Y &=& T^A (T^B T^C + (T^B)^2) + T^B (T^C)^2 + 2 (T^B)^2 T^C + \frac{4}{3} (T^B)^3  \\
{\cal V}_{\rm K3} &=&  (T^B)^2 + T^B T^C \\
{\cal V}_{\mathbb P^1_{b}} &=& T^A  \\
{\cal V}_{C_{k,l}} &=& l (T^B + T^C) + k T^B \,,
\eea
where in the last line we have defined the curve class
\be
C_{k,l} = k T^2 + l ({}{\mathbb P}^1_f  + T^2) \,,\qquad \text{with} \, \, C_{k,l} \cdot_{\rm K3} C_{k,l} = 2 k \, l  \,.
\ee

We are now in a position to investigate the possible finite volume limits at infinite distance in moduli space.
Since the only K\"ahler cone generators with $J_i^3=0$ are $J_A$ and $J_C$, the following infinite distance limits can be taken:
\bea
&\text{$J$-class A}:& \qquad \quad T^C \sim \lambda \,, \qquad T^A \precsim \lambda \,, \\
&\text{$J$-class B}:& \qquad \quad T^A \sim \lambda \,, \qquad T^C \prec  \lambda \label{JBex} \,. 
\eea

\subsubsection*{$J$-class A}

Let us first consider the $J$-class A in more detail and illustrate that it indeed corresponds to a limit of Type $T^2$.
In agreement with the general definitions, we set $J_0 = J_C$, and more precisely identify the index sets appearing in (\ref{classA-sec2}) as 
\be
\cI_0  = \{C \} \,, \qquad   \cI_1  = \{B \} \,,   \qquad   \cI_3  = \{A \} \,. 
\ee
Taking the K\"ahler parameter $T^C = \lambda$, finiteness of ${\cal V}_Y$ requires that $T^B \sim \lambda^{-2}$. By definition of {$J$-class A}, $T^A \precsim \lambda$.
This reproduces precisely the general scaling behaviour (\ref{classA-details}). 
The resulting curve volumes 
\be
{\cal V}_{T^2} \sim \lambda^{-2} \,, \qquad {\cal V}_{\mathbb P^1_f} \sim \lambda \,, \qquad {\cal V}_{\mathbb P^1_b} \sim \lambda
\ee
and the fiber volume
\be
{\cal V}_{\rm K3} \sim {\lambda}^{-1} 
\ee
fall into the pattern (\ref{Ellipticextracond}) defining the Type $T^2$ limit.

\subsubsection*{$J$-class B}

We next turn to $J$-class B, for which
\be
\cI_0 = \{ A \} \,,\qquad \quad \cI_2 = \{ B, C \} \,.
\ee
Let us make a corresponding ansatz
\be
J = \la J_A + c_B J_B + c_C J_C \,.
\ee
If we parametrise 
\be
c_B = c_B' \, \la^{-a_B} \,,\quad \quad c_C = c_C' \, \la^{-a_C} \,, \qquad \quad c_r'  \quad \text{finite for} \,\, \la \to \infty \,,
\ee
then finiteness of ${\cal V}_Y$ requires that
\be
2 a_B \geq 1 \,, \qquad  a_B + a_C \geq 1 \, , \qquad a_B + 2 a_C \geq 0 \,,
\ee
where at least one of the  inequalities must be saturated so that ${\cal V}_Y$ does not vanish.

First, note that the last inequality cannot by saturated because this would be in conflict with $- a_C < 1$. However this is required so that $T^C \prec \lambda$ as in (\ref{JBex}).
The remaining two different ways to keep ${\cal V}_Y$ finite but non-zero, in compliance with (\ref{JBex}), are
\bea
  i) \qquad && a_B = \frac{1}{2}   \qquad a_C =   \frac{1}{2}  + x \, \qquad \quad \quad  x \geq 0   \qquad {\rm or} \qquad  \\
 ii)  \qquad && a_B = \frac{1}{2} + x \,  \qquad a_C = \frac{1}{2}  - x    \qquad 3/2 > x > 0 \,. 
\eea
The upper bound on $x$ in the last line ensures again that $-a_c < 1$.
In both cases 
\be
{\cal V}_{K3} \sim \lambda^{-1} \,, \qquad \quad {\cal V}_{\mathbb P^1_b} \sim \lambda  
\ee
while the fibral curves scale as follows:
\bea
i) \qquad                  &&  {\cal V}_{C_{k,l}} = \la^{-1/2} (k c_B' + l (c_B' + c_C' \la^{-x}))  \\
 ii)  \qquad               &&  {\cal V}_{C_{k,l}} = \la^{-1/2} (k c_B' \la^{-x}+ l (c_B' \la^{-x}+ c_C' \la^{x}))   \,.
\eea

Case i) realizes a Type K3 limit: Indeed,
every fibral curve $C_{ k l }$ with $k l \geq 0$ scales as  
\be
{\cal V}_{C_{k,l}} \sim \la^{-1/2}   
\ee
 for $\la \to \infty$ as in the definition (\ref{VCK3Type}).
The only fibral curve that may shrink at a faster rate is the (contractible) rational base curve $\mathbb P^1_f$ of self-intersection $-2$ in $K3$.

Case ii), on the other hand, exemplifies how a Type $T^2$ limit can arise from in $J$-class B:
In particular, the volume of the $T^2$ scales  as
\be
{\cal V}_{T^2} \sim \lambda^{-\frac{1}{2} - x} 
\ee
for non-zero $x$.
To compare this to  (\ref{Ellipticextracond}), we introduce the parameter $\mu$ via $\mu^{-2} = \lambda^{-1/2 -x}$ and find
\be
{\cal V}_{T^2} \sim \mu^{-2} \,, \qquad {\cal V}_{K3} \sim \mu^{-\frac{4}{1 + 2x}} \,, \qquad {\cal V}_{\mathbb P^1_b} \sim \mu^{\frac{4}{1 + 2x}} \,.
\ee
From the allowed range of $x$ we conclude that this means
\be
{\cal V}_{T^2} \sim \mu^{-2} \,, \qquad \quad \mu^{-4} \prec {\cal V}_{K3} \prec \mu^{-1} \,, \qquad  \quad {\cal V}_{\mathbb P^1_b}  = \frac{1}{{\cal V}_{K3}} \,,
\ee
in agreement with the characteristic property (\ref{Ellipticextracond})  of a limit of Type $T^2$.

\subsection{Large distance limits on K\"ahler surfaces }\label{App_Surface-Limits}

We recall and extend some results from \cite{Lee:2018urn} which are needed at various places in this article.

\begin{Lemma} \label{lemmaBlimit}
Consider a K\"ahler surface $B$ with positive anti-canonical class $\bar K >0$ and a limit in its K\"ahler moduli space 
such that ${\cal V}_{C_0} \to 0$ for some curve $C_0$ with $C_0 \cdot C_0 \geq 0$,
while simultaneously the total volume of $B$ stays finite. 
Then $B$ is (the blowup of) a fibration
\be \label{Hirzebruch}
 \ba
 s :\quad \mathbb {\cal F} \ \rightarrow & \  \ B \cr 
 & \ \ \downarrow \cr 
 & \ \  \mathbb P^1_b
 \ea
 \ee
where the generic fiber ${\cal F}$  is a rational curve  $\mathbb P^1_f$ or a genus-one curve $T^2_f$.
The curve $C_0$ that shrinks in the limit is the fiber of this fibration. In particular, $C_0 \cdot C_0 = 0$.
If $C_0 = T^2_f$, $B$ is a rational elliptic surface dP$_9$.
\end{Lemma}

To see this, note first that a curve with $C_0 \cdot C_0 \geq 0$ is not contractible on $B$. Hence any limit in K\"ahler moduli space where ${\cal V}_{C_0} \to 0$, with ${\cal V}_{B}$ remaining finite, is at infinite distance, and there must exist a curve $C$ for which ${\cal V}_{C} \sim \lambda \to \infty$ in the limit.
As shown in Appendix B of \cite{Lee:2018urn}, this means that the K\"ahler form of $B$ can be expanded as 
\be
J = \lambda J_0 + \sum_\nu s_\nu J_\nu
\ee
 with $J_0 \cdot J_0 = 0$ and $s_\nu (J_0 \cdot J_\nu) \precsim \frac{1}{\lambda}$ such that $J_0 \cdot C \neq 0$ and ${\cal V}_{C} \sim \lambda$.
In the limit, the volume of the curve $\hat C = J_0$ with $\hat C \cdot \hat C =0$ vanishes as ${\cal V}_{\hat C} = J_0 \cdot J \sim \frac{1}{\lambda}$.
Furthermore, any other curve $\hat C'$ which shrinks in the limit satisfies $C' \cdot C' \leq 0$ and $C' \cdot C' = 0$ if and only if $C' =\alpha \,  \hat C$ for some $\alpha$.
This applies in particular to the curve $C_0$, which is therefore to be identified with $\hat C$, i.e. $C_0$ is in the class of the K\"ahler cone generator $J_0$ and in particular $C_0 \cdot C_0 = 0$.

As further discussed in Appendix B.2 of \cite{Lee:2018urn}, there are now two possibilities: 
Either $C_0 \cdot \bar K =2$ or $C_0 \cdot \bar K =0$, where $\bar K$ is the anti-canonical divisor of $B$.

\subsubsection*{1) $C_0 \cdot \bar K =2$} 
If $C_0 \cdot \bar K =2$, $C_0$ is a rational curve and, in fact, the generic rational fiber of a Hirzebruch surface (or a blowup of a Hirzebruch surface). 
Given a Calabi-Yau three-fold $Y$ which is genus-one fibered over $B$, this fibration has the structure of a  K3-fibration over $\mathbb P^1_b$, and the generic K3 fiber is genus-one fibered over $\mathbb P^1_f$. 
In fact this case was already studied in \cite{Lee:2018urn}, as it is indeed the situation that arises for infinite distance limits in F-Theory where the gauge coupling of a $U(1)$ gauge group asymptotes to zero.
A D3-brane wrapping $\mathbb P^1_f$  gives rise to a weakly coupled tensionless heterotic string in the limit $\lambda \to \infty$.

\subsubsection*{2) $C_0 \cdot \bar K =0$} 
If $C_0 \cdot \bar K =0$, $C_0$ is a genus-one curve and $B$ is a genus-one fibration with generic fiber $C_0 =T^2_f$. The surface $B$ is a rational elliptic surface, i.e. a dP$_9$ surface. 
A Calabi-Yau three-fold $Y$ which is genus-one fibered over $B$ has the structure of a  $T^4$-fibration over $\mathbb P^1_b$, and the generic $T^4$ fiber is in fact a product $T^2 \times T^2_f$, where $T^2$ is the generic  genus-one fiber of $Y$. This is the Schoen Calabi-Yau manifold. 
The possibility that $C_0 \cdot \bar K =0$ was not studied further in \cite{Lee:2018urn}, because it does not correspond to a weak coupling limit of an abelian gauge symmetry, but rather to a weak coupling limit for a 2-form field in the F-Theory effective action.
As we discuss in more detail in section \ref{sec_Dmanifolds}, a D3-brane wrapping $T^2_f$ gives rise to a weakly coupled tensionless  Type II string in the limit $\lambda \to \infty$.
In this sense, the present situation is the $N=(1,0)$
supersymmetric version of the infinite distance limits of Type IIB string theory on K3, which
correspond to weak coupling for the Ramond-Ramond 2-form fields \cite{Lee:2019xtm}.

As an immediate application we can state
\begin{Lemma} \label{lemmaVDblimit}
On a Calabi-Yau three-fold $Y$ in a limit of Type $T^2$ as defined in section \ref{classification-result}, the volume of any vertical divisor
\be \label{DCbdef2}
D_{C_{\rm b}} := \pi^\ast C_{\rm b}   \qquad  \text{with} \quad C_{\rm b} \subset B_2 \quad \text{such that} \quad C_{\rm b} \cdot_{B_2} \cdot  C_{\rm b} \geq 0 
\ee
scales as
\be
\lambda^{-4} \prec {\cal V}_{D_{C_{\rm b}}} \,,
\ee
where $\pi: T^2 \to B_2$ denotes the genus-one fibration that underlies the Type $T^2$ limit.
\end{Lemma}

To see this, note that a vertical divisor is by itself genus-one fibered, and hence its volume is {computed as}
\be \label{VDCbfromfibration}
{\cal V}_{D_{C_{\rm b}}} = a {\cal V}_{T^2}^2 + b {\cal V}_{T^2} {\cal V}_{C_{\rm b}} =  \frac{a'}{\la^4} + b {\cal V}_{T^2} {\cal V}_{C_{\rm b}} \sim   \la^{-4 +\Delta}  \qquad \text{for} \quad \Delta \geq 0 \,.
\ee
Here $a$ and $b$  are positive and depend on the intersection numbers on the divisor, and $a'$ is obtained via~\eqref{VT2} as
\beq
a'=a \sum_{\alpha \in\cI_1} \frac{\hat a_\alpha n_{\alpha}}{n}\,,
\eeq 
which is again a finite positive number. 
The only question we need to address is whether the value $\Delta =0$ can be achieved.
This is the case if and only if there exists a curve on $B_2$ as in (\ref{DCbdef2}) with the property that
\be
{\cal V}_{D_{C_{\rm b}}} \sim \la^{-2-y}  \qquad   \text{for} \qquad y \geq 0 \,.
\ee
It is convenient to scale out a factor of $\la$ from the K\"ahler form of the base $B_2$, {\it i.e.}, to define
\be
{\cal V}_{B_2}  = \la^2 \, {\cal V}'_{B_2} \,, \qquad \quad  {\cal V}_{C_{\rm b}} =  \la \,  {\cal V}'_{C_{\rm b}} \,.
\ee
The rescaled volume  ${\cal V}'_{B_2}$ is parametrically finite as $\la \to \infty$, and the question is if we can achieve that
\be
 {\cal V}'_{C_{\rm b}} = \la^{-3 -y} \,.
\ee
By Lemma \ref{lemmaBlimit}, since $C_{\rm b} \cdot_{B_2} C_{\rm b} \geq 0$, such a curve can shrink on the finite volume base $B_2$
if and only if the three-fold $Y$ admits a K3 or $T^4$-fibration whose generic fiber   is in fact given by $\pi^\ast C_{\rm b}$.
By definition, in a limit of Type $T^2$, we have the scaling
$\la^{-4} \prec {\cal V}_{K3/T^4}$, while the scaling $\la^{-4} \sim {\cal V}_{K3/T^4}$ would correspond to a limit of Type K3/$T^4$. This concludes the proof. \\

Lemma \ref{lemmaBlimit} generalizes to K3-surfaces embedded into a Calabi-Yau three-fold:

\begin{Lemma} \label{lemmak3limit}
Consider a  K3 surface with embedding $\iota: K3 \to Y$ into a Calabi-Yau three-fold,  and a limit in the K\"ahler moduli space of $Y$
such that ${\cal V}_{C_0} \to 0$ for a curve $C_0 \in (\iota^*{\rm Pic}(Y))^\vee$ inside the K3 surface  with $C_0 \cdot_{K3} C_0 \geq 0$, while the
 total volume of the K3 remains finite. 
Then the K3 surface is a genus-one fibration
\be \label{Hirzebruch}
 \ba
 p :\quad \mathbb T^2 \rightarrow & \  \ K3 \cr 
 & \ \ \downarrow \cr 
 & \ \  \mathbb P^1_b
 \ea
 \ee
and the curve $C_0$ which shrinks in the limit is the fiber of this fibration. In particular, $C_0 \cdot C_0 = 0$.
\end{Lemma}

This follows by applying the same steps as in Appendix B of \cite{Lee:2018urn} to the K3 surface. Uniqueness of the shrinking curve with non-negative self-intersection follows because  $(\iota^*{\rm Pic}(Y))^\vee$ has signature $(1,r)$, so that the same Cauchy-Schwarz inequality can be readily applied as in Appendix B.3 of  \cite{Lee:2018urn}. See also \cite{Lee:2019xtm}.

\section{Existence and Uniqueness of the Relevant Fibrations}\label{unique-pf}

In this Appendix, we will  rigorously prove the existence and uniqueness of the relevant types of fibrations that
necessarily appear in the infinite distance limits as claimed in Theorem~\ref{classify}. Such existence and uniqueness results are crucial for various duality arguments given in the main text, as they guarantee the emergent critical strings to be of a definite type, be it Type II or Heterotic. 
For the precise statements see Propositions
\ref{Prop-uniqueT2}, \ref{prop-uniqueK3} and \ref{propT2fromB} in the subsequent sections.

Before delving into the detailed discussion of the fibrations, let us recall the following property of two non-trivial classes $D_1, D_2 \in H^{1,1}(S)$ on a K\"ahler surface $S$: 
\beq\label{prop-S}
D_1 \cdot D_2 = 0 \quad \text{and}\quad D_a\cdot D_a \geq 0\quad\text{for}\quad a=1,2 \, \quad \Rightarrow  \quad
D_2 = \kappa D_1 \quad \text{for some}\quad \kappa >0 \,.
\eeq
Then, it is an immediate consequence of~\eqref{prop-S} that any pair of non-proportional divisors $D_1$ and $D_2$ sitting in the K\"ahler cone closure of $S$ do intersect non-trivially in $S$. See Appendix D of Ref.~\cite{Lee:2019tst} for the proof of these two properties of a K\"ahler surface.\footnote{Note that the first property~\eqref{prop-S} is an obvious variation of Lemma 1 in Ref.~\cite{Lee:2019tst} and the second property is precisely Lemma 2 therein.} 
Building upon them, we now claim and prove two Lemmas in turn, which will be used repeatedly throughout this Appendix:

\begin{Lemma}\label{lemma1} On a K\"ahler three-fold $Y$, any pair of non-proportional nef divisors $(D_1, D_2)$ intersect non-trivially, {\it i.e.}, 
\beq
D_1 \cdot D_2 \neq 0 \,.
\eeq
\end{Lemma}  

\paragraph{Proof:} 
Consider the very ample divisor $\cS$ with class
\beq
[\cS] = m \sum_{k\in \cI} J_k\,, \qquad \text{for a large enough}~~m \in \IZ \,,
\eeq
where the sum is taken over the entire set of the K\"ahler cone generators. Then, the surface $\cS$ is irreducible and connected due to Bertini's theorem (see e.g.~\cite{Lazarsfeld}). Now, because the volumes 
\beq
D_1 \cdot [\cS]^2 \,,\qquad D_2 \cdot [\cS]^2 
\eeq
of the divisors $J_i$ and $J_j$, measured with respect to the positive class $[\cS]$, have to be positive, the restrictions
\beq
\cC_1 := D_1|_\cS \,,\qquad \cC_2:=D_2|_\cS 
\eeq
are non-trivial curves in $\cS$. Note that the two classes $[\cC_1]$ and $[\cC_2]$ sit in the K\"ahler cone closure of $\cS$, as $D_1$ and $D_2$ sit in that of $Y$. Furthermore, it is guaranteed by Lefschetz's hyperplane theorem that $[\cC_1]$ and $[\cC_2]$ are not proportional. We thus have
\beq
D_1 \cdot D_2 \cdot [\cS] = \cC_1 \cdot_\cS \cC_2 \neq 0 \,,
\eeq
where the non-vanishing is a consequence of the property~\ref{prop-S} (as discussed a line below~\ref{prop-S}). Therefore we must necessarily have 
\beq
D_1 \cdot D_2 \neq 0\,,
\eeq
as desired.  \hfill$\blacksquare$

\begin{Lemma}\label{lemma2} On a K\"ahler three-fold $Y$, suppose that a given triple of nef divisors $(D_1, D_2, D_3)$ with the properties
\beq\label{ij-jk-ki}
D_1 \cdot D_2 \neq 0 \,,\qquad D_2 \cdot D_3 \neq 0 \,,\qquad D_3 \cdot D_1 \neq 0 
\eeq
have a vanishing triple intersection
\bea \label{JijkApp}
D_1 \cdot D_2 \cdot D_3 = 0 \,.
\eea
Then, all the three non-trivial four-forms~\eqref{ij-jk-ki} are proportional. That is, there exists a positive number $\kappa$ such that
\beq\label{ij-jk}
D_1 \cdot D_2 = \kappa D_2 \cdot D_3 \,, \qquad\text{with}\quad \kappa >0 \,,
\eeq
and analogous relations hold for any permutations of the indices.  
\end{Lemma}

\paragraph{Proof:}
Consider the divisor $S_1$ with class
\beq
[S_1] = D_1 \,,
\eeq
and the (non-trivial) restrictions
\beq
C_{2,1}:=D_2|_{S_1} \,,\qquad C_{3,1}:=D_3|_{S_1}\,.
\eeq
Then, their intersection vanishes because
\beq
C_{2,1} \cdot_{S_1} C_{3,1} = D_2 \cdot D_3 \cdot D_1 =0 
\eeq
by assumption (\ref{JijkApp}).
Since the self-intersections of $C_{2,1}$ and $C_{3,1}$ are non-negative as
\bea
C_{2,1} \cdot_{S_1} C_{2,1} = D_2^2 \cdot D_1 \geq 0 \,,\\
C_{3,1} \cdot_{S_1} C_{3,1} = D_3^2 \cdot D_1 \geq 0 \,,
\eea
we have
\beq
C_{2,1} = \kappa C_{3,1} \,,\qquad\text{for some}\quad \kappa >0 \,,
\eeq
by the property~\eqref{prop-S} of K\"ahler surfaces. Therefore, the two non-trivial classes $D_2 \cdot D_1$ and $D_3 \cdot D_1$ are proportional to each other. Similarly, considering the divisor $S_2$ with class $[S_2] = D_2$ and applying the same logic, we learn also that the two non-trivial classes $D_1 \cdot D_2$ and $D_3 \cdot D_2$ are proportional. This proves that  relations of the form~\eqref{ij-jk} must hold.   \hfill$\blacksquare$ \\

We are now ready to start analyzing various fibrations. In the following we will discuss them in each of the three types of the infinite distance limits, which necessarily arise from K\"ahler forms of $J$-class A or B. 

\subsection{Type $T^2$ Limits with a K\"ahler form of $J$-class A} \label{AppT2ClassAunique}

A given Calabi-Yau three-fold may have multiple distinct genus-one fibrations (see e.g.~\cite{Anderson:2016cdu, Anderson:2017aux}), each of which can be associated, in the shrinking fiber limit, to the axio-dilaton profile of an F-Theory background. Here, we prove
\begin{proposition} \label{Prop-uniqueT2}
Every infinite distance limit with a K\"ahler form of $J$-class A
gives rise to a 
 limit of Type $T^2$. It is associated with a unique genus-one fibration, namely the unique $T^2$-fibration one whose fiber shrinks in the limit.
\end{proposition}
Recall that a K\"ahler form of $J$-class A takes the form~\eqref{classA-sec2}, 
\beq\label{classA-again}
J = \la J_0 +\sum_{\alpha \in \cI_1} {a_\alpha} J_\alpha +  \sum_{r\in \cI_3} c_rJ_r \,,\qquad \text{with}\quad J_0^2 \neq 0 \,,
\eeq
and necessarily leads to a Type $T^2$ limit, as discussed in Appendix~\ref{AppA_subTypeT2}.\footnote{Type $T^2$ limits which arise from K\"ahler forms of $J$-class B are studied in Appendix \ref{T2Unique-JClassB}.} 

Now, according to Oguiso's criterion, the (co)homology class of any elliptic fiber in a Calabi-Yau three-fold has to be proportional to the square of a nef divisor $D$ with a vanishing cubic self-intersection. We will thus analyze all possible nef divisors, i.e. all possible non-negative linear combinations of the K\"ahler cone generators $J_0$, $J_\alpha$ and $J_r$, with $\alpha \in \cI_1$ and $r \in \cI_3$. 

Consider first a nef divisor of the form
\beq\label{restricted}
D= p_0 J_0 + \sum_{r \in \cI_3} p_r J_r \,, \qquad p_0, p_r \geq 0 \,,
\eeq
where $J_\alpha$ are not involved. As proven in Appendix D.1 of~\cite{Lee:2019tst} (see Proposition 2 therein), the classes of $J_0 \cdot J_r$ and $J_r \cdot J_s$ are proportional to $J_0 \cdot J_0$ as
\bea \label{nr>0}
J_0 \cdot J_r &=& n_r J_0 \cdot J_0 \,,\qquad n_r >0 \,,\\
J_r \cdot J_s &=& n_{rs} J_0 \cdot J_0 \,, \qquad n_{rs} \geq 0 \,,
\eea
and furthermore,
\beq
J_r \cdot J_s \cdot J_t = 0 \,,\quad \forall r,s,t \in \cI_3 \,.
\eeq
This leads to the vanishing of the cubic self-intersection of $D$, and also implies that the class of $D^2$ is proportional to $J_0 \cdot J_0$. Therefore, any genus-one fibration associated to $D$ has the same fiber class as $J_0 \cdot J_0$ and hence there exist no new fibrations from the divisors of the form~\eqref{restricted}.\footnote{The fact that $D^2$ is proportional to $J_0^2$ does not necessarily mean that the two associated genus-one fibrations are the same with isomorphic bases. Even when the bases are not isomorphic, however, they are still birational; see Section~1.1 of Ref.~\cite{Anderson:2017aux} for relevant discussions. Since our main concern is in fact the uniqueness of a {\it generic} fiber class, throughout this paper, we will nevertheless view such fibrations over birational bases the same.}

Let us now consider a nef divisor of the most general complementary form
\beq\label{genD}
D=p_0 J_0 + \sum_{\alpha \in \cI_1} p_\alpha J_\alpha + \sum_{r \in \cI_3} p_r J_r \,,\quad p_0, p_\alpha, p_r  \geq 0 \,, 
\eeq
with at least one $p_\alpha$ strictly positive, that is,
\beq
p_{\alpha_0} > 0 \,,
\eeq
for some $\alpha_0 \in \cI_1$. 
We must then have $p_0 = 0$ in order for $D$ to correspond to a genus-one fibration, since, otherwise, $D^3 >0$; indeed, if $p_0 >0$, then the expansion of $D^3$ has a strictly positive contribution
\beq
p_0^2\, p_{\alpha_0} (J_0 \cdot J_0 \cdot J_{\alpha_0}) \,,
\eeq 
where positivity of the intersection $J_0 \cdot J_0 \cdot J_{\alpha_0}$ follows from~\eqref{alpha}, while every other expansion term in $D^3$ contributes non-negatively. 
Now, suppose further that $p_{r_0} >0$ for some $r_0 \in \cI_3$. Then, the volume of the $T^2$ fiber associated to $D$ has the same scaling behavior as that of $D^2$, where the latter is computed as 
\bea\label{JD2}
J \cdot D^2 &=& (\la J_0 +  \sum_{\alpha \in \cI_1}  a_\alpha J_\alpha + \sum_{r \in \cI_3} c_r J_r ) \cdot (\sum_{\alpha \in \cI_1} p_\alpha J_\alpha + \sum_{r \in \cI_3} p_r J_r)^2 \\  \label{expterm}
&\geq & \la p_{r_0} p_{\alpha_0} (J_0 \cdot J_{r_0} \cdot J_{\alpha_0})  \succsim \la \,.
\eea
Here, we have made use of the fact that the expansion of~\eqref{JD2} only contains non-negative terms, one of which is~\eqref{expterm}, and also that the triple intersection therein is positive because 
\beq
    J_0 \cdot J_{r_0} \cdot J_{\alpha_0}= n_{r_0} J_0 \cdot J_0 \cdot J_{\alpha_0} >0 \,,
\eeq
by the proportionality~\eqref{nr>0}. Therefore, the fiber volume diverges in the limit and such a genus-one fibration is not of our interest. 

We may thus restrict to the nef divisor of the form,
\beq\label{final-form}
D= \sum_{\alpha \in \cI_1} p_\alpha J_\alpha \,,\qquad p_\alpha \geq 0 \,,
\eeq
with all the $p_r$ as well as $p_0$ in~\eqref{genD} turned off. 
In order for $D$ to correspond to a non-trivial genus-one fibration, we must have $D^2 \neq 0$ and hence, there must exist a pair $(\alpha_0, \beta_0)$ for which 
\beq\label{non-trivial}
J_{\alpha_0} \cdot J_{\beta_0}  \neq 0\,, \qquad \text{with}\quad p_{\alpha_0}, p_{\beta_0}>0 \,,
\eeq
where $\beta_0$ may in principle coincide with $\alpha_0$. 
Once again, the volume $J \cdot D^2$ is bounded from below by each of the terms in the expansion and hence, we have
\beq\label{JD2-again}
J \cdot D^2= (\la J_0 +  \sum_{\alpha \in \cI_1}  a_\alpha J_\alpha + \sum_{r \in \cI_3} c_r J_r ) \cdot (\sum_{\alpha \in \cI_1} p_\alpha J_\alpha )^2 \geq \la p_{\alpha_0} p_{\beta_0} (J_0 \cdot J_{\alpha_0} \cdot J_{\beta_0})  \,,
\eeq
which diverges unless 
\beq\label{mayassume}
J_0 \cdot J_{\alpha_0} \cdot J_{\beta_0} =0\,.
\eeq
However, since any pair of the K\"ahler cone generators $(J_0, J_{\alpha_0}, J_{\beta_0})$ intersect by Lemma~\ref{lemma1} and~\eqref{non-trivial}, the vanishing~\eqref{mayassume} of the triple intersection leads, via Lemma~\ref{lemma2}, to the proportionality
\beq
J_0 \cdot J_{\alpha_0} = \kappa J_{\alpha_0} \cdot J_{\beta_0} \qquad\text{for some}\quad \kappa>0 \,,
\eeq
and hence we have
\beq
J_0 \cdot J_0 \cdot J_{\beta_0} = \kappa J_0 \cdot J_{\alpha_0} \cdot J_{\beta_0} = 0\,,
\eeq
which contradicts~\eqref{alpha}.
Therefore, the nef divisor of the form~\eqref{final-form} cannot lead to a shrinking $T^2$ fiber either, which completes our uniqueness proof of the shrinking $T^2$ fiber.

\subsection{Type $K3$/$T^4$ Limits with a K\"ahler form of $J$-class B}\label{K3unique}

Just as for the genus-one fibrations, a given Calabi-Yau three-fold may in general admit multiple distinct $K3$ fibrations (see e.g.~\cite{Anderson:2016cdu, Anderson:2017aux}), and similarly, by the same algebro-geometric reasoning, multiple Abelian surface fibrations can occur. Here, we prove
\begin{proposition}\label{prop-uniqueK3}
An infinite distance limit of  Type $K3$ or $T^4$ can be constructed only from 
a K\"ahler form of  $J$-class B with $|\cI_\la|=1$, where $\cI_\la$ is defined in (\ref{TiIlambdadef}).
Every limit of Type $K3$ or $T^4$ is associated with a unique K3 or, respectively, $T^4$-fibration structure with the property that its generic surface fiber shrinks in the limit at a  rate parametrically faster than the shrinking rate of the fiber of any other $K3$- or $T^4$-fibration which may be admitted by the Calabi-Yau three-fold.
\end{proposition}

Note that by definition, for every infinite distance limit of Type $K3$ or $T^4$, the shrinking rate of the surface fiber dominates over the shrinking rate of any 
 $T^2$-fiber, if present. Hence it indeed must only be guaranteed that there is no competing $K3$ or $T^4$-fiber.

For simplicity of presentation, throughout this section, we will proceed with Type $K3$ limits and will also refer to any surface fibers arising from Oguiso's criterion as a $K3$ fiber, while some of such surface fibers could as well be a $T^4$ fiber. It is nevertheless evident that precisely the same logic applies to Type $T^4$ limits and/or $T^4$ fibers, since the fibers of the topological types $K3$ and $T^4$ are both characterized in the same intersection-structural manner by Oguiso's criterion, which is all that matters for our proof. 

Recall that the limits of Type $K3$ necessarily have a K\"ahler form of $J$-class B, expanded as in~\eqref{classB-sec2} as 
\beq\label{classB-again}
J = \la J_0 + \sum_{\mu \in \cI_2} b_\mu J_\mu \,.
\eeq
It turns out that the constraints on $b_\mu$ can be more concretely phrased in case the K\"ahler form~\eqref{classB-again} of $J$-class B has more than one large coefficients of order $\la$. For this purpose, we further decompose the index set $\cI = \cI_0 \cup \cI_2$ ($=\{0\} \cup \cI_2$) as the disjoint union of three subsets,
\beq\label{decomp}
\cI = \cI_\la \cup \cI_2' \cup \cI_2'' \,, 
\eeq
in such a way that 
\begin{itemize}
\item $A \in \cI_\la$ label all the generators with the coefficients of order $\la$;
\item $\mu' \in \cI_2'$, those with $J_{\mu'} \cdot J_{A} \cdot J_{B} = 0$ for all $A, B \in \cI_\la$;
\item $\mu'' \in \cI_2''$, those with $J_{\mu''}\cdot J_A \cdot J_B >0$ for all $A \neq B \in \cI_\la$.  
\end{itemize}
Note that the RHS of the decomposition~\eqref{decomp} does not miss any generators, that is, the index subsets $\cI_\la$, $\cI_2'$ and $\cI_2''$, as defined above satisfy the following relation, 
\beq
\cI_\la^c = \cI_2' \cup \cI_2''\,.
\eeq
In order to see this, we will have to show that, for $J_i$ with $i\in \cI_\la^c$, 
\beq
\exists A_0, B_0 \in \cI_\la ~~\text{with}~~ A_0 \neq B_0 ~~\text{and}~~ J_i \cdot J_{A_0} \cdot J_{B_0} =0 \quad \Rightarrow \quad J_i \cdot J_A \cdot J_B = 0 \quad \forall A, B \in \cI_\la \,.
\eeq
The assumed vanishing of $J_i \cdot J_{A_0}\cdot J_{B_0}$ leads, via Lemma~\ref{lemma2}, to  
\beq
J_i \cdot J_{A_0} = \kappa J_{A_0} \cdot J_{B_0} \,,\qquad \text{for some}\quad \kappa >0\,,
\eeq
and hence,
\beq\label{iA0B}
J_i \cdot J_{A_0} \cdot J_B = \kappa J_{A_0} \cdot J_{B_0} \cdot J_B = 0 \,,\qquad \forall B \in \cI_\la \,,
\eeq
where the last step has used the finiteness of $\cV_Y$ in the limit. Similarly, via Lemma~\ref{lemma2}, the vanishing~\eqref{iA0B} of $J_i \cdot J_{A_0} \cdot J_B$ implies, for all $B \neq A_0$, that
\beq
J_{i} \cdot J_B = \kappa_{BA_0} J_{A_0} \cdot J_B \,, \qquad \text{for some}\quad \kappa_{BA_0} > 0\,,
\eeq
and hence we have
\beq\label{iAB}
J_i \cdot J_A \cdot J_B = \kappa_{BA_0} J_{A_0} \cdot J_B \cdot J_A = 0 \,,\qquad\forall A, B \in \cI_\la ~~\text{with}~~ B \neq A_0 \,,
\eeq
and this vanishing~\eqref{iAB} obviously extends to the case $B=A_0$ by~\eqref{iA0B}.

Note also that $J_A^2 = 0$ for all $A \in \cI_\la$ as the limit would be of $J$-class A otherwise.
Since $J_0$ is of course labeled by $\cI_\la$, we have either of the two possibilities, $|\cI_\la|=1$ or $|\cI_\la| >1$, which we will analyze in turn.

\subsubsection{$|\cI_\la|=1$}\label{la=1}

With $|\cI_\la| =1$, the decomposition~\eqref{decomp} is rather trivial in that
\beq
\cI_\la=\{0\} \,, \qquad \cI_2' = \cI_2 \,,\qquad \cI_2'' = \emptyset \,.
\eeq
We will thus keep using the original notation of ~\eqref{classB-again}. 
The first statement we will prove is that there exists at least one $K3$ fibration whose fiber volume shrinks as 
\beq \label{VK3required}
\cV_{K3}  \sim \la^{-1} \,\qquad \text{for}\quad \la \to \infty \,.
\eeq

As discussed in Section~\ref{SubsecTypeB}, the volume of the obvious $K3$-fiber with class $J_0$ is
\beq\label{VK3-behavior}
\cV_{K3, 0} = \frac12 J_0 \cdot J^2 = \frac12\sum_{\mu, \nu \in\cI_2}b_\mu b_\nu J_{0} \cdot J_\mu \cdot J_{\nu} \,,
\eeq
in terms of which the volume of $Y$ can be expressed as
\beq\label{VY-K3-again}
\cV_Y = \frac16 J^3 = 
 \la \cV_{K3,0} + \frac16 \sum_{\mu, \nu, \rho\in\cI_2}b_\mu b_\nu b_\rho J_{\mu}\cdot J_\nu \cdot J_{\rho} \,. 
\eeq
To show the required behavior~\eqref{VK3required} for this $K3$ fiber, it suffices to prove that the second term in the RHS of~\eqref{VY-K3-again} tends to $0$ for $\la \to \infty$ as we keep $\cV_Y \sim 1$ in the limit. 

In order to prove this, suppose instead that there exist a triple $(\mu, \nu, \rho)$ with
\beq\label{triple-sim}
b_\mu b_\nu b_\rho \sim 1 \quad \text{and}\quad J_{\mu}\cdot J_{\nu}\cdot J_{\rho} >0 \,.
\eeq
Note first that for such a triple not all of the three intersection numbers, $(J_{0}\cdot J_\mu \cdot J_\nu)$, $(J_{0}\cdot J_\nu \cdot J_\rho)$ and $(J_{0}\cdot J_\rho \cdot J_\mu)$, can be non-zero: If they were all non-zero, we would have 
\beq\label{pair-supp}
b_{\mu} b_{\nu} \precsim \la^{-1} \,  \qquad b_{\nu} b_{\rho} \precsim \la^{-1} \,,\qquad b_{\rho} b_{\mu} \precsim \la^{-1}\,,
\eeq
as, otherwise, the $K3$ volume~\eqref{VK3-behavior} would behave as $\cV_{K3,0} \succ \la^{-1}$ and hence the three-fold volume~\eqref{VY-K3-again} as $\cV_Y \succ 1$, which is a contradiction. The parametric behavior~\eqref{pair-supp} would then lead to
\beq
b_\mu b_\nu b_\rho \precsim \la^{-3/2} \,,
\eeq
which is in contradiction with~\eqref{triple-sim}. Therefore, without loss of generality, we may assume that $J_{0}\cdot J_\mu \cdot J_\nu=0$. Then, by Lemma~\ref{lemma2}, we have
\beq
J_{0}\cdot J_{\nu}  = \kappa J_{\mu}\cdot J_{\nu} \,, \qquad \text{for some}\quad \kappa >0 \,,
\eeq
and hence it follows that
\beq
J_{0} \cdot J_\nu \cdot J_\rho =  \kappa\, J_{\mu}\cdot J_\nu\cdot J_\rho \neq 0 \,,
\eeq
where we used (\ref{triple-sim}).
Thus, comparing again the two volumes~\eqref{VK3-behavior} and~\eqref{VY-K3-again}, we learn that
\beq\label{pair-supp1}
b_\nu b_\rho \precsim \la^{-1} \,. 
\eeq
Combining the asymptotic behaviors~\eqref{triple-sim} and \eqref{pair-supp1}, we then have
\beq
b_\mu \succsim \la \,,
\eeq
which contradicts the fact that $\cI_\la = \{0\}$. We have thus proven that any order $1$ contribution to $\cV_Y$ in~\eqref{VY-K3-again} comes from $\la \cV_{K3,0}$ and hence that 
\beq\label{K30-fastest}
\cV_{K3,0} \sim \la^{-1} \,.
\eeq

We will now prove that if $Y$ admits any additional $K3$ fibration, the volume of its generic fiber is parametrically bigger than $\la^{-1}$ or the limit is of Type $T^2$ rather than of Type K3. The scaling behaviour of the genus-one fiber of such limits will be discussed in Appendix \ref{T2Unique-JClassB}.  This will establish the uniqueness of the most rapidly shrinking $K3$ fiber in a limit of Type K3 in the parametric sense.

Let us first recall that the class of a $K3$ fiber is proportional to a K\"ahler cone generator. This is because a nef divisor is a linear combination of the K\"ahler cone generators; if it involved more than one such generators, its square would necessarily be non-zero since any cross term in the square between distinct generators is, by Lemma~\ref{lemma1}, a non-trivial class, which cannot be cancelled by any other terms. 

Let us now suppose that there exists another $K3$ fiber with the property that it shrinks at a faster rate than or equal to~\eqref{K30-fastest}. That is, suppose that there exists some $\mu_0 \in \cI_2$ for which 
\beq\label{oguisoK3}
J_{\mu_0}^2 = 0\,,
\eeq
and with a corresponding $K3$ fiber volume
\beq\label{VK3-other}
\cV_{K3, \mu_0} \precsim \la^{-1} \,. 
\eeq
Under this assumption we will show in the following that the asymptotic behavior~\eqref{VK3-other} has to saturate $\la^{-1}$, that is, 
\beq\label{VK3-other-sat}
\cV_{K3, \mu_0} \sim \la^{-1} \,, 
\eeq
and also that the limit in scrutiny is necessarily of Type $T^2$. 

We start by assuming that the parametric comparison~\eqref{VK3-other} holds strictly, that is, 
\beq\label{contra?}
\cV_{K3, \mu_0} \prec \la^{-1}\,.
\eeq 
Since the volume is written as
\beq\label{VK3-mu0}
\cV_{K3, \mu_0} =  \la \sum_{\nu \in \cI_2} b_\nu J_{0}\cdot J_{\mu_0}\cdot J_{\nu}  + \frac12 \sum_{\nu, \rho \in \cI_2}b_\nu b_\rho J_{\mu_0}\cdot J_\nu \cdot J_\rho\,,
\eeq
this implies that 
\beq\label{nuconstraint}
b_\nu \prec \la^{-2} \,,\qquad \forall \nu ~~\text{with}~~ J_{0} \cdot J_{\mu_0} \cdot J_\nu>0 \,.
\eeq
Now, since $\cV_{K3,0}$ in~\eqref{VK3-behavior} has to be of order $\la^{-1}$, there must exist at least one pair $(\nu_0, \rho_0)$ for which 
\beq\label{nu0rho0}
b_{\nu_0} b_{\rho_0} \sim \la^{-1}\,,\quad \text{and}\quad J_{0}\cdot J_{\nu_0}\cdot J_{\rho_0} >0\,.
\eeq
The constraint~\eqref{nu0rho0} cannot be fulfilled, however, if $J_{0}\cdot J_{\mu_0}\cdot J_{\nu_0}>0$ (because in this case $b_{\nu_0} \prec \la^{-2}$ by~\eqref{nuconstraint}, which would require $ b_{\rho_0} \succ \lambda$ for (\ref{nu0rho0})). Therefore, we have 
\beq
J_{0}\cdot J_{\mu_0}\cdot J_{\nu_0}=0\,,
\eeq
and similarly, $J_{0}\cdot J_{\mu_0}\cdot J_{\rho_0}$ also vanishes. 
{Let us now observe that $\mu_0$ is different at least from one of $\nu_0$ and $\rho_0$. This is obvious if $\nu_0 \neq \rho_0$, and furthermore, if $\nu_0 =\rho_0$, it also follows that $\mu_0 \neq \nu_0$ because $J_{\nu_0}\cdot J_{\nu_0} \neq 0$ by~\eqref{nu0rho0}, while $J_{\mu_0}\cdot J_{\mu_0}=0$ by~\eqref{oguisoK3} (for the divisor $J_{\mu_0}$ to define a K3 fiber). Therefore, we may assume without loss of generality that $\mu_0 \neq \nu_0$.}
We can thus apply Lemma~\ref{lemma2} to the triple $(0, \mu_0, \nu_0)$ and learn that 
\beq
J_{\mu_0}\cdot J_{\nu_0} = \kappa J_{0} \cdot J_{\nu_0} \,,\qquad\text{for some}\quad \kappa>0\,.
\eeq
Therefore, the second sum in~\eqref{VK3-mu0} contains the positive contribution
\beq
b_{\nu_0} b_{\rho_0} J_{\mu_0}\cdot J_{\nu_0}\cdot J_{\rho_0} = \kappa b_{\nu_0} b_{\rho_0} J_0 \cdot J_{\nu_0}\cdot J_{\rho_0} \,,
\eeq
which is of order $\la^{-1}$ by~\eqref{nu0rho0} and hence, a contradiction to the assumption~\eqref{contra?}. 

Having proven~\eqref{VK3-other-sat}, we have learnt that the limit in scrutiny has at least two leading shrinking $K3$ fibers with classes $J_0$ and $J_{\mu_0}$ whose volumes both behave as $\la^{-1}$. This may at first look like it contradicts uniqueness of leading shrinking $K3$ fibers. However, in this situation, we can easily find a $T^2$ fiber whose volume is suppressed by $\la^{-2}$, which will show that the limit is actually of Type $T^2$, not of Type $K3$. For this purpose, consider the nef divisor
\beq
D_{\mu_0} = J_0 + J_{\mu_0} \,,
\eeq
and observe that its square is non-trivial while its cube is trivial:
\bea
D_{\mu_0}^2 &=& 2 J_0 \cdot J_{\mu_0} \neq 0 \,,\\
D_{\mu_0}^3 &=& J_0^2 (J_0 + 3 J_{\mu_0}) + J_{\mu_0}^2 (J_{\mu_0} + 3 J_0) =0 \,,
\eea
where $J_0^2 = 0 = J_{\mu_0}^2$ has been used. Therefore, there exists a genus-one fibration with its fiber class proportional to $J_0 \cdot J_{\mu_0}$, and hence, the volume of this $T^2$ fiber scales as
\beq
J_0 \cdot J_{\mu_0} \cdot J = \sum_{\nu \in \cI_2} b_\nu J_{0}\cdot J_{\mu_0}\cdot J_ {\nu} \precsim \la^{-2} \,,
\eeq
where the last step follows from the expression~\eqref{VK3-mu0} for the $K3$ volume and its scaling behavior~\eqref{VK3-other-sat}. Thus, the limit is indeed of Type $T^2$. Uniqueness of this $T^2$ fiber, in the sense that it is the fiber shrinking at the fastest rate, will be proven in Appendix \ref{T2Unique-JClassB}.

\subsubsection{$|\cI_\la| >1$} 

We will now show that every K\"ahler form of $J$-class B with $|\cI_\la| >1$ leads to a limit of Type $T^2$, and hence this case is not relevant for studying uniqueness of the surface fibers which shrink
at the fastest rate.

Provided that 
$|\cI_\la| >1$, 
we will start by showing that the K\"ahler class takes the form
\beq\label{classB-final}
J=  \sum_{A\in \cI_\la} l_A J_A + \sum_{\mu' \in \cI_2'} b_{\mu'} J_{\mu'} + \sum_{\mu'' \in \cI_2''} b_{\mu''} J_{\mu''}\,, 
\eeq
where 
\beq \label{classB-details}
l_A \sim  \la\,, \quad b_{\mu'} \prec \la \,,\quad b_{\mu''}\precsim \la^{-2} \,, \qquad \text{for}\quad \la \to \infty \,.
\eeq  
Furthermore, all the intersection numbers amongst the generators labelled by $\cI_2'$ turn out to vanish, {\it i.e.}, 
\beq\label{classB-vanishing}
J_{\mu'} \cdot J_{\nu'} \cdot J_{\rho'} = 0 \,,  \qquad \forall \mu', \nu', \rho' \in \cI_2' \,.
\eeq 

Note first that the parametric behavior of $l_A$ and $b_{\mu'}$ in~\eqref{classB-details} is obvious by definition. The scaling of $b_{\mu''}$ in~\eqref{classB-details} also follows easily by imposing the finiteness of $\cV_Y$ since, for each $\mu'' \in\cI_2''$, there exist, by definition, $A \neq B\in \cI_\la$ such that $J_A\cdot J_B \cdot J_{\mu''} > 0$. 

Next, the vanishing~\eqref{classB-vanishing} can be seen as follows. 
Let us consider a pair of distinct indices $A \neq B \in \cI_\la$ and an arbitrary triple $(\mu', \nu', \rho')$ in $\cI_2'$. By definition, we have
\beq
J_{A} \cdot J_B \cdot J_{\mu'}=0\,,
\eeq
and hence, by Lemma~\ref{lemma2}, 
\beq\label{mu'A=AB}
J_{\mu'} \cdot J_A = \kappa J_A \cdot J_B \,,\qquad \text{for some}\quad \kappa >0\,, 
\eeq
from which it follows that
\beq
J_{A} \cdot J_{\mu'} \cdot J_{\nu'}=\kappa J_A \cdot J_B \cdot J_{\mu'} = 0\,. 
\eeq
Lemma~\ref{lemma2} can then be applied again, leading to
\beq\label{kappa'}
J_{\mu'} \cdot J_{\nu'} = \kappa' J_{A} \cdot J_{\mu'}  \,,\qquad \text{for some}\quad \kappa' \geq 0\,, 
\eeq
where $\kappa'$ is potentially zero for $\mu' =\nu'$. 
Therefore, 
\beq\label{vanish-triple-prime}
J_{\mu'}\cdot J_{\nu'}\cdot J_{\rho'} = \kappa' J_A \cdot J_{\mu'} \cdot J_{\rho'} = 0 \, \qquad \forall \mu', \nu', \rho' \in \cI_2' \,.
\eeq

We are now ready to prove existence and uniqueness of the leading shrinking $K3$ fiber. We start with the existence of a shrinking $K3$. As already explained in Subsection~\ref{la=1}, according to Oguiso's criterion, the (co)homology class of any $K3$ fiber in a Calabi-Yau three-fold has to be proportional to a nef divisor $D$ with a vanishing double self-intersection, and hence, it has to be one of the K\"ahler cone generators. Obvious such $K3$ fibers are associated with $J_A$ for $A\in \cI_\la$, with the respective volumes,
\beq\label{VK3A}
\cV_{K3, A} = \frac12 J_A \cdot J^2 =\frac12 J_A \sum_{B \in \cI_\la}  \sum_{\mu'' \in \cI_2''} l_B b_{\mu''} \, J_B \cdot J_{\mu''}  + \frac12 J_A \cdot ( \sum_{\mu' \in \cI_2'} b_{\mu'} J_{\mu'} + \sum_{\mu'' \in \cI_2''} b_{\mu''} J_{\mu''})^2 \,.
\eeq
We can then estimate the volume $\cV_Y$ of $Y$ in terms of $\cV_{K3,A}$ as follows,
\beq\label{VY-K3-asymp}
\cV_Y = \frac16 J^3 \sim \la \sum_A \cV_{K3,A} +  \frac16 \sum_{\mu, \nu, \rho \in \cI_\la^c} b_\mu b_\nu b_\rho J_\mu \cdot J_\nu \cdot J_\rho \,,
\eeq
where the symbol $\sim$ has been used to indicate equality of the parametric behaviors.\footnote{As per a few lines of algebra, one can verify the following inequality,
\beq
\frac12 \la \sum_{A\in\cI_\la} \cV_{K3,A} < \cV_Y - \frac16 \sum_{\mu, \nu, \rho \in \cI_\la^c} b_\mu b_\nu b_\rho J_{\mu}\cdot J_\nu \cdot J_\rho < \la \sum_{A \in \cI_{\la}}\cV_{K3,A}\,,
\eeq
which justifies the asymptotic comparison~\eqref{VY-K3-asymp}.
}
In showing the existence of a $K3$ fiber with the volume of order $\la^{-1}$, it suffices to prove that the second sum in the RHS of~\eqref{VY-K3-asymp} tends to $0$ for $\la \to\infty$ as we keep $\cV_Y \sim 1$ in the limit. 
Note that this second sum expands as
\beq\label{munurho}
\sum_{\mu, \nu, \rho \in \cI_\la^c} b_\mu b_\nu b_\rho J_\mu \cdot J_\nu \cdot J_\rho = \sum_{\mu', \nu', \rho' \in \cI_2'} b_{\mu'} b_{\nu'} b_{\rho'} J_{\mu'} \cdot J_{\nu'} \cdot J_{\rho'} + \cdots
\eeq
where the ellipsis denote only terms of order strictly less than $1$, since $b_{\mu'} \prec \la$ for all $\mu' \in \cI_2'$ and $b_{\mu''} \precsim \la^{-2}$ for all $\mu'' \in \cI_2''$. It is thus enough to show that 
\beq
J_{\mu'} \cdot J_{\nu'} \cdot J_{\rho'}=0\,, \qquad  \forall \mu',\nu',\rho' \in \cI_2'\,,
\eeq
but this is precisely the relation (\ref{classB-vanishing}) which we have derived above.
This completes the proof that at least one $K3$ fiber has volume of order $\la^{-1}$.

As a byproduct, it immediately follows that the volume of {\it every} $K3$ fiber associated with $J_A$ is suppressed as
\beq\label{allK3A-prelim}
\cV_{K3, A} \precsim \la^{-1} \qquad \forall A\in \cI_\la\,.
\eeq 
In fact, we can make a stronger statement that the relation~\eqref{allK3A-prelim} has to saturate for all $A$, that is,  
\beq\label{allK3A}
\cV_{K3, A} \sim \la^{-1}  \qquad \forall A \in \cI_{\la} \,.
\eeq
This is because the first sum in the expression~\eqref{VK3A} for $\cV_{K3,A}$ contains the following terms,
\beq
\frac12 l_B b_{\mu''} J_A \cdot J_B \cdot J_{\mu''} \,, 
\eeq
where the triple intersection $J_A \cdot J_B \cdot J_{\mu''}$ is strictly positive for all $B \in \cI_\la$ with $B\neq A$ and for all $\mu'' \in \cI_2''$. Note that there exists at least one $B \neq A$ since $|\cI_\la|>1$ and also that 
\beq\label{somemu''}
b_{\mu''} \sim \la^{-2}\,,\qquad \text{for some}\quad \mu'' \in \cI_2''\,,
\eeq
as order $1$ contributions to $\cV_Y$ only arise from the cross terms of the form,
\beq
l_A l_B b_{\mu''} J_A \cdot J_B \cdot J_{\mu''}\,,
\eeq
in the expansion of $J^3$. 
The volume suppression~\eqref{allK3A-prelim} must thus be saturated as in~\eqref{allK3A}.

Now, there may in principle exist $K3$ fibers associated with $J_{\mu'}$ or $J_{\mu''}$, but it turns out that no such fibers can shrink at a faster rate than $\la^{-1}$: The volume $\cV_{\mu''}$ of the latter type fibers diverges since
\beq
\cV_{\mu''} = \frac12 J_{\mu''} \cdot J^2 = \frac12 \sum_{A,B \in \cI_\la} l_A l_B J_{\mu''}\cdot J_A \cdot J_B  + \cdots \,,
\eeq
where the sum is of order $\la^2$ due to~\eqref{classB-details}, as well as due to the non-vanishing of the triple intersections involved, and the ellipsis only contains non-negative terms. The volume $\cV_{\mu'}$ of the former type fibers on the other hand tends to zero precisely as $\la^{-1}$. This follows from the expansion of $\cV_{\mu'}$, 
\bea\label{Vmu'} \nn
\cV_{\mu'} &=& \frac12 J_{\mu'} \cdot J^2 \\
 &=& \sum_{A \in \cI_\la} \sum_{\rho'' \in \cI_2''} l_A b_{\rho''} J_{\mu'}\cdot J_A \cdot J_{\rho''} +  \sum_{\nu'\in \cI_2'} \sum_{\rho''\in \cI_2''} b_{\nu'} b_{\rho''} J_{\mu'}\cdot J_{\nu'}\cdot J_{\rho''} \\ \nn
 &&+\, \frac12 \sum_{\nu'', \rho'' \in \cI_2''} b_{\nu''} b_{\rho''} J_{\mu'}\cdot J_{\nu''}\cdot J_{\rho''} \,,
\eea
where we have used vanishing of any triple intersections only involving the generators $J_i$ with $i\in \cI_\la \cup \cI_2'$. Note that every term in the first sum of~\eqref{Vmu'} is of order $\la^{-1}$ or less by~\eqref{classB-details}; in fact, the sum must contain non-trivial terms of order precisely $\la^{-1}$ because $b_{\rho''} \sim \la^{-2}$ for some $\rho''$ by~\eqref{somemu''} and also because
\beq
J_{\mu'}\cdot J_A \cdot J_{\rho''} = \kappa J_{A}\cdot J_B\cdot J_{\rho''} > 0 \,,
\eeq
where the proportionality~\eqref{mu'A=AB} of the four-forms have been used. Since, by~\eqref{classB-details}, the second and the third sums of~\eqref{Vmu'} are subleading compared to the first, we have
\beq\label{potK3mu'}
\cV_{\mu'} \sim \la^{-1} \,, \qquad \forall\mu' \in \cI_2' \,,
\eeq
whether or not the divisor $J_{\mu'}$ corresponds to a $K3$ fiber. 

By now we have established that the limit in scrutiny contains multiple shrinking $K3$ fibers with classes $J_A$ for $A\in\cI_\la$, whose volumes $\cV_{K3,A}$ all scale as $\la^{-1}$ as in~\eqref{allK3A}. Furthermore, we have also seen that additional shrinking $K3$ fibers with class $J_{\mu'}$ for some $\mu' \in \cI_2'$ may potentially be present, whose volumes $\cV_{\mu'}$ scale at the same rate as $\cV_{K3,A}$ as in~\eqref{potK3mu'}. No further shrinking $K3$ fibers can be admitted by the three-fold $Y$, and therefore, these form a complete set of leading shrinking $K3$ fibers. 

This statement, however,  does not contradict our claim of uniqueness of the  shrinking $K3$ fiber in a Type $K3$ limit. Just as in Section~\ref{la=1}, the limit under consideration is actually of Class $T^2$, since there exists a genus-one fibration where the fiber volume is of order $\la^{-2}$, while the leading shrinking $K3$ fibers all have volumes of order $\la^{-1}$. 
In order to see this, let us consider a nef divisor $D$,
\beq\label{nef-pa}
D =\sum_{A\in\cI_\la} J_A   \,,
\eeq
which satisfies
\beq
D^3 =0\,, 
\eeq
as $J_A J_B J_C = 0$ for all $A, B, C \in \cI_\la$ in order for ${\cal V}_Y$ to remain finite.
Furthermore, $D^2$ is expanded as
\beq
D^2= \sum_{A\neq B \in \cI_\la} J_A \cdot J_B \,,
\eeq
and hence is non-zero since every pair of distinct indices $A,B\in\cI_\la$ leads to a non-trivial intersection $J_A \cdot J_B$ by Lemma~\ref{lemma1}. Therefore, $D^2$ corresponds to the class of a (multiple of a) $T^2$ fiber, the volume of which scales as
\beq\label{Dsq-estimate}
D^2 \cdot J = \sum_{A,B \in \cI_\la}  \sum_{\mu'' \in \cI_2''} b_{\mu''} J_{A}\cdot J_B\cdot J_{\mu''}  \sim \la^{-2} \,,
\eeq
where, in the last step, we have used ~\eqref{somemu''}. 
Therefore, the volume of this $T^2$ fiber is parametrically much smaller than the square root of the $K3$ volume and hence, the limit is of Class $T^2$. 

\subsection{Type $T^2$ Limits with a K\"ahler form of $J$-class B}\label{T2Unique-JClassB}
As emphasized several times, while K\"ahler forms of $J$-class A only lead to Type $T^2$ limits, those of $J$-class B could lead either to Type $K3/T^4$ limits or to Type $T^2$ limits as well. The aim of this section is to prove 

\begin{proposition} \label{propT2fromB}
For limits of Type $T^2$ realized by K\"ahler forms of $J$-class $B$, the three-fold $Y$ admits a unique genus-one fibration with the property that its generic fiber shrinks at a rate parametrically faster than the fiber of 
any other genus-one fibration on $Y$. 
\end{proposition}
We first show that if a K\"ahler form of $J$-class B,
\beq\label{classB-again2}
J = \la J_0 + \sum_{\mu \in \cI_2} b_\mu J_\mu \,,
\eeq
gives rise to a limit of Type $T^2$, at least one K\"ahler coefficient is parametrically strictly bigger than $\la^{-1/2}$, {\it i.e.},
\beq\label{mu0-exist}
\exists \, \hat \mu \in \cI_2 ~~\text{such that}~~ b_{\hat \mu} \succ \la^{-1/2} \,.
\eeq
In order to see~\eqref{mu0-exist}, 
note first that by assumption there must a exist  
a shrinking genus-one fiber, which we will call $\hat T^2$, with a parametric volume suppression
\beq\label{hatT2supp}
\cV_{\hat T^2} \prec \la^{-1/2} \,.
\eeq
This is because otherwise, as discussed in Section~\ref{K3unique}, the limit in scrutiny would be of Type K3 (or $T^4$) due to the guaranteed existence of a shrinking K3 (or $T^4$) fiber whose volume is suppressed as $\la^{-1}$. Thus, there must exist a nef divisor $\hat D$ associated to the $\hat T^2$-fibration 
which satisfies Oguiso's criteria,
\beq\label{oguisoDhat}
\hat D^2 \neq 0 \,,\qquad  \hat D^3 = 0\,,
\eeq
and which leads to a shrinking $\hat T^2$ fiber whose volume is parametrically smaller than $\la^{-1/2}$.
We can expand this divisor as
\beq\label{Dhat}
\hat D=p_0 J_0 + \sum_{\hat \mu \in \hat \cI_2} p_{\hat \mu}^+ J_{\hat \mu} \,,\qquad p_0 \geq 0\,,~~ p^+_{\hat \mu} >0 \,,
\eeq
where the index subset $\hat \cI_2 \subset \cI_2$ is non-empty because, otherwise, the criteria~\eqref{oguisoDhat} cannot be met.

Then, since the volume of $\hat T^2$,  
\beq
\cV_{\hat T^2} \sim \hat D^2 \cdot J = (p_0 J_0 + \sum_{\hat \mu \in \hat \cI_2} p_{\hat \mu}^+ J_{\hat \mu})^2 \cdot   (\la J_0 + \sum_{\mu \in \cI_2} b_\mu J_\mu) \,,
\eeq
must tend to zero in the limit, we know that
\beq\label{vanhathat}
J_0 \cdot J_{\hat \mu} \cdot J_{\hat \nu} =0\,,\qquad \forall \hat{\mu}, \hat{\nu} \in \hat \cI_2\,.
\eeq
By Lemma~\ref{lemma2}, this implies that the classes $J_0 \cdot J_{\hat \mu}$ and $J_{\hat \mu} \cdot J_{\hat \nu}$ are all proportional for any $\hat \mu, \hat \nu \in \hat \cI_2$. Hence, 
\beq\label{prophats}
J_{\hat \nu} \cdot J_{\hat \rho} = \kappa_{\hat \nu \hat \rho}^{\hat\mu} J_0 \cdot J_{\hat \mu} \,,\quad  \qquad \forall \hat \mu, \hat \nu, \hat \rho \in \hat \cI_2\,,
\eeq
for some non-negative constants $ \kappa_{\hat \nu \hat \rho}^{\hat\mu}$. Note that $ \kappa_{\hat \nu \hat \rho}^{\hat\mu}=0$ is possible for $\nu = \rho$, but this does not affect our argument.
With this we can express $\hat D^2$ in terms of any product $J_0 \cdot J_{\hat \mu}$ as 
\beq\label{Dhatsquare}
\hat D^2 = \kappa_{\hat \mu} J_0 \cdot J_{\hat \mu} \,, \qquad \forall \hat\mu \in \hat \cI_2\,,
\eeq
with $\kappa_{\hat \mu}>0$. 

Let us now group the K\"ahler cone generators as the disjoint union
\beq
\cI_2  = \hat \cI_2 \cup  \bar \cI_2 \cup \check \cI_2 \,,
\eeq
such that 
\bea\label{const1}
\hat D^2 \cdot J_{\bar \mu} &=& 0\,, \qquad \forall \bar \mu \in \bar \cI_2 \,,\\ \label{const2}
\hat D^2 \cdot J_{\check \mu} &>&0 \,,\qquad \forall \check \mu \in \check \cI_2 \,.
\eea
We now show that without loss of generality we can assume that $\bar\cI_2 = \emptyset$ by proving that any two Ooguiso divisors satisfying (\ref{oguisoDhat}) which differ only by elements in $\bar\cI_2 = \emptyset$ define the same $T^2$-fibration. To see this, note first
that
 the vanishing~\eqref{const1}, when combined with~\eqref{Dhatsquare}, implies, by Lemma~\ref{lemma2}, that 
\beq\label{Dhat0bar}
\hat D^2 = \kappa_{\bar \mu} J_0 \cdot J_{\bar \mu} = \kappa_{\bar \mu\hat \nu} J_{\bar  \mu} \cdot J_{\hat \nu} \,, \qquad \forall \bar \mu \in \bar \cI_2 \,,~\forall \hat \nu \in\hat \cI_2 \,,
\eeq 
{with $\kappa_{\bar \mu}, \kappa_{\bar \mu \hat \rho} >0$. Similarly, ~\eqref{const1} and the first part of~\eqref{Dhat0bar}
imply that
\beq
J_0\cdot J_{\bar \mu} \cdot J_{\bar \nu} = 0 \,.
\eeq
Again by Lemma~\ref{lemma2} we deduce from this that
\beq\label{propbars}
J_{\bar \nu} \cdot J_{\bar \rho} = \kappa_{\bar \nu \bar \rho}^{\bar\mu} J_0 \cdot J_{\bar \mu} \,,  \qquad \forall \bar \mu, \bar \nu, \bar \rho \in \bar \cI_2\,,
\eeq
with $\kappa^{\bar \mu}_{\bar \nu \bar \rho} \geq 0$. Here, $\kappa^{\bar \mu}_{\bar \nu \bar \rho}$ is potentially zero for $\bar \nu = \bar \rho$. 
The relations~\eqref{const1},~\eqref{Dhat0bar} and~\eqref{propbars} then lead to the vanishing
\beq\label{bartriple}
J_{\bar \mu} \cdot J_{\bar \nu} \cdot J_{\bar \rho} =0 \,,\qquad \forall \bar \mu, \bar \nu, \bar \rho \in \bar \cI_2 \,. 
\eeq

Combining all of this, we can now deduce that 
\beq\label{Dhatnew}
\hat D_{\rm new} = \hat D + \sum_{\bar \mu \in \bar \cI_2} p^+_{\bar \mu} J_{\bar \mu}\,,\qquad p^+_{\bar \mu} >0 \,,
\eeq
defines the same genus-one fibration as $\hat D$. In order to see this, let us first note that $\hat D \cdot J_{\bar \mu}$ is proportional to $\hat D^2$ due to the vanishing~\eqref{const1} and Lemma~\ref{lemma2}, that is,
\beq\label{prop-DhatJmubar}
\hat D \cdot J_{\bar \mu} = \kappa_{\hat D \mu} \hat D^2 \,, \quad \kappa_{\hat D \mu} >0 \,. 
\eeq
Now, the cubic self-intersection of the divisor $\hat D_{\rm new}$ has the following expansion,
\beq\label{Dhatnewcube}
\hat D_{\rm new}^3 = \hat D^3 + 3 \sum_{\bar \mu \in \bar \cI_2} p^+_{\bar \mu} \hat D^2 \cdot J_{\bar \mu} + 3 \sum_{\bar \mu, \bar \nu \in \bar \cI_2} p^+_{\bar \mu} p^+_{\bar \nu} \hat D \cdot J_{\bar \mu} \cdot J_{\bar \nu} + \sum_{\bar \mu, \bar \nu, \bar \rho \in \bar \cI_2} p^+_{\bar \mu} p^+_{\bar \nu} p^+_{\bar \rho} J_{\bar \mu} \cdot J_{\bar \nu} \cdot J_{\bar \rho} \,,
\eeq 
and it vanishes as each of the four parts in the RHS of~\eqref{Dhatnewcube} does; The first vanishes by~\eqref{oguisoDhat}, the second by~\eqref{const1}, the third by the proportionality~\eqref{prop-DhatJmubar} combined with~\eqref{const1}, and the last by~\eqref{bartriple}. It still remains to see that $\hat D_{\rm new}^2$ and $\hat D^2$ are proportional as four-forms, which easily follows from the expansion of the former,
\beq\label{Dhatnewsquare}
\hat D_{\rm new}^2 = \hat D^2 + 2 \sum_{\bar \mu \in \bar \cI_2} p^+_{\bar \mu} \hat D \cdot J_{\bar \mu} + \sum_{\bar \mu, \bar \nu \in \bar \cI_2} p^+_{\bar \mu} p^+_{\bar \nu} J_{\bar \mu} \cdot J_{\bar \nu} \,.
\eeq
In this expansion, we immediately observe that the second part is proportional to $\hat D^2$ by~\eqref{prop-DhatJmubar} and the third also to $\hat D^2$ by~\eqref{propbars} and~\eqref{Dhat0bar}. This completes the proof that $\hat D_{\rm new}$ and $\hat D$ define the same genus-one fibration and we may therefore assume, without loss of generality, that 
\beq
\bar \cI_2 = \emptyset \,.
\eeq

}

We are now ready to constrain the K\"ahler parameters $b_{\hat \mu}$ and $b_{\check \mu}$. For this purpose, let us first compute the scaling behavior of the $\hat T^2$ volume,
\beq
\cV_{\hat T^2} \sim \hat D^2 \cdot J \sim J_0 \cdot J_{\hat \mu} \cdot (\la J_0 + \sum_{\hat \nu \in\hat  \cI_2} b_{\hat \nu} J_{\hat \nu} + \sum_{\check \nu \in\check \cI_2} b_{\check\nu} J_{\check\nu}) \,,
\eeq
where $\hat \mu$ has been chosen arbitrarily from $\hat \cI_2$ using the proportionality~\eqref{Dhatsquare}. 
By the vanishing~\eqref{vanhathat}, we thus have
\beq
\cV_{\hat T^2} \sim \sum_{\check \nu \in\check \cI_2} b_{\check\nu} J_0 \cdot J_{\hat \mu} \cdot J_{\check\nu} \sim \sum_{\check \nu \in\check \cI_2} b_{\check\nu} \hat D^2 \cdot  J_{\check\nu} \,,
\eeq
and hence,
\beq\label{checknu}
b_{\check \nu} \prec \la^{-1/2} \,,\qquad \forall \check\nu \in \check \cI_2 \,,
\eeq
where the non-vanishing~\eqref{const2}, as well as the volume suppression~\eqref{hatT2supp}, has been used. 
In the meantime, the volume of the $K3$ fiber with class $J_0$ can be expanded, upon using the vanishing~\eqref{vanhathat}, as
\beq\label{K30-TypeT2}
\cV_{K3, 0} = \frac12 J_0 \cdot J^2 =\frac12 \sum_{\hat \mu \in \hat \cI_2} \sum_{\check \nu \in \check \cI_2} b_{\hat \mu} b_{\check \nu}  J_0 \cdot J_{\hat \mu} \cdot J_{\check \nu}+ \sum_{\check \mu,\check \nu \in \check \cI_2} b_{\check \mu} b_{\check \nu} J_0 \cdot J_{\check \mu} \cdot J_{\check \nu} \,,
\eeq
which must scale precisely as $\la^{-1}$ as proven in Section~\ref{K3unique}. With the suppression of the $b_{\check \nu}$~\eqref{checknu}, we thus conclude, as claimed in~\eqref{mu0-exist}, that at least one $b_{\hat \mu}$ must be parametrically bigger than $\la^{-1/2}$.

Finally we are ready to prove the uniqueness of the parametrically leading shrinking $T^2$ fiber. Specifically, we will show that there exists a unique $T^2$ fibration whose fiber shrinks at a rate faster than $\la^{-1/2}$. Given one such fiber $\hat T^2$ associated with the nef divisor $\hat D$ as in~\eqref{Dhat}, let us suppose that there is another nef divisor $D$, not proportional to $\hat D$, with the property
\beq
D^2 \neq 0\,,\qquad D^3 =0\,,
\eeq
whose associated $T^2$ fiber also shrinks as
\beq\label{T2aswell}
\cV_{T^2} \prec \la^{-1/2} \,.
\eeq
Then, we have the following relations for the triple intersections, 
\beq\label{DhatDJ0}
D \cdot \hat D \cdot J_0 = \sum_{\hat\mu\in\hat \cI_2}p_{\hat \mu}^+  \frac{\kappa_{\hat \mu_0}}{\kappa_{\hat \mu}}D \cdot J_0 \cdot J_{\hat \mu_0}  = \kappa^+_{\hat \mu_0} D^2 \cdot J_{\hat \mu_0} \,, \qquad \text{for some}\quad \hat \kappa^+_{\hat \mu_0} >0  \,,
\eeq
where $\hat \mu_0 \in \cI_2$ is an arbitrary index that we may choose to be one of those that obey~\eqref{mu0-exist}. That is,
\beq
b_{\hat \mu_0} \succ \la^{-1/2} \,.
\eeq
Note that in obtaining the relations~\eqref{DhatDJ0}, the expression~\eqref{Dhat} for $\hat D$ and the proportionality~\eqref{Dhatsquare} have been used in the first step, and in the second step, the fact that $D \cdot J_0$ and $D^2$ are proportional to each other, 
\beq\label{DJ0-D2}
D\cdot J_0 = \kappa_D D^2 \,, \qquad \kappa_D >0 \,.
\eeq
Hence, 
\beq
\kappa_{\hat \mu_0}^+ = \kappa_D \sum_{\hat\mu\in\hat \cI_2}p_{\hat \mu}^+  \frac{\kappa_{\hat \mu_0}}{\kappa_{\hat \mu}} \,.
\eeq
The proportionality~\eqref{DJ0-D2} follows from Lemma~\ref{lemma2} with the vanishing intersection
\beq
D^2 \cdot J_0 = 0 \,,
\eeq
which is an obvious consequence of the assumption that the $T^2$ fiber associated with $D$ also shrinks in the limit as~\eqref{T2aswell}. As another consequence of the volume suppression~\eqref{T2aswell}, we also know that
\beq
D^2 \cdot J_{\hat \mu_0} = 0\,,
\eeq
which then leads, by~\eqref{DhatDJ0}, to
\beq\label{DhatDJ0-2}
D \cdot \hat D \cdot J_0 = 0 \,.
\eeq
Here, the two nef divisors $D$ and $\hat D$ do intersect non-trivially by Lemma~\ref{lemma1} and they are not proportional to $J_0$ either, since their squares are non-trivial. Therefore, by Lemma~\ref{lemma2}, we have 
\beq
D \cdot J_0 = \kappa_{D\hat D} \hat D \cdot J_0 \,,\qquad \text{for some}\quad \kappa_{D \hat D} >0\,, 
\eeq
and hence,
\beq\label{DDhatDhatD}
D \cdot D = \frac{1}{\kappa_D} D \cdot J_0 = \frac{\kappa_{D\hat D}}{\kappa_D} \hat D \cdot J_0 = \frac{\kappa_{D\hat D}}{\kappa_D} \kappa_{\hat D} \hat D\cdot \hat D \,,
\eeq
where in the first step we used~\eqref{DJ0-D2} and the last step, the similar relation
\beq
\hat D \cdot J_0 = \kappa_{\hat D} \hat D^2 \,, \qquad \kappa_{\hat D} > 0\,.
\eeq
This follows, for the same reason as for~\eqref{DJ0-D2}, from the vanishing of $\hat D^2 \cdot J_0$, which is necessary for the shrinking of $\hat T^2$. 

From the proportionality of $D\cdot D$ and $\hat D \cdot \hat D$, as established in~\eqref{DDhatDhatD}, we finally conclude that the fiber $\hat T^2$ associated with $\hat D$ is indeed the unique $T^2$ fiber that shrinks at a rate faster than $\la^{-1/2}$, and hence, in particular that the leading shrinking $T^2$ fiber is unique.

\section{Mirror Symmetry} \label{app_mirror}

We  recall some well-known facts with emphasis on quantum volumes and BPS masses.
One basic tenet of mirror symmetry \cite{Cox:2000vi,Hori:2003ic}  
is that the quantum corrected K\"ahler moduli space, ${\QMK}(Y)$, pertaining to some
 Calabi-Yau three-fold, $Y$, is  equivalent to the classical moduli space, ${\MCS}(X)$, of complex structures of the mirror Calabi-Yau three-fold, $X$.
This implies a map between even-dimensional and middle-dimensional cohomologies on $Y$ and $X$, 
$H^{2p}(Y)  \leftrightarrow H^3(X)$, $p=0,..,3$, and an analogous map
between integral, symplectic bases of the even-dimensional and the middle-dimensional cycles of $Y$ and $X$, respectively:
\be\label{mirrorcycles}
C\in H_{2p}(Y)    \qquad    \xleftrightarrow[\text{map}]{ \text{mirror} }   \qquad \gamma   \in H_3(X)   \,.
\ee
This allows to compute the exact BPS masses of D2p-branes that wrap holomorphic cycles $C_{2p}$ in Type IIA string theory on $Y$ 
even deep in the regime of quantum geometry, by mapping these to
 D3-branes wrapping special Lagrangian 3-cycles in Type IIB string theory on $X$. 
 
Specifically, consider first the mirror three-fold $X$, equipped with K\"ahler form $J_X$ and holomorphic 3-form $\Omega_X$.
In terms of an integral symplectic basis $\Gamma=(\gamma^A, \gamma_B)^T$ of $H_3(X,\mathbb Z)$ with
polarization
\be\label{gammadelta}
\gamma^A \cap \gamma_B = \delta^A_B \,,
\ee
one defines the periods $\Pi=(X^A,F_B)^T$ via
\be\label{periods}
X^A: = \int_{\gamma^A} \Omega_X \,, \qquad \quad F_B: = \int_{\gamma_B} \Omega_X \,.
\ee
In terms of the Poincar\'e dual basis of $H^3(X)$ defined by
\be
\int_{\gamma^A} \alpha_B = \delta^A_B \,, \qquad \quad \int_{\gamma_B} \beta^A = - \delta^A_B
\ee
the 3-form hence enjoys the expansion
\be
\Omega_X = X^A \alpha_A -  F_B \beta^B \,.
\ee
The periods determine the BPS masses of D3-branes wrapping special Lagrangian 3-cycles
 $\Gamma = n^A \gamma_A + n_B \gamma^B$ as follows:
\bea
\frac{M_\Gamma}{M_{\rm Pl}} =  \sqrt{\frac{2}{\pi}}\frac{ |Z_\Gamma|}{ { (i \int_X  \Omega_X \wedge \bar \Omega_X)^{1/2}} } \,,  
\eea
where the central charge is given by
\be
Z_\Gamma = \int_\Gamma \Omega = n_A X^A + n^B F_B \,.
\ee
This formula originates from the calibration condition 
\be \label{calibration}
{\cal V}_\Gamma = \int_\Gamma \sqrt{{\rm det}(g)} =  \frac{ (8 \, {\cal V}_X)^{1/2}}{(i \int_X \Omega_X \wedge \bar \Omega_X)^{1/2}} \,   {\rm Im} \, \int_\gamma  e^{-i \theta} \Omega_X 
\ee
for the volume of a special Lagrangian 3-cycle $\Gamma$, together with the relation $\frac{M^2_{\rm Pl}}{M_s^2} = \frac{4 \pi}{g^2_{\rm IIB}} {\cal V}_X$ with ${\cal V}_X = \frac{1}{6} \int_X {J_X}^3$.\footnote{Here we are assuming that we are working at large K\"ahler volume of $X$ so that the classical expression for ${\cal V}_X$ is valid. The angle $\theta$ depends on the precise calibration of the special Lagrangian.}

Now we switch to the K\"ahler picture via mirror symmetry. 
{\it A priori} the mirror map is defined near a large complex structure point (LCS) of $X$, which we put
at the origin of ${\MCS}(X)$ by suitable choice of coordinates.
Near such a point of unipotent monodromy, the periods $X^A$ 
split into a unique power series plus single logs, ie., $X^A = (X^0, X^a)$, \, $a = h^{2,1}(X)$, which we can take as
\be \label{splitXvar}
X^0(z) = 1 + {\cal O}(z^a) \,, \qquad \quad X^a(z) = \frac{1}{2 \pi i} \, {\rm log}(z^a) + {\cal O}(z^a) \,, \qquad \text{near }\,\,   z=0 \,.
\ee
Then one can define inhomogenous, flat coordinates on $\MCS$ as follows:
\be\label{flatco}
t^a(z)  =  \frac{X^a}{X^0} = \frac{1}{2 \pi i} \,  {\rm log}(z^a) + {\cal O}(z^a) \,.
\ee
The statement of mirror symmetry is that these flat coordinates near $z=0$
coincide with the classical, complexified  K\"ahler parameters, 
\bea \label{tadefinition}
t^a(z\rightarrow0) &=& \int_{C_2^a}  {\bf J}_Y :=   \int_{C_2^a} (B + i  J_Y) \ , \qquad C_2^a \in H_{2}(Y)\,,\\
&=&i T^a+ \int_{C_2^a} B\,,\nn
\eea
near the large volume limit of the mirror manifold, $Y$. Away from $z=0$, the
$t^a(z)$  then define, via analytic continuation, the (multi-valued) coordinates over the full quantum K\"ahler moduli space, $\QMK(Y)$.

In the regime of large $t^a=t^a(z)$, where classical geometry applies, the prepotential of $N=2$ special geometry
can be written as 
\bea \label{calFcoeff}
 {\cal F} &=& (X^0)^2 \,F(t^a)   \,,    \nn\\
F(t^a)  &=& - \frac{1}{6} d_{a b c} t^a t^b t^c + d_{a b} t^a t^b + c_a t^a   + c +
 {\cal O}(e^{2 \pi i t^a})\,
\eea
where in terms of a dual basis $\{D^a\}$ of $H^2(Y,\mathbb Z)$
\be
d_{a b c} = D_a \cdot D_b \cdot D_c \,, \qquad c_a = \frac{1}{24} c_2(Y) \cdot D_a\,,   \qquad c = \frac{i \zeta(3)}{(2\pi)^3}\chi(Y) \,.
\ee
The $d_{a b}$ are in general not specified geometrically but can be determined (up to monodromy) by
requiring good symplectic transformation properties of the period vector
\be \label{Pivector}
\Pi(t) = \left(\begin{matrix} X^0 \\ X^a\\   \partial_a  {\cal F}  \\  \partial_0  {\cal F} \end{matrix} \right)  = X^0  \left(\begin{matrix} 1 \\ t^a\\   \partial_a {F}  \\  2 {F} - t^a \partial_a{F} \end{matrix} \right) \,.
\ee
Via the correspondence (\ref{mirrorcycles}), the components of $\Pi$ can be
 interpreted as the complexified quantum volumes of the 0-cycle $C_0$, the 2-cycles $C^a_2$, the dual divisors $D_a\equiv C_{4,a}$ and the 6-cycle $C_6$, respectively, as probed by respective D2p-branes that can wrap them.
At large K\"ahler volume, the leading terms in each entry indeed matches the classical K\"ahler volume with respect to $J_Y$, while the subleading terms  reflect the  lower-brane charges induced on D-branes that wrap curved cycles on $Y$. Importantly, away from the large K\"ahler regime, the analytic continuation of the period vector then defines the quantum volumes of the respective cycles over the full moduli space.

Note that at an arbitrary point in moduli space, the quantum volume of the 6-cycle as defined in this way
is not identical to the volume of the Calabi-Yau $Y$.\footnote{For example, the quantum volume of the 6-cycle 
may stay finite while a formally lower-dimensional cycle blows up, so it is not a good measure for volume. The expression (\ref{VYqvolume}) treats all quantum cycle volumes on equal footing, apart from singling out a frame at large volume via the choice of $X^0$. See 
also the main text. }
 The latter is globally defined via mirror symmetry as  
 \be \label{VYqvolume}
{\cal V}_Y(z)  = \frac18 \frac{ i\int_{X} \Omega_X \wedge \bar\Omega_X}{ |X^0|^2}(z) \,,
\ee
in terms of the coordinates on $\MCS(X)$.
Only in the regime of large K\"ahler  volume of $Y$, which corresponds to $z\simeq0$,
does this coincide with the volume of the 6-cycle.
This can be easily checked with the help of 
\be
i \int_X \Omega_X \wedge \bar \Omega_X = i |X^0|^2 \left(2 (F- \bar F) - (t^a - \bar t^a) (\partial_a F + \partial_a \bar F) \right)
\ee

 Indeed it is (\ref{VYqvolume})
which globally defines the Planck mass in terms of the 10d string scale as
\be
\frac{M^2_{\rm Pl}}{M^2_s}(z) = \frac{4 \pi}{\gA^2} {\cal V}_Y(z)  = \frac{4 \pi i}{ \gA^2}  \frac{\int_{X} \Omega_X \wedge \bar\Omega_X}{8 | X^0|^2}(z) \,.
\ee
Correspondingly, we have two notions of masses of BPS states, depending on whether we normalize them with respect to the 10d string scale, $M_s$, or the four-dimensional Planck scale. Explicitly, 
the BPS masses for a bound state $n_0$ D0-branes, $n_{2,a}$ D2-branes along the curves $C_2^a$, $n^a_4$ D4-branes along $C_{4,a}$ and $n_6$ D6-branes wrapping the entire Calabi-Yau are respectively given as
\bea\label{BPStension}
\frac{M_{n_{2p}}}{M_s}           &=& \frac{1}{g_{\rm IIA}} \left| n_0 + n_{2,a} t^a + n^a_4 \partial_a { F} + n_6 (2 { F} - t^a \partial_a{ F})   \label{M2p1}     \right| \,, \\
\frac{M_{n_{2p}}}{M_{\rm Pl}} &=& \sqrt{\frac{2}{{\pi}} }  \frac{|X^0|}{(i \int_{X} \bar\Omega \wedge \Omega )^{1/2}}    \left| n_0 + n_{2,a} t^a + n^a_4 \partial_a { F} + n_6 (2 {F} - t^a \partial_a{F})    \label{M2p2}    \right| \,.
\eea

\section{Analytic Continuation for the K3 fibration  $\IP^5_{1,2,2,2,6}[12]$  } \label{App_K3quantum}

In Section~\ref{sec_QK3IIA} we consider as example the K3-fibered Calabi-Yau three-fold in $\IP^5_{1,2,2,2,6}[12]$,
which has been extensively studied over the years, starting from ref.~\cite{Candelas:1993dm} to which we refer for details.
Here we present some further aspects of the analytic continuation of the periods from the large complex structure limit
($L_1$ in Figure~\ref{f:Fig1MCS})), to the point of classically vanishing fiber volume, $L_2$.  
This specific analytic continuation had been discussed in the context of deriving
 the Seiberg-Witten model for gauge group $SU(2)$ \cite{Kachru:1995fv}, and was worked out in refs.~\cite{LMunpubl,Curio:2000sc}.

For a convenient patch near $L_1$ in complex structure moduli space, parametrized by $z_{1,2}$, 
the three-fold under consideration is defined by the vanishing of
$$
W_Y={x_1}^{12}+{x_2}^{12}+{x_3}^{6}+{x_4}^{6}+{x_5}^{2}+ {z_1}^{-1/6}{z_2}^{-1/12} x_1x_2x_3x_4x_5+
 {z_2}^{-1/2}{x_1}^{6}{x_2}^{6}\,.
$$
In this patch the Picard-Fuchs system comprises the following differential operators:
\bea \label{PFops}
\CL_1&=&{\theta_1}^2(\theta_1-2 \theta_2)-8z_1(6 \theta_1+5)(6 \theta_1+3)(6 \theta_1+1)\,,
\\
\CL_2&=&{\theta_2}^2-z_2(2 \theta_2-\theta_1+1)(2 \theta_2-\theta_1)\,, \qquad  \theta_i\equiv z_i\partial_i\,.\nn
\eea
The solutions are well-known \cite{Candelas:1993dm,Hosono:1993qy} and consist of the unique power series
solution, $X^0\sim 1+120 z_1+\CO(z) $, plus further linear, quadratic and cubic logarithmic solutions. 
Expressing these in terms of the flat coordinates (\ref{flatco}),
one can asymptotoically match appropriate linear combinations against the integral symplectic basis of periods defined by
(\ref{Pivector}), where 
\be\label{FK3}
F= -\frac23 {t_1}^3-{t_1}^2t_2+b_1t_1+b_2t_2+ \xi+\CO(e^{-t})\,,
\ee
with  $b_i=\frac1{24}\int  c_2\wedge J_i$ so $b_1=\frac{13}6$, $b_2=1$, and 
$\xi=\frac i{2(2\pi)^3}\chi\zeta(3)$ for $\chi=-252$.

The conifold locus is given by the vanishing of $\Delta_c=(1728z_1-1)^2-4(1728z_1)^2z_2$, and we will be 
interested in the intersection of it with the locus $z_2=0$ of large base $\IP^1_b$.
In order to have well-defined normal crossings of divisors at $L_2: z_1=1/1728,  z_2=0$, one needs to blow up the point of 
quadratic tangency by introducing suitable coordinates \cite{Kachru:1995fv}
\bea
{\hat z}_1 &=& 1-1728 z_1\,,\\
{\hat z}_2 &=& \frac{4z_2{1728 z_1}^2 }{(1-1728 z_1)^2}\,.\nn
\eea
In terms of these, the Picard-Fuchs system reads
\bea \label{PFops2}
\hat \CL_1&=&
5 {\hat z}_1{}^2+2 \left(113 {\hat z}_1-77\right) {\hat z}_1{}^2 {\hat\partial}_1+108 \left({\hat z}_1-1\right) \left(3 {\hat z}_1-1\right) {\hat z}_1{}^2
   {\hat\partial}_1{}^2+72 \left({\hat z}_1-1\right){}^2 {\hat z}_1{}^3 {\hat\partial}_1{}^3
\nn   \\ &&
   +4 {\hat z}_2 \left(5 {\hat z}_1-54\right)
   {\hat\partial}_2+288 {\hat z}_2 {\hat z}_1 \left(1+{\hat z}_1\right){\hat\partial}_2 {\hat\partial}_1+288 {\hat z}_2
   \left({\hat z}_1-1\right) {\hat z}_1{}^2 {\hat\partial}_2 {\hat\partial}_1{}^2-144 {\hat z}_2{}^2 {\hat\partial}_2{}^2
 \nn  \\&&
   +288 {\hat z}_1 {\hat z}_2{}^2 {\hat\partial}_1 {\hat\partial}_2{}^2\,,
\\
\hat \CL_2&=&
{\hat z}_1{}^2 {\hat\partial}_1{}^2+2 \left(3 {\hat z}_2-2\right) {\hat\partial}_2-4 {\hat z}_2 {\hat z}_1 {\hat\partial}_2
   {\hat\partial}_1{}+4 \left({\hat z}_2-1\right) {\hat z}_2 {\hat\partial}_2{}^2\,,\nn
\eea
where ${\hat\partial}_i\equiv\partial_{\hat z_i}$. 
With hindsight we pick the following basis of solutions:
\bea \label{hatPisolution}
\hat{\Pi}_1 &=& 
\frac1\pi\left(1+\frac{5 {\hat z}_1}{36}+\frac{295 \left(2+{\hat z}_2\right) {\hat z}_1{}^2}{7776}\right)
+\CO({\hat z}^3),\nn\\
\hat{\Pi}_2 &=& 
\frac1\pi\left({\hat z}_1+\frac{77}{216} \left(2+{\hat z}_2\right) {\hat z}_1{}^2\right)
+\CO({\hat z}^3),\nn\\
\hat{\Pi}_3 &=& 
-\frac{\sqrt{{\hat z}_1}}\pi \left(1-\frac{{\hat z}_2}{16}+\frac{23}{864} \left(16+3 {\hat z}_2\right) {\hat z}_1-\frac{15 {\hat z}_2{}^2}{1024}\right)
+\CO({\hat z}^3),\\
\hat{\Pi}_4 &=& 
\frac 1{i\pi}\left(\hat{\Pi}_3(\log({\hat z}_2)-6\log2+3)-\frac{\sqrt{{\hat z}_1}}\pi
\left(4-\frac{{\hat z}_2}{8}-\frac{47{\hat z}_2{}^2}{1024}\right)\right)
+\CO({\hat z}^3),\nn\\
\hat{\Pi}_5 &=& 
\frac i {2\pi}\left(
\hat{\Pi}_1\log({{\hat z}_1{}^2\hat z}_2)+\frac1\pi\left(5+\frac{25 {\hat z}_1}{36}+\frac{2597 {\hat z}_1{}^2}{11664}+\frac{827 {\hat z}_2 {\hat z}_1{}^2}{23328}\right)\right)
+\CO({\hat z}^3),\nn\\
\hat{\Pi}_6 &=& 
\frac i {2\pi}\left(
\hat{\Pi}_2\log({{\hat z}_1{}^2\hat z}_2)+\frac1\pi\left({\hat z}_1+\frac{787 {\hat z}_1{}^2}{324}+\frac{325}{648} {\hat z}_2 {\hat z}_1{}^2\right)\right)
+\CO({\hat z}^3).\nn
\eea      
The analytic continuation of the basis of integral symplectic periods from $L_1$ to $L_2$ can then be determined \cite{LMunpubl} by
matching expansions along the fixed locus of the involution $L_1\leftrightarrow L_2$:
$$
\hat\Pi\big\vert_{\substack{\hat z_1=1/2\\ \hat z_2=4 z_2\phantom{i}}} = \CN\cdot  {\Pi}(z_2)\big\vert_{\substack{z_1=1/2(1728)^{-1}}}\,,
 $$
and we find as a result
 \bea \label{calNmatrix}
 \CN = 
\left(
\begin{array}{cccccc}
 0 & -i X & 0 & \frac{X}{2} & 0 & 0 \\
 0 & \frac{i}{X} & 0 & \frac{1}{2 X} & 0 & 0 \\
 2 & 0 & 0 & -1 & 0 & 0 \\
 0 & 0 & 0 & 0 & 0 & \frac{1}{2} \\
 0 & \xi _2 & -X & i \xi _1 & -\frac{i X}{2} & \frac{X}{2} \\
 0 & \xi _3 & -\frac{1}{X} & i \xi _4 & \frac{i}{2 X} & \frac{1}{2 X} \\
\end{array}
\right)\,.
\eea
Here,
 $$
X= \frac{ \Gamma \left(\frac{3}{4}\right)^4}{\sqrt{3} \pi ^2}, \ \ \
 \xi_1=\left(\xi _4-\frac{\xi _3}{2}\right) X^2-\frac{\xi _2}{2}\,,
  $$
 and
$\xi_2\approx2.251234070$, $\xi_3\approx -23.36100861$, $\xi_4\approx3.346738000$ are numerical constants
that are not important for us. The important feature is the third row of $\CN$, which specifies the parti\-cular
linear combination of integral periods that vanishes at $L_2$. 
Note that the integers arise from a non-trivial resummation in the process of analytical continuation.
 
The map $\CN$ preserves the symplectic inner product. As quick consistency check, 
note that the monodromies induced from encircling ${\hat z}_i=0$ are:
$$
M_1=\left(
\begin{array}{cccccc}
 1 & 0 & 0 & 0 & 0 & 0 \\
 0 & 1 & 0 & 0 & 0 & 0 \\
 0 & 0 & -1 & 0 & 0 & 0 \\
 0 & 0 & 0 & -1 & 0 & 0 \\
 -2 & 0 & 0 & 0 & 1 & 0 \\
 0 & -2 & 0 & 0 & 0 & 1 \\
\end{array}
\right)
,\qquad
 M_2=\left(
\begin{array}{cccccc}
 1 & 0 & 0 & 0 & 0 & 0 \\
 0 & 1 & 0 & 0 & 0 & 0 \\
 0 & 0 & 1 & 0 & 0 & 0 \\
 0 & 0 & 2 & 1 & 0 & 0 \\
 -1 & 0 & 0 & 0 & 1 & 0 \\
 0 & -1 & 0 & 0 & 0 & 1 \\
\end{array}
\right)\,.
$$
Indeed one finds that $M_\infty^{SW}=M_1\cdot{M_2}^{-2}$ 
correctly reproduces the semi-classical monodromy of the Seiberg-Witten model \cite{Seiberg:1994rs}.
Note that in contrast to $M_1$, $M_2$ is unipotent, which signals \cite{Grimm:2018ohb,Grimm:2018cpv,Corvilain:2018lgw}
 the long distance limit for the degeneration ${\hat z}_2\rightarrow0$.

\end{appendix}

\newpage

\bibliography{papers}
\bibliographystyle{JHEP}

\end{document}